\documentclass[onecolumn,12pt]{article}
\usepackage{jheppub}
\usepackage[font={small}]{caption}
\usepackage{subfigure}
\usepackage{graphicx}% Include figure files
\usepackage{dcolumn}% Align table columns on decimal point
\usepackage{float}
\usepackage[normalem]{ulem}

\usepackage{mathrsfs} 
\usepackage{multicol}
\usepackage{cancel}
\usepackage{mathtools}
\usepackage{hyperref}

\usepackage{subfigure}
\usepackage{xcolor}
\def\({\left(} \def\){\right)}
\def\[{\left[} \def\]{\right]}

\newcommand{\eg}{{\it e.g.,}\ }
\newcommand{\ie}{{\it i.e.,}\ }

\newcommand{\beq}{\begin{equation}}
\newcommand{\eeq}{\end{equation}}
\newcommand{\bea}{\begin{eqnarray}}
\newcommand{\eea}{\end{eqnarray}}

\def\le{\left(}
\def\ri{\right)}

\renewcommand{\eqref}[1]{(\ref{#1})}

\begin{document}

\title{The Cosmological Switchback Effect}

\author{Stefano Baiguera$^{1}$, Rotem Berman$^{1}$, Shira Chapman$^{1}$, Robert C. Myers$^2$}

\affiliation{$^1$Department of Physics, Ben-Gurion University of the Negev, \\ David Ben Gurion Boulevard 1, Beer Sheva 84105, Israel \\
$^2$Perimeter Institute for Theoretical Physics, \\
Waterloo, ON N2L 2Y5, Canada}

\emailAdd{baiguera@post.bgu.ac.il}
\emailAdd{bermar@post.bgu.ac.il}
\emailAdd{schapman@bgu.ac.il}
\emailAdd{rmyers@perimeterinstitute.ca}

\abstract{\sloppy The volume behind the black hole horizon was suggested as a holographic dual for the quantum computational complexity of the boundary state in AdS/CFT. This identification is strongly motivated by the switchback effect: a characteristic delay of complexity growth in reaction to an inserted perturbation, modelled as a shockwave in the bulk. 
Recent proposals of de Sitter (dS) holography suggest that a dual theory could be living on a stretched horizon near the cosmological horizon. 
We study how the spacetime volume behind the cosmological horizon in Schwarzschild-dS space reacts to the insertion of shockwaves in an attempt to characterize the properties of this dual theory. We demonstrate that a switchback effect can be observed in dS space. That is, the growth of complexity is delayed in reaction to a perturbation. This delay is longer for earlier shocks and depends on a scrambling time which is logarithmic in the strength of the shockwave and proportional to the inverse temperature of the cosmological dS horizon. This behavior is very similar to what happens for AdS black holes, albeit the geometric origin of the effect is somewhat different.}

\maketitle

\setcounter{tocdepth}{1}

\section{Introduction}
\label{sec:introduction}

The holographic principle states that the physical description of certain gravitational systems is encoded in properties of their boundary \cite{tHooft:1993dmi,Susskind:1994vu}. 
Many developments in quantum information and gravity support this statement, starting from the Bekenstein-Hawking entropy associated with the horizon area of a black hole \cite{Bekenstein:1972tm,Hawking:1975vcx}.
 A concrete framework for holography was set in the context of the AdS/CFT correspondence \cite{Maldacena:1997re}.
This framework provides a precise relation between quantum gravity in Anti-de Sitter (AdS) space and a conformal field theory (CFT) living on its boundary.
Despite the success and importance of this program, it would be desirable to investigate the existence of a holographic description of spacetimes with a positive cosmological constant, which models the universe where we live.
In this work, we will pursue this task by studying quantum information properties of asymptotically de Sitter (dS) spacetimes, following the program recently developed in \cite{Susskind:2021esx,Shaghoulian:2021cef,Shaghoulian:2022fop,Jorstad:2022mls}. 

\paragraph{Holographic nature of dS space.}
The interpretation of area as a thermodynamic entropy also applies to geometries with a cosmological horizon, as first pointed out by Gibbons and Hawking \cite{PhysRevD.15.2738}.
More precisely, the region causally accessible to an observer moving along a timelike worldline in dS space is delimited by past and future cosmological horizons, which have an associated Hawking temperature and horizon entropy \cite{PhysRevD.15.2738} given by
\beq
T_{\rm dS} = \frac{1}{2 \pi L} \, , \qquad
S_{\rm dS} = \frac{A}{4 G_N} = \frac{\Omega_{d-1} L^{d-1}}{4 G_N} \, ,
\eeq
where $L$ is the curvature scale and $\Omega_{d-1}=2\pi^{d/2}/\Gamma(d/2)$ is the volume of the $(d-1)$--dimensional unit sphere. Here and in the following, we refer to de Sitter space in $d+1$ dimensions, to facilitate the comparison with results in anti-de Sitter space where this convention for the dimension is very common, with $d$ specifying the dimension of the boundary theory.

The notion of entropy in statistical mechanics is related to the counting of the microstates of a system.
In the context of black holes, we may ask ourselves, what does their entropy count? 
The \emph{central dogma} of black holes \cite{Almheiri:2020cfm}
states that black holes can be described from the outside as quantum systems containing $A/4G_N$ degrees of freedom that evolve unitarily in time.
A foremost achievement of string theory was the verification of this statement in the case of certain extremal black holes \cite{Strominger:1996sh}.

In this regard, cosmological horizons are more puzzling than event horizons of black holes, because they are observer-dependent.
While one may expect that a black hole is a localized object with quantum microstates and an observer can be located unambiguously at infinity, instead
in spacetimes with positive cosmological constant, different observers can influence different regions of spacetime, and they even experience different horizons.
The importance of role of the observer was recently highlighted in \cite{Chandrasekaran:2022cip,Witten:2023qsv}.

The region in causal contact with an observer in dS space is called \emph{static patch}.
The proposed extension of the central dogma to dS 
is that the cosmological horizon can be described from the inside as a unitary quantum system with $\exp(A/4G)$ states 
\cite{Bousso:1999dw,Banks:2000fe,Bousso:2000nf,Banks:2001yp,Banks:2002wr,Parikh:2002py,Dyson:2002nt,Dyson:2002pf,Banks:2005bm,Banks:2006rx,Anninos:2011af,Banks:2018ypk,Banks:2020zcr,Susskind:2021omt,Susskind:2021dfc,Susskind:2021esx,Shaghoulian:2021cef,Shaghoulian:2022fop}.
The consequence is that the quantum mechanical theory dual to dS space has a finite number of degrees of freedom and it is described by a maximally mixed state \cite{Dong:2018cuv,Lin:2022nss,Chandrasekaran:2022cip}.
This theory is typically assumed to live on a constant radial slice surface, right inside the cosmological horizon. We refer to the idea described here as \emph{static patch holography}.
In three bulk dimensions, a concrete dual theory was formulated in terms of a $T\bar{T}$ deformation plus a cosmological constant term \cite{Lewkowycz:2019xse}. This setting allows for a counting of the microstates encoded by the cosmological horizon \cite{Coleman:2021nor}, in agreement with \cite{Anninos:2020hfj}. Generalizations to other dimensions have been discussed in \cite{Silverstein:2022dfj}.

Let us briefly mention that there are a few other approaches to dS holography.
One is the so-called dS/CFT correspondence, that relates quantum gravity in
asymptotically $(d+1)$--dimensional de Sitter (dS) spacetimes to a Euclidean $d$--dimensional field theory living at timelike infinity \cite{Strominger:2001pn}. 
Alternatively, one can probe cosmological horizons by embedding dS space in a spacetime with a timelike boundary, where the tools of AdS/CFT correspondence can still be applied \cite{Freivogel:2005qh,Lowe:2010np,Fischetti:2014uxa,Anninos:2017hhn,Anninos:2018svg,Mirbabayi:2020grb,Anninos:2020cwo}.
These models are called \emph{flow} or \emph{centaur} geometries. 
In this work, we will focus on the program of static patch holography.

\paragraph{Holographic complexity.}
The profound relation between entanglement entropy and geometry, initiated with the seminal work by Ryu and Takayanagi \cite{Ryu:2006bv}, was recently extended to static patch holography in dS space \cite{Susskind:2021esx,Shaghoulian:2021cef,Shaghoulian:2022fop,Sybesma:2020fxg,Aalsma:2021bit,Aalsma:2022swk}.
It is natural to consider a similar extension for holographic complexity, which was proposed as the dual quantity associated with the time evolution of the Einstein-Rosen bridge (ERB) \cite{Susskind:2014moa,Stanford:2014jda,Susskind:2014rva,Brown:2015bva,Brown:2015lvg} (see \cite{Chapman:2021jbh} for a review).
Quantum computational complexity is a well-known quantity used in quantum information science \cite{https://doi.org/10.48550/arxiv.0804.3401,Aaronson:2016vto}. Heuristically, it counts the minimal number of unitary operations necessary to move from a reference state to a given target state. 
In holography, complexity measures the difficulty to create a target state in the boundary theory starting from a product (\ie unentangled) state.
The two main proposals for the dual quantity on the gravity side are the Complexity=Volume (CV) and Complexity=Action (CA) conjectures.
The former associates complexity to the spatial volume of an extremal slice anchored to the boundary \cite{Susskind:2014moa,Susskind:2014rva,Stanford:2014jda}, while the latter evaluates the gravitational action in the Wheeler-De Witt (WDW) patch, \ie the causal domain of dependence of the ERB \cite{Brown:2015bva,Brown:2015lvg}. 
In mathematical terms:
\beq
\mathcal{C}_V  = \frac{\mathcal{V}_{\rm max}}{G_N \ell_{\rm bulk}} \, , \qquad\qquad
\mathcal{C}_A = \frac{I_{\rm WDW}}{\pi \hbar} \, ,
\eeq
where $\ell_{\rm bulk}$ is a certain length scale (usually identified with the cosmological constant curvature scale) and $I_{\rm WDW}$ the on-shell action of the WDW patch, including boundary terms \cite{Lehner:2016vdi}.
The CV2.0 conjecture is another proposal that associates complexity to the spacetime volume of the WDW patch \cite{Couch:2016exn}
\beq
\mathcal{C}_{2.0 V} = \frac{V_{\rm WDW}}{G_N \ell_{\rm bulk}^2} \, .
\eeq
This proposal is often easier to compute technically and gives similar characteristics to the CA and CV proposals. We have therefore focused our efforts on this approach in the present paper.
The different proposals for a dual quantity of complexity have been studied in different spacetimes with AdS asymptotics or deformations thereof (\eg see \cite{Lehner:2016vdi,Carmi:2016wjl,Chapman:2016hwi,Cai:2016xho,Reynolds:2016rvl,Carmi:2017jqz,Alishahiha:2018tep,Bolognesi:2018ion,Auzzi:2018pbc,Auzzi:2018zdu,Bernamonti:2019zyy,Goto:2018iay,Bernamonti:2021jyu,Auzzi:2022bfd}).
All three of these approaches can be seen as special cases of a much broader conjecture for holographic complexity which was recently proposed in \cite{Belin:2022xmt,Belin:2021bga,Jorstad:2023kmq}.

The previously mentioned attempts to extend holography to dS space naturally lead us to ask if holographic complexity can be of use in this context. Recently the complexity proposals have been studied using static patch holography in general dimensions for the empty dS background \cite{Jorstad:2022mls}.\footnote{ Complexity conjectures in dS have been proposed in \cite{Susskind:2021esx}.
Reference \cite{Anegawa:2023wrk} computed the volume and action conjectures in two-dimensional dS space, and \cite{Chapman:2021eyy} studied volume complexity in the framework of two-dimensional flow geometries. The setting with a bubble of dS space contained inside AdS geometry (introduced in \cite{Freivogel:2005qh}) was considered in \cite{Auzzi:2023qbm}. }
In this context, the maximal slice is anchored to the boundary of the static patch, the so-called \emph{stretched horizon}. The stretched horizon is the place where the holographic degrees of
freedom are conjectured to live and is typically taken to be a surface with constant radial coordinate close to the cosmological horizon.

\begin{figure}[ht]
    \centering
    \includegraphics[scale=0.5]{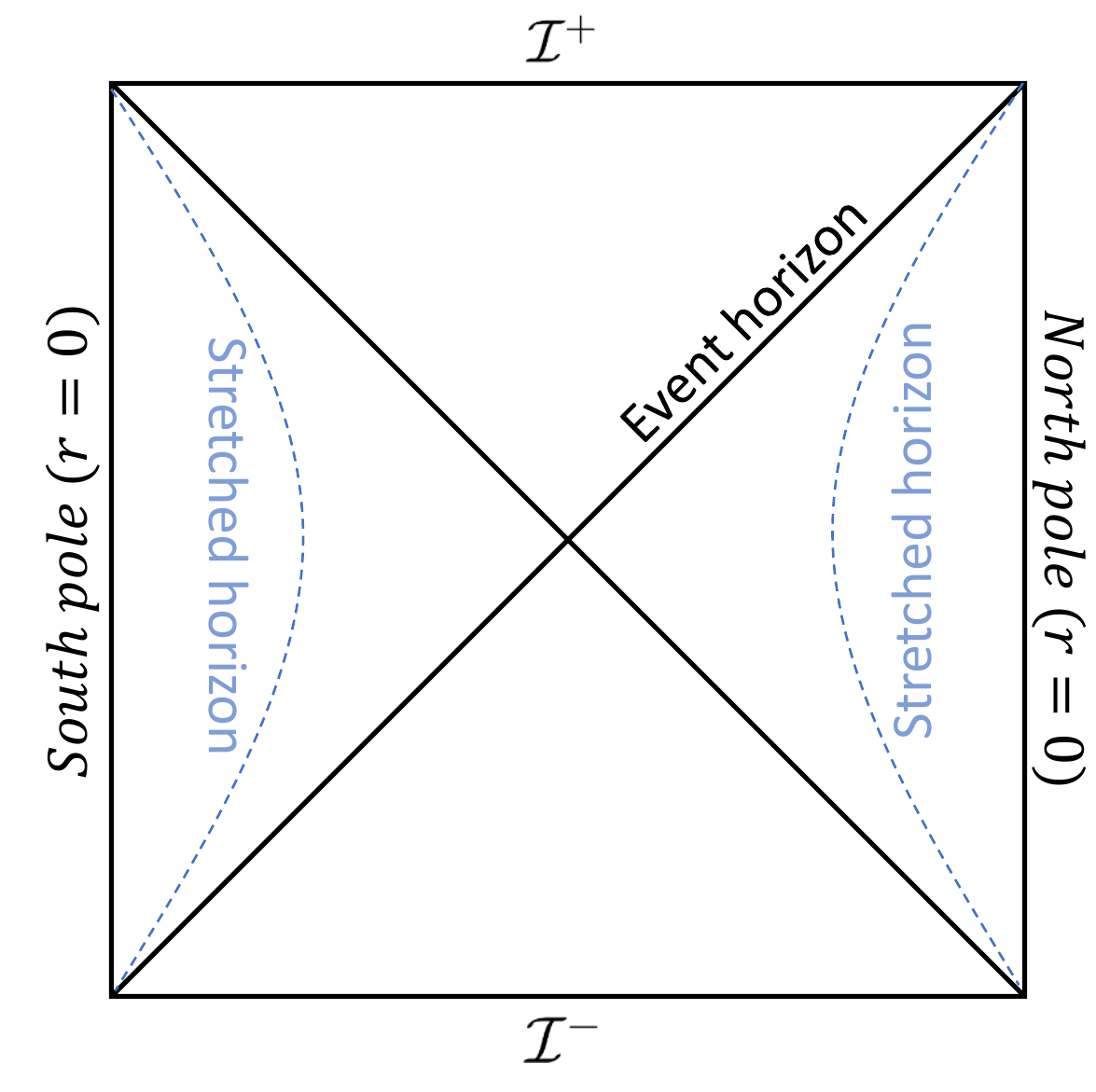}
    \caption{Penrose diagram of de Sitter space. The diagonal black lines represent the cosmological event horizons, and the horizontal black lines are the future and past timelike infinity $\mathcal{I}^{\pm}$ (at $r=\infty$). The vertical black lines represent the worldline of an observer, located at $r=0$ and the blue line represents the stretched horizon that we defined to be a constant r surface.
    }
    \label{fig:Penrose_dS_intro}
\end{figure}

As time progresses along the stretched horizon, the different geometric objects associated with complexity (for instance the volume, or the WDW patch) include contributions from spacetime regions near $\mathcal{I}^+$ and diverge.
This was already noticed in \cite{Susskind:2021esx}. One way to deal with this divergence is to introduce a regulator near future and past timelike infinity as shown by the dashed top and bottom lines in fig.~\ref{fig:WDW_patches}. 
In this way, one recovers the linear late-time behavior typical of quantum computational complexity in quantum circuits \cite{Jorstad:2022mls}.

One of the famous tests of holographic complexity, in the context of AdS black holes, is the switchback effect, which is a delay in the growth of complexity in a perturbed background including a shockwave \cite{Stanford:2014jda,Chapman:2018dem,Chapman:2018lsv}. In the boundary picture, this delay arises from cancellations between gates which commute with the perturbation at early times after its insertion \cite{Susskind:2014jwa} 

In this work, we will investigate the reaction of asymptotically de Sitter spacetimes to the insertion of a gravitational shockwave \cite{Hotta_1993,PhysRevD.47.3323,Sfetsos:1994xa}. 
We will try to understand how the shockwave influences the volume behind the dS horizon by focusing on the computation of CV2.0 only, since it is already a non-trivial geometric quantity that presents several non-trivial features of complexity in dS. In particular, we find that dS exhibits a delay in the growth of the volume, similar to the switchback effect seen for AdS black holes. This is perhaps a general property of geometries with shockwaves and therefore a general theorem may be formulated along the lines of \cite{Engelhardt:2021mju}.

While holographic complexity shows signatures  of scrambling here similar to AdS black holes, we anticipate that the geometric origin of this behaviour is very different in asymptotically dS spaces compared to the AdS case. An important role is played by the fact that the Penrose diagram of dS space grows taller when positive energy sources  are inserted in the bulk \cite{Gao:2000ga}. This leads to a time advance for null rays crossing the shockwaves \cite{Anninos:2018svg} and brings the two stretched horizons on the two sides of the dS spacetime into causal contact. The latter is crucial for the appearance of certain special configurations of the WDW patch discussed in section \ref{ssec:special_configurations_WDW}, and for the evaluation of the scrambling time.
It would be interesting to interpret the causal contact between the two sides of the Penrose diagram from a potential boundary quantum mechanics point of view. Perhaps some inspiration can be drawn from traversable wormholes in AdS \cite{Gao:2016bin,Maldacena:2017axo} where a similar situation occurs.

\paragraph{Outline.}
The paper is organized as follows.
We start in section \ref{sec:geometric_preliminaries} by
introducing the asymptotically dS spacetimes perturbed by shockwaves. This includes a review of the solutions and the analysis of their causal structure, including the check that the null matter supporting the shockwaves satisfies the null energy condition.
We propose in section \ref{sec:stretched_horizon_shocks} two prescriptions to define a stretched horizon in the shockwave geometries.
We define a regularized WDW patch in these geometries and we study its time evolution in section \ref{sec:WDW_patch}.
Then we present the general framework for computing complexity via the CV2.0 conjecture and its rate of change in section \ref{sec:CV20}. We apply both analytic and numerical methods to study the growth rate in specific cases in section \ref{sec:examples_CV20}.
Finally, we summarize and discuss our results and present possible future directions in section \ref{sec:discussion}.
The appendices contain additional technical details. 
For the sake of clarity, appendix \ref{app:comparison_jap} provides a  comparison of our conventions and results with those of reference \cite{Anegawa:2023dad}, which contains a related discussion of holographic complexity in three-dimensional dS spaces with shockwaves.
Appendix \ref{app:ping_pong} presents an alternative prescription to determine the location of the stretched horizon in shockwave geometries.
Appendix \ref{app:formation} provides an alternative derivation of the cosmological scrambling time through a computation of the complexity of formation.

\paragraph{Note added.}
During the preparation of this paper we became aware of \cite{Anegawa:2023dad}, which presents a related discussion on the complexity in the presence of shockwaves in de-Sitter space.

\section{Geometric preliminaries} 
\label{sec:geometric_preliminaries}

We review in section \ref{ssec:geometry_dS_spaces} the main aspects of asymptotically dS backgrounds, including both the cases of empty dS space and of the Schwarzschild-de Sitter (SdS) black holes.
Then we perturb in section \ref{ssec:perturbation_shocks} these geometries with the insertion of spherically symmetric shockwaves, which induce a transition between black holes with different masses. As we will see in section \ref{ssec:NEC}, the null energy condition (NEC) always implies that the mass of the black hole has to decrease after the shock.\footnote{Physically, this should be thought of as a spherically symmetric configuration, a star for instance, where some mass is ejected in a spherically symmetric manner into the cosmological horizon. As a consequence the cosmological horizon will recline from the observer.}

\subsection{Asymptotically de Sitter spacetimes}
\label{ssec:geometry_dS_spaces}

We consider the Einstein-Hilbert action in $d+1$ dimensions with positive cosmological constant
\beq
I = \frac{1}{16 \pi G_N} \int d^{d+1} x \, \sqrt{-g} \, \le R - 2 \Lambda \ri \, , \qquad
\Lambda = \frac{d(d-1)}{2 L^2} \, .
\label{eq:action_EOM_gend}
\eeq
The vacuum solutions of the Einstein's equations are described by the class of metrics  \cite{Spradlin:2001pw}
\beq
ds^2 = - f(r) dt^2 + \frac{dr^2}{f(r)} + r^2 d\Omega_{d-1}^2 \, ,  \qquad
f (r) = 1 - \frac{2 m}{r^{d-2}} - \frac{r^2}{L^2} \, ,
\label{eq:asympt_dS}
\eeq
where $L$ is the dS curvature radius and $m$ is a parameter related  to the mass of the solution, and we refer to $f(r)$ as the blackening factor. The case $m=0$ corresponds to empty dS space, while the case with $m\neq 0$ describes a Schwarzschild-de Sitter black hole. 

The previous coordinates only cover a portion of spacetime; the analytic extension behind the cosmological horizon can be achieved by introducing null coordinates
\beq
u = t - r^* (r) \, , \qquad
v = t + r^* (r) \, ,
\label{eq:general_null_coordinates}
\eeq
where the tortoise coordinate is defined in terms of the blackening factor as follows
\beq
r^*(r) = \int_{r_{0}}^r \frac{d r'}{f(r')} \, .
\label{eq:general_tortoise_coordinate}
\eeq
The lower limit of the integral $r_0$  corresponds to an arbitrary constant of integration. We will always select it such that $r^*(r\rightarrow \infty)=0$ (in 3d or in the absence of a mass, equivalently $r^*(r\rightarrow 0)=0$).\footnote{There is a small subtlety in integrating the tortoise coordinate across the horizons where the blackening factor vanishes. Typically the blackening factor can be decomposed as $1/f(r) = c_1 /(r-r_h)+ c_2 /(r-r_C)+\text{regular}$ where $r_h$ stands for the black hole horizon, $r_C$ stands for the cosmological horizon and $c_1, c_2$ are constants. When we integrate this expression, we could choose different integration constants on the two sides of each horizon. However, we will always fix the integration constants to be equal on the two sides in the sense that $r^*(r) = c_1 \log|r-r_h|+ c_2 \log|r-r_C|+\int \text{regular}$.} 
The metric \eqref{eq:asympt_dS} expressed in Eddington-Finkelstein (EF) coordinates reads
\beq
ds^2 = - f(r) du^2 - 2 du dr + r^2 d\Omega_{d-1}^2 \, .
\label{eq:generic_null_metric_noshock}
\eeq
The Penrose diagram is built using the following null compact coordinates
\beq
U = \tan \le e^{\frac{u}{L}} \ri \, , \qquad
V = - \tan \le e^{-\frac{v}{L}} \ri \, ,
\eeq
where the above convention refers to null coordinates defined in the right quadrant of the conformal diagram; the other regions are achieved by changing the overall sign of $U$ and $V$ in all the possible combinations, providing a natural Kruskal extension of spacetime.

The above blackening factor in eq.~\eqref{eq:asympt_dS} typically has one or two physical roots (\ie $f(r)$ vanishes at one or two positive values of the radius), defining the positions of the cosmological and black hole event horizons. We will refer to these values of the radial coordinate as $r_C$ and $r_h$, respectively ($r_h<r_C$). Due to the thermal emission from their respective horizons, both the black hole and cosmological horizons have an associated temperature and entropy, given by \cite{PhysRevD.15.2738} 
\beq
     T_{h(C)} = \frac{1}{4 \pi} \left|\frac{\partial f(r)}{\partial r}\right|_{r=r_{h(C)}}  
    \qquad
    S_{h(C)} =  \frac{\Omega_{d-1} r_{h(C)}^{d-1}}{4 G_N} \, .
\label{eq:temp_entropy_SdS_general}  
\eeq
Below we give explicit expressions for the tortoise and null coordinates as well as the thermodynamic quantities and a short description of the Penrose diagram for different choices of the metric mass parameter and dimension.

\subsubsection{Empty de Sitter space} 

\begin{figure}[ht]
    \centering
    \includegraphics[scale=0.5]{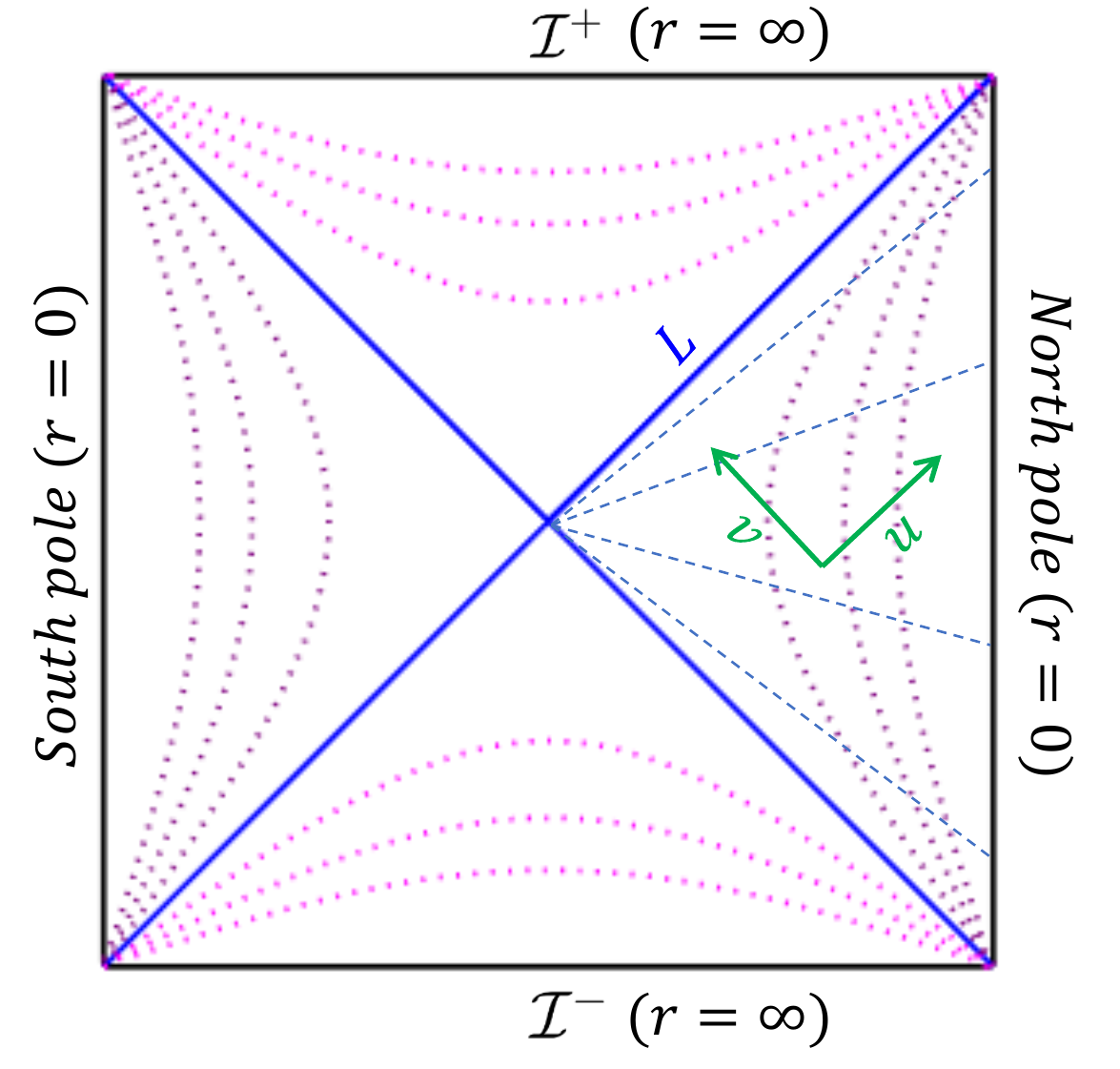}
    \caption{Penrose diagram of de Sitter space. The blue lines represent the cosmological horizons ($r=L$), the horizontal black lines are the future and past timelike infinity $\mathcal{I}^{\pm}$ (at $r=\infty$) 
    and the vertical black lines represent the north pole (right) and the south pole (left), located at $r=0$ along the world line of an observer.     }
    \label{fig:Penrose_dS}
\end{figure}

We start with empty de Sitter space in $d+1$ dimensions, which is a maximally symmetric solution of Einstein's equations with positive cosmological constant.
The topology of the geometry is $\mathbb{R} \times S^d,$ and the  Penrose diagram is depicted in fig.~\ref{fig:Penrose_dS}.

The worldline of an inertial observer in this geometry is represented by the vertical black lines on the right and left sides of the diagram, corresponding to the north and south poles (of the spatial $S^d$), respectively.
The metric in empty dS space corresponds to eq.~\eqref{eq:asympt_dS} with $m=0$, \ie in any number of dimensions the blackening factor reads
\beq
f(r) = 1 - \frac{r^2}{L^2} \, .
\label{eq:metric_dS_gend}
\eeq
The coordinate system \eqref{eq:asympt_dS} describes the static patch of de Sitter space, which is the region in causal contact with an inertial observer at the north (or south) pole, located at $r=0.$
The geometry presents a cosmological horizon at $r=L,$ as a consequence of the positive cosmological constant. 
The dS background has a non-vanishing global temperature and entropy associated with the cosmological horizon. From eq.~\eqref{eq:temp_entropy_SdS_general},  they read
\beq
T_{\rm dS} = \frac{1}{2 \pi L} \, , \qquad
S_{\rm dS} = \frac{\Omega_{d-1} L^{d-1}}{4 G_N} \, .
\label{eq:Hawking_temperature_dS}
\eeq
The tortoise coordinate \eqref{eq:general_tortoise_coordinate} in empty dS space reads
\beq
r^*(r)  =  \frac{L}{2} \log \left| \frac{L+r}{L-r} \right| \, .
\label{eq:rstar_dS}
\eeq
This quantity is used to define the null coordinates \eqref{eq:general_null_coordinates} and build the causal diagram in fig.~\ref{fig:Penrose_dS}.
In the figure, we depicted several curves at constant radial or constant time coordinate. The constant radial coordinate is depicted both outside and inside the cosmological horizon. 
At some large $r$, the former will be used as a cutoff surface that regularizes geometric quantities reaching the timelike infinities $\mathcal{I}^{\pm}$; the latter will be related to the location where the holographic degrees of freedom live, as we now describe.

\paragraph{Stretched horizon.}
It is conjectured that there exists a holographic description of the static patch of empty dS space whose degrees of freedom live at its boundary, \ie the stretched horizon  \cite{Dyson:2002pf,Susskind:2011ap,Susskind:2021esx,Susskind:2021dfc,Shaghoulian:2021cef,Susskind:2021omt,Shaghoulian:2022fop}.
This is identified as a surface with constant radial coordinate
\beq
r_{\rm st} \equiv  \rho L \, , \qquad
\rho \in [0,1]  \, ,
\label{eq:definition_stretched_horizon}
\eeq
defined on both sides of the Penrose diagram.
While we are mainly interested in the limit $\rho \rightarrow 1,$ we will allow $\rho$ to vary along all the range $\rho \in [0,1],$ which corresponds to interpolating between the poles and the cosmological horizon.
We identify the time coordinates running upwards along the left and right stretched horizons as the boundary times $t_L, t_R$ associated with the dual field theory.
Due to the existence of a Killing vector $\partial_t$ for the metric \eqref{eq:metric_dS_gend}, there is a boost symmetry such that the conjectured dual state is invariant under the shift
\beq
t_L \rightarrow t_L + \Delta t \, , \qquad
t_R \rightarrow t_R - \Delta t \, .
\label{eq:time_shift_symmetry}
\eeq
In principle, these time coordinates are independent, but one can synchronize them by considering an extremal codimension-one surface running through the bifurcation surface and connecting the two stretched horizons. This geometric construction connects the time slice at time $t_R$ on the right stretched horizon with the one at time $t_L=-t_R$ on the left horizon (the relative minus sign is a consequence of our choice of the upward orientation of the boundary time coordinate on both sides of the geometry).
Throughout this work, a symmetric time evolution will correspond to the case 
\beq
\frac{t}{2} \equiv t_R = t_L \, .
\label{eq:symmetric_times}
\eeq

\subsubsection{Schwarzschild-de Sitter black hole}
\label{sec:SdS_black_hole}

Schwarzschild-de Sitter spacetime (sometimes also called the Kottler background \cite{Kottler}), is a neutral, static, and spherically 
 symmetric solution of Einstein equations in asymptotically dS$_{d+1}$ spacetime.
We will restrict to the case $d\geq 2.$
The number of dimensions greatly affects the conformal structure of the space.
In particular, the case $d=2$ only admits a cosmological horizon, while for $d\geq 3$ there is also a black hole horizon in general. 
For this reason, we treat the two classes of solutions separately.

\paragraph{Three dimensions.}

In three dimensions (\ie $d=2$), the blackening factor in the metric \eqref{eq:asympt_dS} becomes 
\beq
f (r) = 1 - 8 G_N \mathcal{E} - \frac{r^2}{L^2}  \, ,
\label{eq:metric_SdS3}
\eeq
where we have defined $m\equiv4 G_N \mathcal{E},$ and we refer to  $\mathcal{E}$ as the energy of the solution
\cite{DESER1984405,Spradlin:2001pw}.
When $\mathcal{E}=0,$ we recover empty dS space.
The geometry only admits a cosmological horizon at
\beq
r_C = a L  \, , \qquad
a \equiv \sqrt{1 - 8 G_N \mathcal{E}} \, ,
\label{eq:cosmological_horizon_SdS3}
\eeq
and there is no black hole horizon.
This is a consequence of working in three dimensions, where the SdS$_3$ is locally identical to that of empty dS$_3$ and the former can be produced by making a global identification of dS$_3$. 
By introducing the following coordinates
\beq
\tilde{t} = a \, t \, , \qquad 
\tilde{r} = \frac{r}{a} \, , 
\qquad  \tilde{\theta} = a \, \theta \, ,
\eeq
it is indeed possible to show that the SdS$_3$ metric \eqref{eq:metric_SdS3} simply becomes empty dS$_3$ with cosmological horizon $L.$
For this reason, the three-dimensional conformal diagram of the black hole solution is the same as for empty dS, which was depicted in fig.~\ref{fig:Penrose_dS}. Since the new angular coordinate is identified with period $2 \pi a,$ there is a conical singularity with positive deficit angle $\alpha_{\rm def} = 2 \pi \le 1 - a \ri $ at the origin. The Hawking temperature and the entropy, determined at the cosmological horizon, as computed from eq.~\eqref{eq:temp_entropy_SdS_general} read
\beq
T_{\rm SdS_3} = \frac{a}{2 \pi L} \, , \qquad
S_{\rm SdS_3} = \frac{\pi a L}{2 G_N} \, .
\label{eq:Hawking_temperature_SdS3}
\eeq
Null directions are defined according to eq.~\eqref{eq:general_null_coordinates}, where the tortoise coordinate is  given by
\beq
r^*(r) = \frac{L}{2 a} 
\log \left| \frac{a L+ r}{a L  - r} \right| \, .
\label{eq:rstar_SdS}
\eeq
The stretched horizon is a surface with a constant value of the radial coordinate given by
\beq\label{eq:definition_stretched_horizon3D}
r_{\rm st} = \rho \, a L \, , \qquad
\rho \in [0,1]  \, .
\eeq

\paragraph{Higher dimensions.}

\begin{figure}[ht]
    \centering
    \includegraphics[scale=0.45]{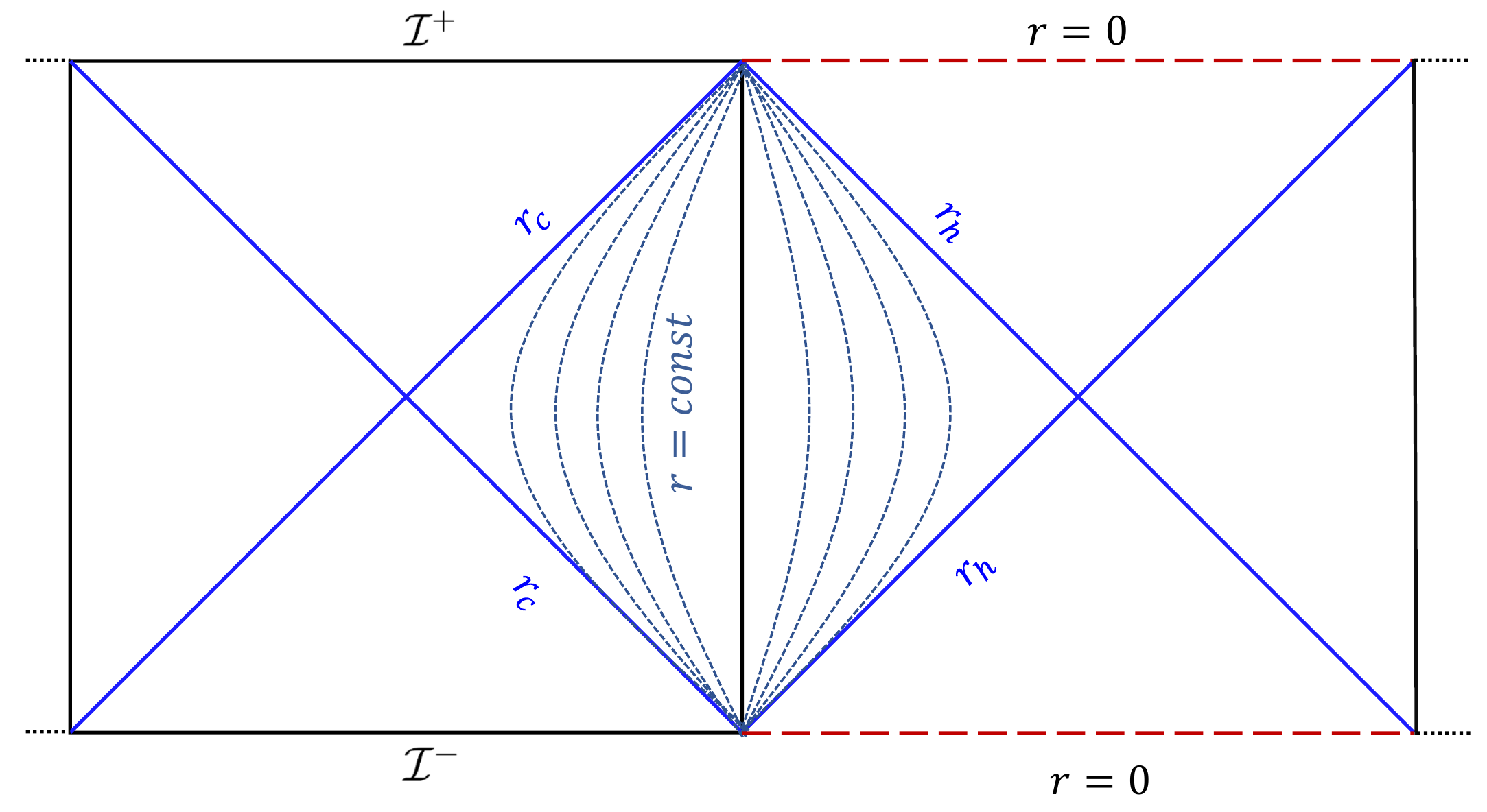}
    \caption{Penrose diagram of SdS$_{d+1}$ space in dimensions $d\geq 3,$ when the mass parameter satisfies $0<m<m_{\rm cr}.$ 
    $r_h$ denotes the black hole horizon and $r_C$ the cosmological horizon.}
    \label{fig:Penrose_SdS}
\end{figure}

In higher dimensions $d\geq 3$, the blackening factor of the metric \eqref{eq:asympt_dS} takes the general form \cite{Spradlin:2001pw}
\beq
f (r) = 1 - \frac{2 m}{r^{d-2}} - \frac{r^2}{L^2} \, ,\label{blacken3}
\eeq
where $m$ is a parameter related to the mass of the black hole (\eg see \cite{PhysRevD.15.2738,Ghezelbash:2001vs}) for different proposals on how to define the masses in this setup).
When the mass parameter is in the range $0<m<m_{\rm cr}$ 
\beq
m_{\rm cr} \equiv \frac{r_{\rm cr}^{d-2}}{d}  \, , \qquad
r_{\rm cr} \equiv L \sqrt{\frac{d-2}{d}} \, ,
\label{eq:critical_mass_SdS}
\eeq
then the geometry admits two horizons $r_C > r_h$, the larger value being a cosmological horizon and the smaller one representing a black hole horizon.
The Penrose diagram for the SdS geometry in the regime $0<m<m_{\rm cr}$ is illustrated in fig.~\ref{fig:Penrose_SdS}.

In the limit, $m\rightarrow m_{\rm cr}$, the black hole and cosmological horizon approach each other towards the common critical value $r_h=r_C=r_{\rm cr}.$
However, their proper distance does not vanish because the blackening factor is also going to zero near $r \sim r_{\rm cr}$. 
This limit is known as the 
Nariai solution, see \eg \cite{Anninos:2012qw}. A simple coordinate transformation demonstrates that in this limit, the static patch   effectively becomes dS$_2\times S^{d-2}$, see \eg \cite{Maldacena:2019cbz,Anninos:2012qw}.

While it is not possible to find a closed form for the solution to the equations $f(r_h)=f(r_C)=0$ in general, they can be used to re-express the mass and dS curvature scale in terms of the two horizon radii, see \eg \cite{Morvan:2022ybp}:
\beq
m = \frac{1}{2} \frac{r_C^d r_h^{d-2} - r_h^d r_C^{d-2}}{r_C^d - r_h^d} \, , \qquad
L^2 = \frac{r_C^d - r_h^d}{r_C^{d-2} - r_h^{d-2}} \, .
\label{eq:general_relation_rhrcm}
\eeq
Substituting these results into the blackening factor \eqref{blacken3} yields
\beq
f(r) = \frac{1}{r_C^d - r_h^d} 
\left[ r_C^d \le 1 - \frac{r^2}{r_C^2} - \frac{r_h^{d-2}}{r^{d-2}}  \ri
- r_h^d \le 1 - \frac{r^2}{r_h^2} - \frac{r_C^{d-2}}{r^{d-2}}  \ri
\right] \, .
\label{eq:general_blackening_SdS_rcrh}
\eeq
Using eq.~\eqref{eq:temp_entropy_SdS_general}, the associated temperatures for the black hole and cosmological horizons can be written as
\beq 
T_h  = d \, \frac{r_{\rm cr}^2 - r_h^2 }{4 \pi r_h L^2}  \, , \qquad
    T_C  = d \, \frac{r_C^2 -r_{\rm cr}^2 }{4 \pi r_C L^2}  \, .
   \label{eq:hor_cosm_temp_SdS}
\eeq
For arbitrary mass parameter in the range $(0,m_{\rm cr}),$ the Hawking temperatures at the two horizons satisfy $T_h > T_C,$ so that the system is out of equilibrium.
It is relevant to notice that when the horizon radius reaches the critical value $r_{\rm cr}$, defined in eq.~\eqref{eq:critical_mass_SdS}, 
corresponding to the Nariai limit, both temperatures approach zero and the configuration approaches equilibrium \cite{Bousso:1997wi,Anninos:2012qw}. In this case, the black hole reaches an extremal configuration. 
Another case where the background is in thermal equilibrium corresponds to $m=0$, which of course corresponds to empty dS space. In this case, the dS horizon is in equilibrium with the emitted radiation and therefore it does not evaporate, see \eg \cite{Anninos:2012qw}.

To define the holographic prescription for a dual field theory associated with this spacetime, we need to introduce a stretched horizon in this background.
Since $r_C$ is the cosmological horizon for observers on worldlines of constant radial coordinate between $r \in [r_h, r_C]$, a natural prescription is to locate the stretched horizon at
\beq
r_{\rm st} = (1-\rho) r_h + \rho r_C \, , \qquad
\rho \in [0,1] \, ,
\label{eq:stretched_horizon_SdS4}
\eeq
such that it is a surface at constant $r$ interpolating between the two horizons.

\paragraph{Four dimensions.}

\begin{figure}[ht]
    \centering
    \includegraphics[scale=0.66]{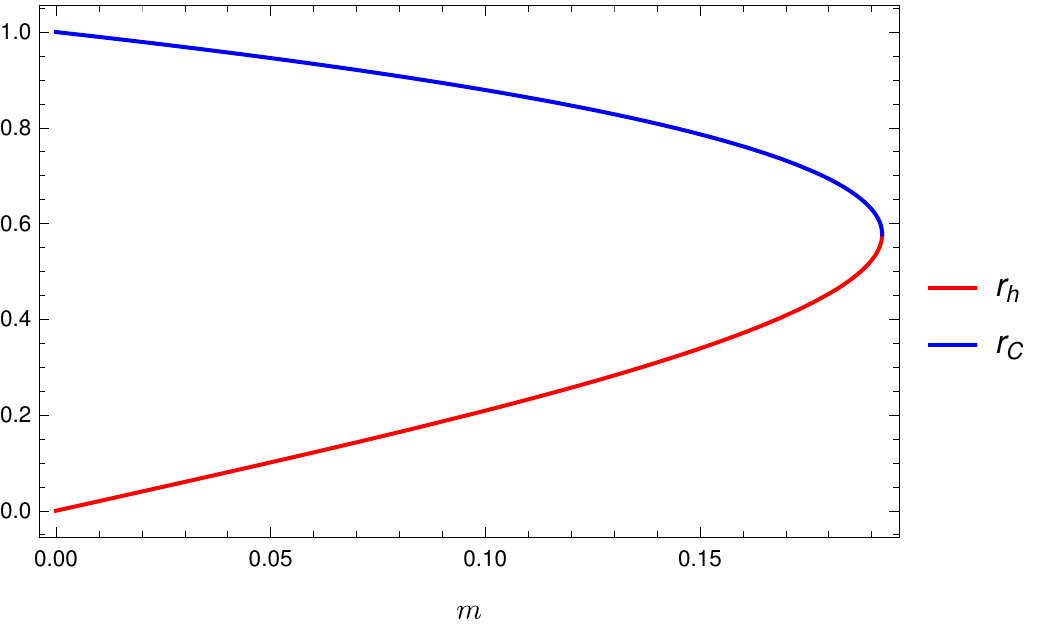}
    \caption{Plot of the black hole and cosmological horizons for SdS$_4$ \eqref{eq:metric_SdS4} as functions of the mass, in units of $L=1$.}
    \label{fig:plotrcrh_SdS4}
\end{figure}

Let us focus on the special case of four dimensions, \emph{i.e.}, we set $d=3$.  
The blackening factor becomes
\beq
f_{\rm SdS_4} (r) = 1 - \frac{2m}{r} - \frac{r^2}{L^2} \, .
\label{eq:metric_SdS4}
\eeq
The critical mass \eqref{eq:critical_mass_SdS} implies that under the condition $m^2/L^2 <1/27$, we have two different roots of the blackening factor.
The relations \eqref{eq:general_relation_rhrcm} become
\beq
m = \frac{r_h r_C (r_h + r_C)}{2 (r_C^2 + r_h r_C + r_h^2)} \, , \qquad
L^2 = r_C^2 + r_h^2 + r_C r_h \, .
\label{eq:relations_ML_SdS_higherd}
\eeq
Inverting these identities, one can express the horizons in terms of the mass.
In four dimensions it is possible to find the solutions in the closed form \cite{Shankaranarayanan:2003ya,Choudhury:2004ph,Visser:2012wu}:
\beq
   r_h = r_{\rm cr} \le \cos \eta - \sqrt{3} \sin \eta \ri \, , \quad
r_C = r_{\rm cr} \le \cos \eta + \sqrt{3} \sin \eta \ri \, , \quad  
\eta \equiv \frac{1}{3} \, \mathrm{arccos}  \le \frac{m}{m_{\rm cr}} \ri \, ,
\label{eq:analytic_rh_rc_SdS4}
\eeq
where $r_{\rm cr}=L/\sqrt{3}$ and $m_{\rm cr}=r_{\rm cr}/3=L/(3\sqrt{3})$, with the definitions in eq.~\eqref{eq:critical_mass_SdS}.
We depict these functions in fig.~\ref{fig:plotrcrh_SdS4}.
It is evident that increasing the mass of the black hole leads to increasing $r_h$ and decreasing $r_C,$ until they approach each other when the parameters satisfy $m^2/L^2 = 1/27,$ which is the critical mass leading to the Nariai solution.
In this case, the black hole and cosmological horizon both reach the critical value $r_{\rm cr}.$
When the mass vanishes, we recover empty dS$_4$ space without a black hole horizon ($r_h=0$), and with $r_C=L.$

\noindent
The blackening factor can alternatively be expressed using the two horizon radii rather than the mass and dS length, as in eq.~\eqref{eq:general_blackening_SdS_rcrh}
\beq
f_{\rm SdS_4} (r) = - \frac{(r-r_h)(r-r_C)(r+r_h+r_C)}{r(r_C^2 + r_h r_C + r_h^2)} \, ,
\eeq
which can be integrated analytically to obtain the tortoise coordinate 
\beq
\begin{aligned}
r^*(r) = &
- \frac{r_C^2 + r_h r_C + r_h^2}{(r_C-r_h)(2r_C+r_h) (r_C+2r_h)} 
\left[ r_C^2 \log \left| \frac{r-r_C}{r+r_C+r_h} \right| \right. \\
& \left. + 2 r_h r_C \log \left| \frac{r-r_C}{r-r_h} \right|
-  r_h^2 \log \left| \frac{r-r_h}{r+r_C+r_h} \right|
\right] \, .
\end{aligned}
\label{eq:tortoise_coord_SdS4}
\eeq
The Hawking temperatures of the two horizons are given by eq.~\eqref{eq:temp_entropy_SdS_general}, which yields
\beq
T_h = \frac{(r_C-r_h)(r_C+2r_h)}{4 \pi L^2 r_h} = \frac{L^2 - 3 r_h^2}{4 \pi L^2 r_h}  \, , \quad
T_C =\frac{(r_C-r_h)(r_h+2r_C)}{4 \pi L^2 r_C} = \frac{3 r_C^2 - L^2}{4 \pi L^2 r_C}  \, . 
\label{eq:Hawking_temperature_SdS4}
\eeq

\subsection{Perturbation with shockwaves}
\label{ssec:perturbation_shocks}

We perturb the class of asymptotically dS spacetimes introduced in eq.~\eqref{eq:asympt_dS} by the inclusion of a shockwave intersecting the right stretched horizon at time $-t_w$. 
The shockwaves are sourced by a null fluid moving along a spherically symmetric null surface. Their effect is to shift the mass parameter from $m_1$ before the shock to $m_2$ after the shock. As a consequence, the black hole horizon is shifted from $r_{h1}$ to $r_{h2}$ and the cosmological horizon of the geometry is shifted from $r_{C1}$ to another value $r_{C2}$.

\begin{figure}[ht]
    \centering
   \subfigure[]{  \includegraphics[scale=0.43]{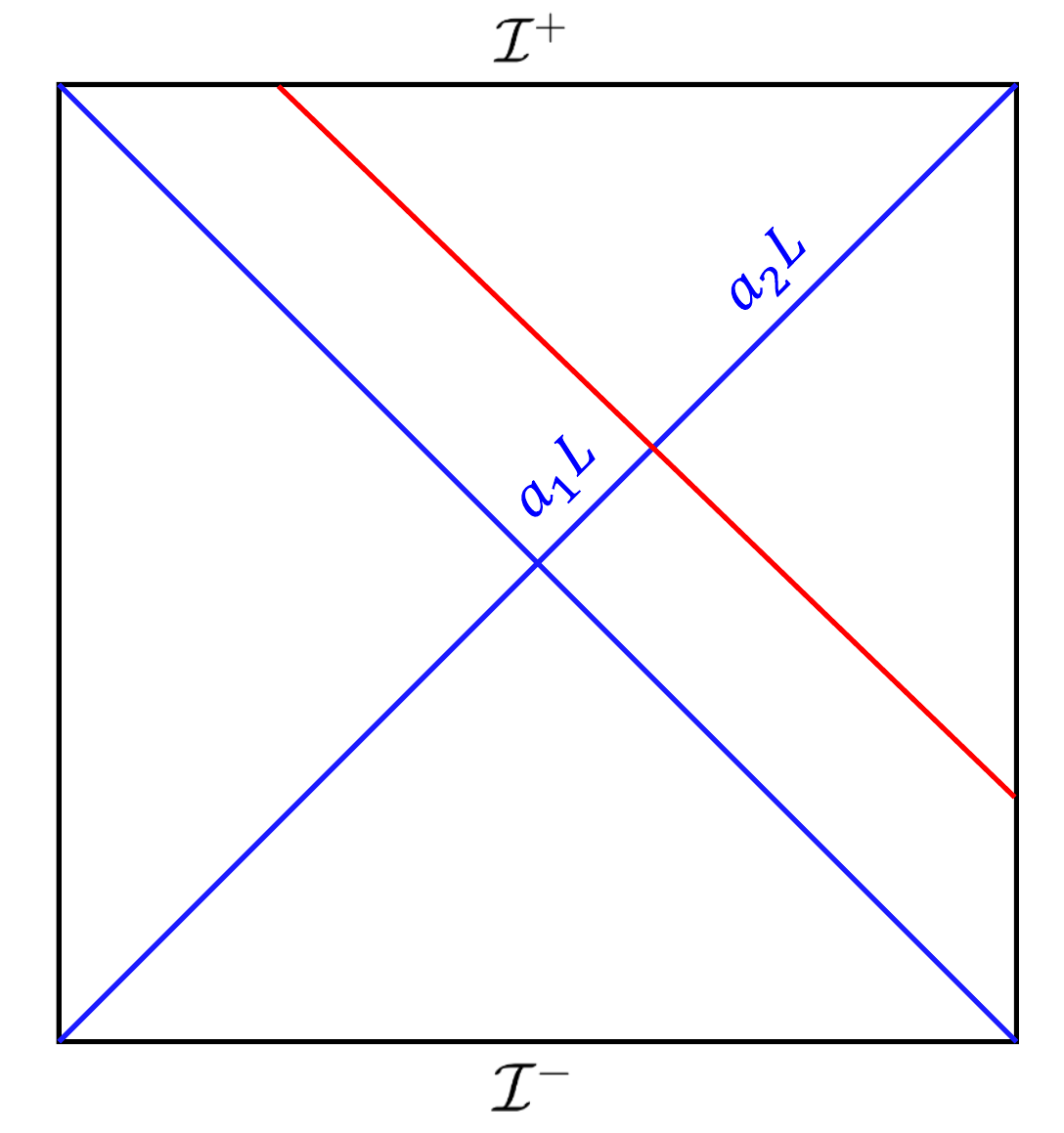}} \qquad
    \subfigure[]{ \includegraphics[scale=0.42]{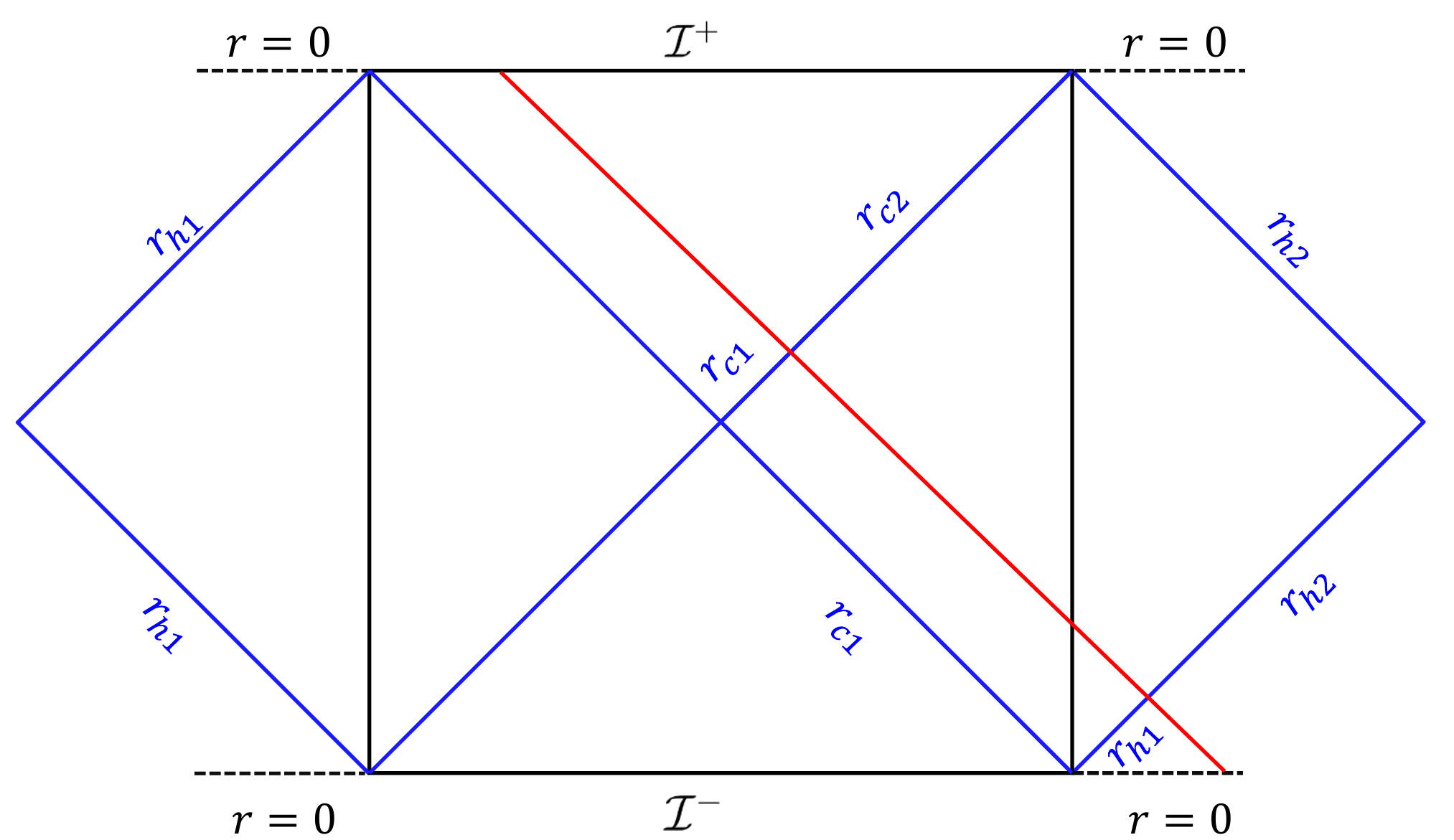}}
    \caption{Penrose diagrams of SdS black hole in the presence of a shockwave (a) SdS$_3$ space and (b) SdS in higher dimensions. }
    \label{fig:Penrose_shock} 
\end{figure}

\paragraph{Geometric setting.} 
A shockwave deformation of SdS spacetimes can be described by the following metric
\bea
& ds^2 = - F(r,u) du^2 - 2 dr du + r^2 d\Omega_{d-1}^2 \, , &
\label{eq:dS_metric_shock_wave}
\\
& F(r,u) = f_1 (r) \le 1-\theta (u-u_s) \ri + 
f_2 (r) \theta(u-u_s) \, , &
\eea
where $u_s$ is the constant null coordinate parametrizing the location of the shockwave and $\theta$ is the Heaviside step function.
The perturbation induces a jump in the blackening factor given by
\beq
\begin{aligned}
 &   u < u_s \, : \quad
    F(r,u) = f_1(r) = 1 - \frac{2m_1}{r^{d-2}} - \frac{r^2}{L^2} \, , & \\
    &   u > u_s \, : \quad
    F(r,u) = f_2(r) = 1 - \frac{2m_2}{r^{d-2}} - \frac{r^2}{L^2} \, . &
\end{aligned}
\label{eq:blackening_factor_shock_SdS}
\eeq
The tortoise coordinate is also affected by this jump and reads\footnote{As before, the constants of integration are fixed by requiring a vanishing tortoise coordinate at future/past timelike infinity.}
\beq
r^*(r)  = \int_{r_{0,1}}^r \frac{dr'}{f_1(r')} \le 1-\theta (u-u_s) \ri + \int_{r_{0,2}}^r \frac{dr'}{f_2(r')}  \theta (u-u_s)  \, .
\label{eq:general_def_tortoise_coordinate_shock}
\eeq
The Penrose diagram corresponding to this geometric setting is depicted in fig.~\ref{fig:Penrose_shock} for both the SdS$_3$ and for SdS$_{d+1}$ ($d\geq 3$) backgrounds.
In these geometries, the radial coordinate $r$ and the ingoing null coordinate $u$ is continuous along the entire shockwave geometry, instead the time $t$ and the outgoing null coordinate $v$ are discontinuous across the shockwave, as a consequence of the jump in the blackening factor.
We stress that the diagrams in fig.~\ref{fig:Penrose_shock} are obtained with a cut-and-paste procedure. That is, the diagrams are depicted in such a way that the cosmological horizons above and below the shockwave (\ie at $r_{C2}$ and $r_{C1}$, respectively) are glued to one another across the shockwave. We will often employ a slight abuse of notation and still refer to those diagrams as ``Penrose'' diagrams.\footnote{For further comments, see the discussion of fig.~\ref{fig:constant_redshift_SdSall} at the beginning of section \ref{sec:stretched_horizon_shocks}.}

In general, this setting corresponds to the transition between black holes with different masses $m_1,m_2$, which also includes the possibility to evolve from a black hole solution to empty dS (when $m_2=0$).
Previous studies of shockwaves in asymptotically dS space can be found in \cite{Hotta_1993,PhysRevD.47.3323,Sfetsos:1994xa,Aalsma:2021kle}.
We will show in section \ref{ssec:NEC} that the NEC imposes $m_1 \geq m_2$, therefore such transitions always correspond to the shockwave carrying mass away from the black hole.
Interpreting the SdS background as describing the exterior geometry of a spherically symmetric star-like object, this process corresponds to the (spherically symmetric) ejection of mass into the cosmological horizon, and the consequence is that the cosmological horizon is pushed further away from an observer sitting at some fixed radius.

According to the static patch holography proposal (as discussed in section \ref{sec:introduction}), the dual theory lives on the stretched horizon, where the boundary time is measured.
In particular, the null coordinate of the shockwave and the right boundary time $-t_w$ of its insertion are related by
\beq
u_s = - t_w - r^*(r_{\rm st})  \, ,
\label{eq:constant_us_shock}
\eeq
where $r_{\rm st}$ is the radial coordinate of the stretched horizon.
So far, we did not specify how the stretched horizon is affected by the presence of the shockwave; this plays an important role in the determination of the boundary time.
We will return to this point in section \ref{sec:stretched_horizon_shocks}.

\paragraph{Geometric data.}
We can relate the strength of the shock to the jump it induces in the location of the cosmological horizon. For this, let us label the cosmological horizons as $r_{Ci}$, where the index $i=1,2$ corresponds to the part of the geometry below or above the shockwave at $u=u_s$, respectively:
\beq
r_{Ci} = \begin{cases}
 a_i L & \mathrm{SdS}_3  \\
r_{\rm cr} \le \cos \eta + \sqrt{3} \sin \eta \ri  & \mathrm{SdS}_{4} ,  \\
\end{cases}
\label{eq:generic_rc_geometries}
\eeq
where
\beq
a_i \equiv \sqrt{1- 8 G_N  \mathcal{E}_i} \, ,
\label{eq:ai_SdS3}
\eeq
and the quantities $r_{\rm cr}, \eta $ were defined in eqs.~\eqref{eq:critical_mass_SdS} and \eqref{eq:analytic_rh_rc_SdS4}.
In higher dimensions, there is no closed form expression for the cosmological horizon, and we need to numerically solve the conditions \eqref{eq:general_relation_rhrcm}.
According to the NEC that we will study in section \ref{ssec:NEC}, a shockwave perturbation induces a transition between geometries such that the cosmological horizon gets pushed away from the observer, \ie $r_{C1} \leq r_{C2}$.

For later convenience, we also define a dimensionless parameter $\varepsilon$ describing the relative variation between the geometry before and after the shockwave insertion.
It reads
\beq
 \varepsilon \equiv  \begin{cases}
1- \mathcal{E}_2/\mathcal{E}_1  & \mathrm{SdS}_3  \\
1- m_2/m_1  & \mathrm{SdS}_{d+1} \quad (d \geq 3) . \\
\end{cases}
\label{eq:generic_epsilon_geometries}
\eeq
The case of an infinitesimal deformation of the original geometry always corresponds to $\varepsilon \ll 1.$
The NEC will impose the conditions
\beq
\mathcal{E}_1 \geq \mathcal{E}_2 \, ,  \qquad
m_1 \geq m_2 \, .
\eeq
Therefore, in the cases of interest, we have $\varepsilon \in [0,1].$
In particular, the maximal value $\varepsilon=1$ corresponds to the transition from the SdS black hole to empty dS space. Of course, we always assume  $m_i \in (0,m_{cr})$ (if this is satisfied before the shock it will also be satisfied after the shock).

\subsection{Null energy condition}
\label{ssec:NEC}

The perturbation of a background geometry through the insertion of a shockwave should be consistent with the NEC.
We analyze the consequences of this constraint on the parameters of the geometry.

The metric \eqref{eq:dS_metric_shock_wave} with blackening factor \eqref{eq:blackening_factor_shock_SdS} solves the Einstein equations arising from the gravitational action \eqref{eq:action_EOM_gend} 
\beq
G_{\mu\nu} + \Lambda g_{\mu\nu} = 8 \pi G_N T_{\mu\nu} \, .
\label{eq:Einstein_eq}
\eeq
Therefore the shockwave insertion is encoded by the presence of a non-vanishing stress tensor
\begin{equation}
    T_{\mu\nu} = \frac{1}{8\pi G_N} \frac{d-1}{2} \frac{m_1-m_2}{r^{d-1}} \, \delta(u-u_s) \ \delta_{\mu u}\, \delta_{\nu u}\, .
    \label{eq:matter_stress_BH}
\end{equation}
That is, the only non-vanishing component corresponds to $T_{uu}$.
However, not all matter can be inserted in a consistent way. We require that the energy-momentum tensor satisfies the null-energy condition, \emph{i.e.}, for any null vector $k^{\mu}$, we impose
\beq
T_{\mu\nu} k^{\mu} k^{\nu}  \geq 0 \, .
\eeq
Since the only non-vanishing component of the stress tensor is $T_{uu},$ we find that the only non-trivial constraint is 
\beq
T_{\mu\nu} k^{\mu} k^{\nu} = \frac{1}{8 \pi G_N} \frac{d-1}{2} \frac{(k^u)^2}{r^{d-1}} (m_1-m_2) \, \delta(u-u_s) \geq 0 \, ,
\label{eq:contraction_Tkk}
\eeq
where $k^u$ is the $u$ component of the null vector $k^\mu$.
This implies that $m_1 \geq m_2,$ so that shockwaves with positive energy are responsible for decreasing the mass of the black hole.\footnote{Of course, we could also consider infalling shockwaves (following surfaces of constant $v$), which would increase the the black hole mass. This behaviour would be similar to the usual holographic studies of asymptotically AdS black holes. Further, let us recall that one encounters shockwaves emerging from the white hole region in the AdS/CFT when considering a perturbed thermofield double state in the boundary theory \cite{Shenker:2013pqa,Shenker:2013yza}. In this scenario, however, the shockwave reflects from the asymptotic boundary to become an infalling shock. In the present case of static patch holography, it is unnatural to consider shockwaves being reflected at the stretched horizon. Rather the holographic screen at the stretched horizon is treated as a transparent surface.}
Since SdS black holes with lower mass have a bigger cosmological horizon, the constraint arising from the null energy condition can be alternatively formulated as $r_{C1} \leq r_{C2}$.

\section{Stretched horizon in the presence of shockwaves}
\label{sec:stretched_horizon_shocks}

Static patch holography essentially states that the holographic dual theory lives on a constant radial slice surface $r=r_{\rm st}$, given by equations \eqref{eq:definition_stretched_horizon}, \eqref{eq:definition_stretched_horizon3D} and
\eqref{eq:stretched_horizon_SdS4} for empty dS space, SdS$_3$ and SdS$_{d+1}$ ($d\geq 3$), respectively. 
However, if we use a constant radius surface as the stretched horizon in the presence of shockwaves, we encounter the somewhat uncomfortable property that a stretched horizon located within the static patch before the shock $r_{\rm st} < r_{C1}$, can never approach the larger cosmological horizon $r_{C2}$ after the shock. 

\begin{figure}[ht]
    \centering
     \subfigure[SdS$_3$~~~~~~~~~~~~~~~~~~~~~]
     {\begin{minipage}[tl]{0.4\linewidth}
    \includegraphics[scale=0.4]{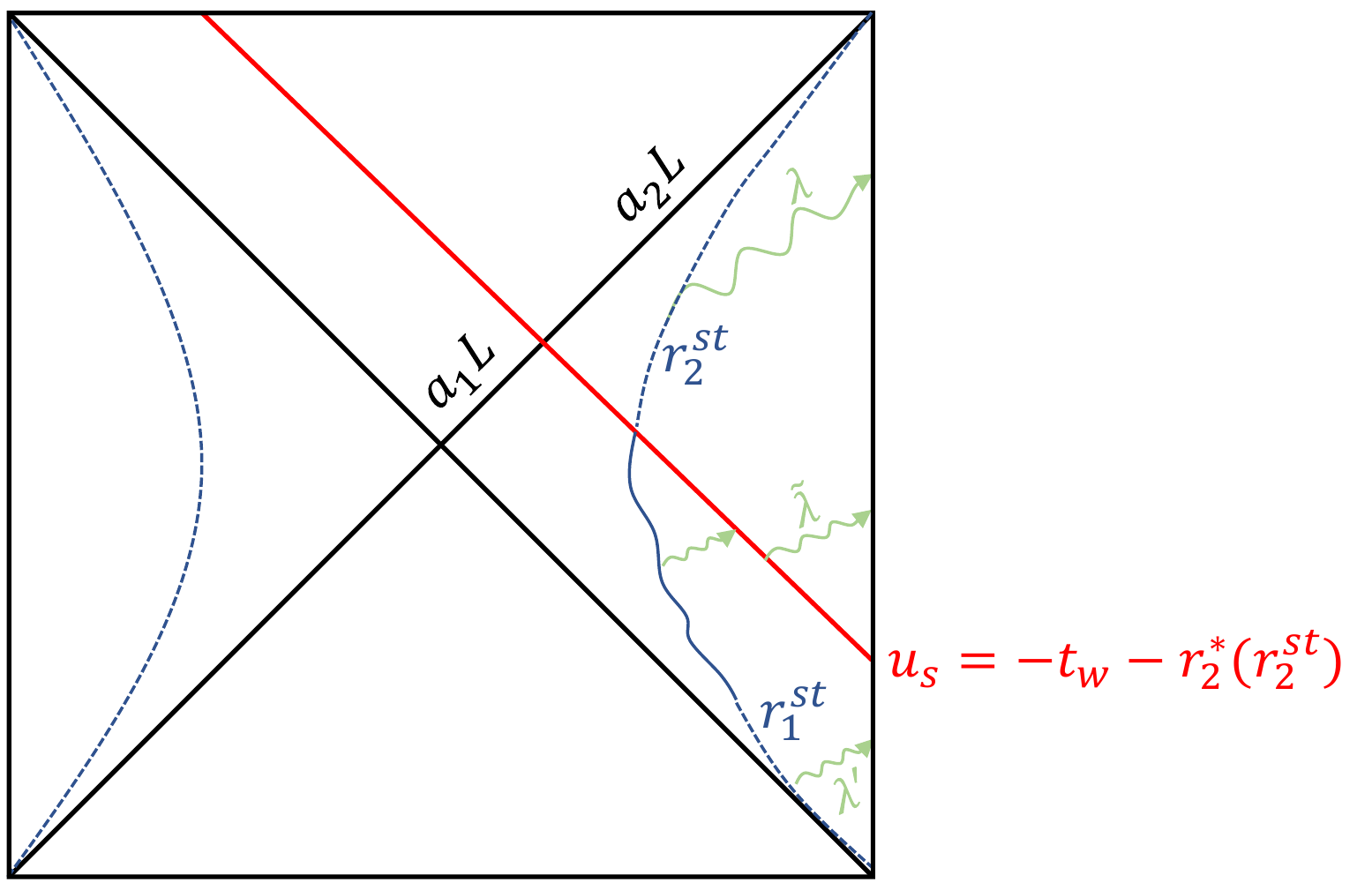}
    \vspace{0.2pt}
     \end{minipage}
     \label{fig:constant_redshift}
     }
      \subfigure[SdS$_{d+1}$ with $d\geq 3$]{
      \begin{minipage}[tr]{0.5\linewidth}
    \includegraphics[scale=0.4]{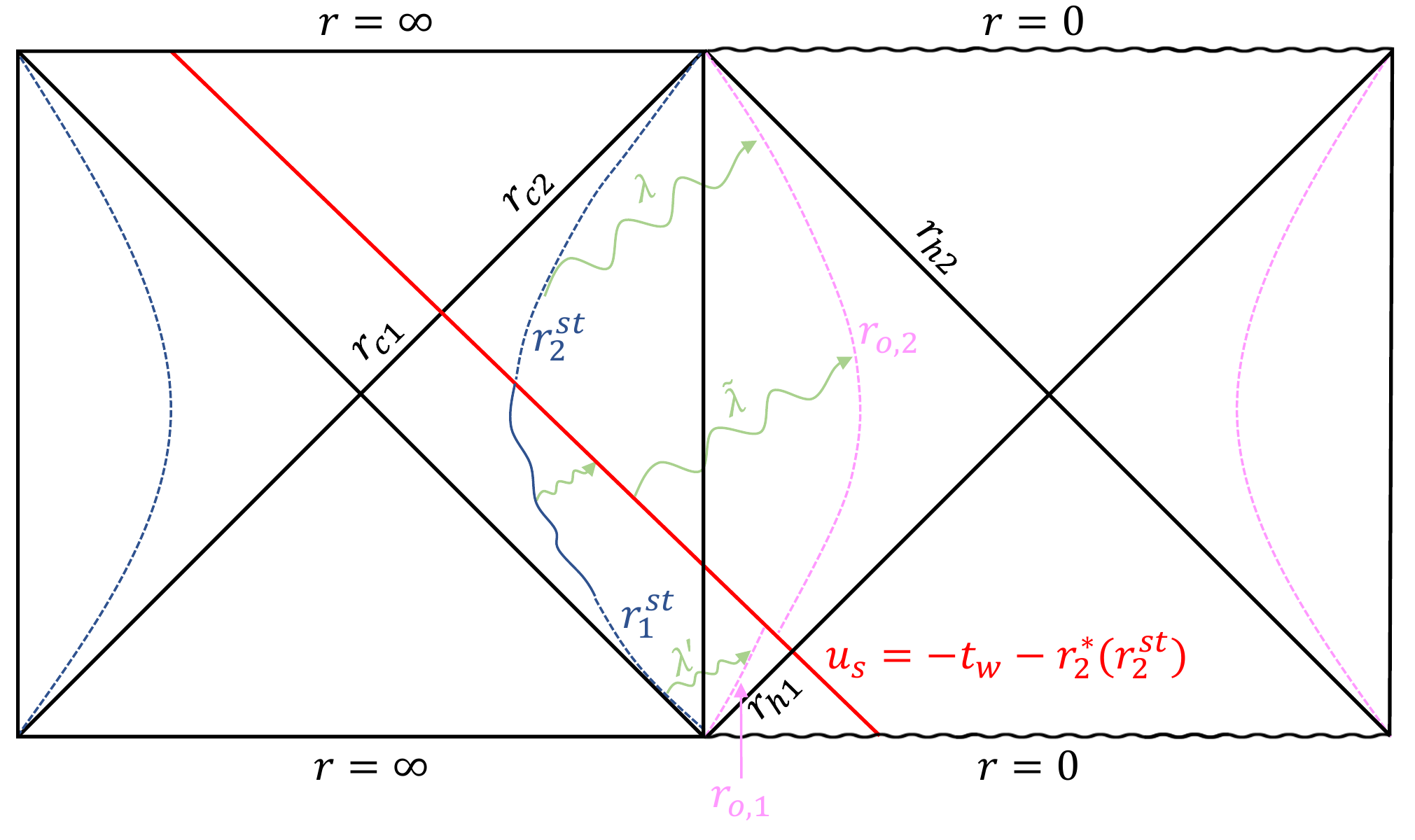}
        \end{minipage}
       \label{fig:constant_redshift_SdS}
      }  
    \caption{
    Light rays at constant $v$ sent from an emission surface parametrized by $r=R_e(t_e).$ The shockwave sits at constant $u_s=-t_w-r^*_2(r^{\rm st}_2).$
For early times $t_e \ll -t_w$, the trajectory satisfies $R_e=r^{\rm st}_1$ and the light ray is denoted with $\lambda'$. 
For late times $t_e \gg -t_w$, we have $R_e = r^{\rm st}_2$ and the light ray is denoted by $\lambda$.     
For intermediate times, the light ray is denoted by $\tilde{\lambda}$ and the trajectory of the stretched horizon is time-dependent. 
The observer sits on a surface at constant radial coordinates $r_{o,1}$ before the shockwave, and $r_{o,2}$ after the shock. In the case of SdS$_3$ the observer is located at $r_{o,1}=r_{o,2}=0$.
The different subfigures depict different dimensions as indicated in the subcaption. }
    \label{fig:constant_redshift_SdSall}
\end{figure}

To avoid this problem, in section \ref{ssec:details_const_red} we propose a different prescription, which locates the stretched horizon at a surface of constant redshift from an observer located at constant radius. The idea is to impose that outgoing light rays emitted from the stretched horizon are received by an observer located at some fixed $r=r_{o,1}$ before the shock and at some fixed $r=r_{o,2}$ after the shock\footnote{We will mostly focus on the case $r_{o,1}=r_{o,2}$ below.}  in such a way that the cosmological redshift is constant along the time evolution, see fig.~\ref{fig:constant_redshift} and \ref{fig:constant_redshift_SdS}. In this context, the terminology \emph{ingoing} and \emph{outgoing} refers to light rays sent towards or outwards from the cosmological horizon, respectively. In SdS$_3$ we will locate the observer at the north pole, \ie $r_{o,1}=r_{o,2}=0$.
As remarked above fig.~\ref{fig:Penrose_shock}, it is worth reminding the reader that in our  diagrams, we rescale the geometry after the shockwave so that the cosmological horizons $r_{C1}$ and $r_{C2}$ are glued together. For this reason, the light ray $\tilde{\lambda}$ jumps in fig.~\ref{fig:constant_redshift_SdS}, even though its radial coordinate is continuous when crossing the shockwave. This is a manifestation of the time advance for a pulse crossing the shock, \eg see \cite{Gao:2000ga, Anninos:2018svg}.

Let us add that one may determine the position of the stretched horizon by any of a number of prescriptions. The choice described above is non-local in character, \ie the position is related to a distant observer (or family of observers near the black hole), but it has the virtue of imposing a physical condition for the observer.  In appendix \ref{app:ping_pong}, a second non-local prescription is examined which again relies on the redshift between the stretched horizon and observers near the black hole, but with slightly different boundary conditions. 
However, one can also devise local approaches to determine the stretched horizon. One simple local approach would be to ask that the proper acceleration of observers following the local Killing flow on the stretched horizon is radially directed and has a fixed magnitude. Another arises from the observation that a consequence of the redshift approach initially described above
is that the SdS time coordinate $t$ along the stretched horizon is discontinuous at the intersection with the shockwave. Hence another approach is to impose the local boundary condition that the time coordinate is continuous across the shock.\footnote{We 
study the consequences of this approach in section \ref{ssec:details_cont_time}. We also return to a discussion of the definition of time in the boundary theory in section \ref{sec:discussion}.} 

In fact, the precise prescription for the stretched horizon will be unimportant for our results in the following. Irrespective of the prescription, the results are qualitatively similar in general, and they will coincide in the limit of light shockwaves, which will be a focus of our discussion.\footnote{
As an example, consider the scrambling time in the SdS$_3$ background. This will be given by eq.~\eqref{eq:duration_plateau_SdS3} using the prescription that keeps a constant cosmological redshift, and by eq.~\eqref{eq:duration_plateau_SdS3_conflux} when requiring that the time coordinate along the stretched horizon is continuous. In general, these two results are qualitatively similar, and they coincide in the limit of light shockwaves with $\varepsilon \rightarrow 0$. }
However, let us add that all of the prescriptions which were mentioned here (both local and non-local) share certain universal features. These are: (1) the stretched horizons are located at constant radial coordinate in the early past and in all the region after the shock wave insertion, and (2) there is a limit of the parameters such that the stretched horizons before and after the shock wave approach the respective cosmological horizons together. Of these, the behaviour of the scrambling time which we find below only requires that the stretched horizon sits at a fixed radius after the shockwave.

Let us remark that the previous discussion only applies to shockwaves intersecting the right stretched horizon at finite coordinate time $-t_w.$
When the shock propagates along the cosmological horizon ($t_w \rightarrow \infty$), the cosmological horizon itself does not change and therefore the stretched horizon keeps the same form throughout the full geometry.

\subsection{Constant redshift prescription}
\label{ssec:details_const_red}

On the right portion of the Penrose diagram, the stretched horizon will be generically time-dependent, because outgoing light rays can intersect the shockwave.
We denote the time-dependent position of the stretched horizon as $r=R_e(t_e),$ where $t_e$ is a time parameter along the stretched horizon that we call \emph{emission time}. 
To recover the case of an unperturbed asymptotically dS space, we require that the stretched horizon is time-independent in the far past and future:
\beq
\lim_{t_e \rightarrow -\infty} R_e (t_e) = r^{\rm st}_1 \, , \qquad
\lim_{t_e \rightarrow \infty} R_e (t_e) = r^{\rm st}_2  \, ,
\label{eq:far_limits_stretched_hor}
\eeq
where $r^{\rm st}_i$ is the location of the stretched horizon in the far past/future and as before it can be re-expressed in terms of a parameter $\rho_i$ ($i=1,2$) 
\beq
r^{\rm st}_i = \begin{cases}
\rho_i \, a_i L & \mathrm{SdS}_3  \\
(1-\rho_i) r_{hi} + \rho_i r_{Ci}  & \mathrm{SdS}_{d+1} \quad (d \geq 3) \, .
\end{cases}
\label{eq:generic_location_rst1}
\eeq
We will soon see that the constant redshift requirement imposes that $\rho_1$ and $\rho_2$ go to 1 together. Hence, when $r_1^{\rm st}$ before the shock approaches $r_{C1}$,   $r_2^{\rm st}$ after the shock will approach $r_{C2}$. As advertised above, this was the original motivation for us to suggest this prescription of the stretched horizon. 

Since light rays sent after the shock do not intersect it, the stretched horizon takes the time-independent value $r^{\rm st}_2$ over all the spacetime region after the shockwave insertion.
Of course, in the case without the shockwave, this definition recovers our previous choices in equations \eqref{eq:definition_stretched_horizon}, \eqref{eq:definition_stretched_horizon3D} and
\eqref{eq:stretched_horizon_SdS4}. We will study the location of the stretched horizon  starting from the far past and moving towards the future direction 
on the right side of the Penrose diagram.  
On the left side of the geometry the shockwave does not intersect the stretched horizon. Therefore the  left stretched horizon will be  located at the fixed value $r^{\rm st}_1$ defined in eq.~\eqref{eq:generic_location_rst1}, which is the same as on the right side in the far past before the shock.

A few comments are in order. First, note that the tortoise coordinate jumps across the shockwave.
Using the definition \eqref{eq:general_null_coordinates} for the null coordinate $u,$ which is continuous across the shockwave, we find a matching condition between the time coordinate defined in the region before and after the null surface at $u=u_s.$
Specifically, when focusing on the place where the stretched horizon crosses the shock, we have 
\beq\label{eq:def_us_appA}
u_s=t_{s1} - r_1^*(R_e(t_{s1})) = -t_w - r_2^*(r_2^{\rm st}) \, ,
\eeq
where $t_{s1}$ denotes the boundary time on the right stretched horizon immediately before the shockwave insertion. We will later see that when using our prescription the time is discontinuous across the shockwave $t_{s1}\neq -t_w$ and furthermore, the stretched horizon jumps across the shock, that is $
R_e(t_{s1}) \ne 
r^{\rm st}_2$.
In the above formula \eqref{eq:def_us_appA}, we measure the insertion time of the shockwave $-t_w$ using an observer located at the stretched horizon in the part of the geometry at $u>u_s$.
\emph{That is, we clarify that eq.~\eqref{eq:constant_us_shock} for the insertion time of the shockwave was actually defined when approaching the shockwave from above.}

In what follows, it will be useful to define the critical time $\tilde{t}_{e1}$ such that light emitted from the stretched horizon at time $\tilde{t}_{e1}\leq t_e\leq t_{s1}$ crosses the shock. The time $\tilde{t}_{e1}$ corresponds to the limiting case of a light ray emitted from the stretched horizon and reaching $r_{o,1}$ at the intersection with the shockwave $u=u_s$. It reads
\beq
\tilde{t}_{e1} = - t_w - r^*_1 (r^{\rm st}_1) -r^*_2(r^{\rm st}_2) +2r_1^*(r_{o,1})\, .
\label{eq:boundary_condition1_redshift}
\eeq

\paragraph{Computation of the redshift.}
We consider the setting with a transition between two SdS$_{d+1}$ ($d\geq 2$) black holes, as depicted in fig.~\ref{fig:constant_redshift_SdSall}.
The gravitational shockwave propagates through the spacetime along the constant null coordinate $u_s$ defined in eq.~\eqref{eq:def_us_appA}, reaching either the north pole or the singularity of the black hole (both located at $r=0$). 
In this setting, light rays at constant $v$ are emitted at time $t_e$ from the surface $r=R_e(t_e)$ towards an observer that receives and collects the radiation at  $r=R_o(t_o)$.
We assume that the observer is always located on a time-independent surface, even at intermediate times, and that $R_o$ remains continuous when crossing the shockwave (therefore $R_0\equiv r_{o,1}=r_{o,2}$). 

\begin{figure}[ht]
    \centering
    \includegraphics[scale=0.55]{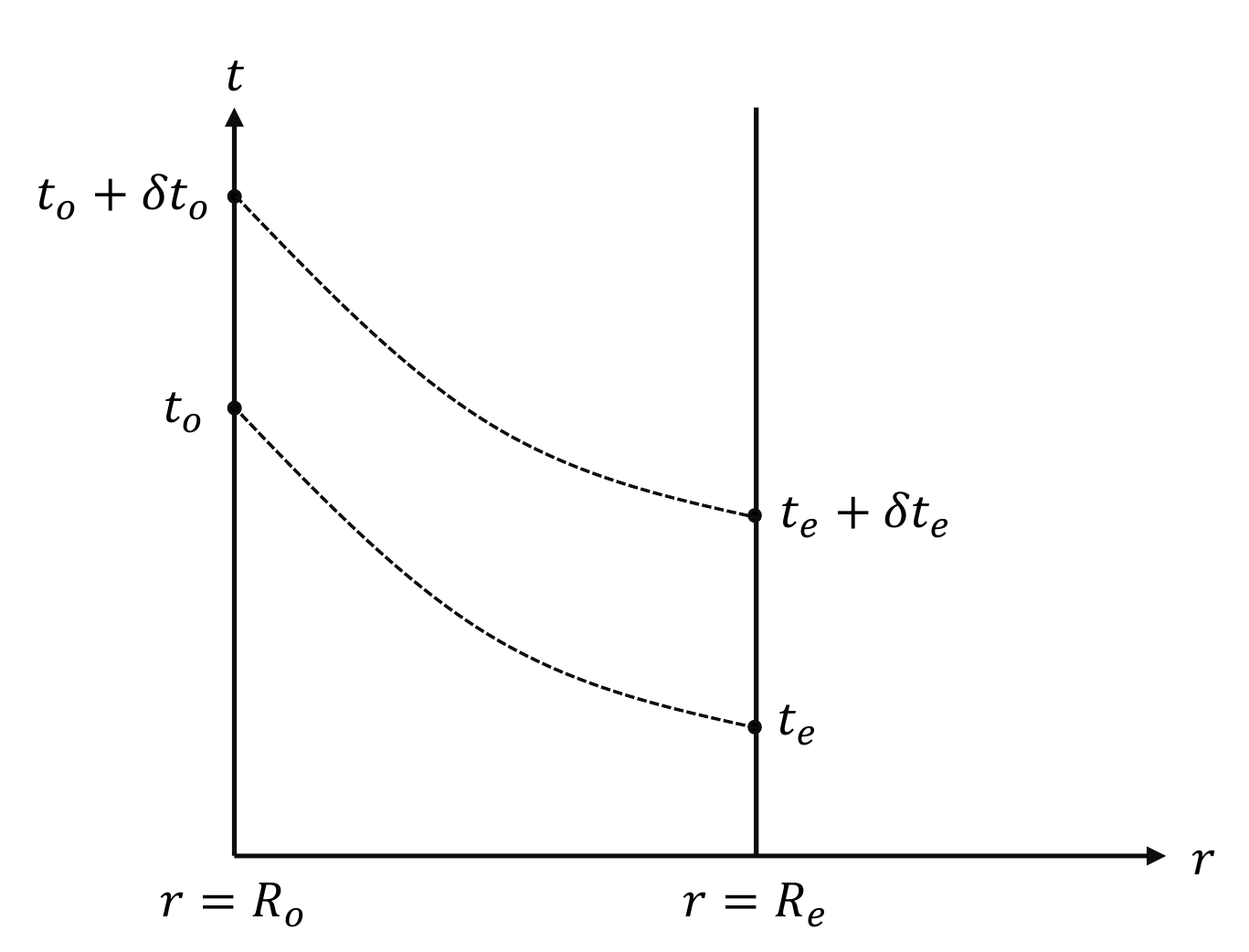}
    \caption{Light rays sent from the emission surface $R_e$ at times $t_e$ and $t_e+\delta t_e$ towards the absorption surface $R_o,$ where they are received at times $t_o$ and $t_o+\delta t_o$ }
    \label{fig:frequency_photons_dS}
\end{figure}

In order to measure a frequency, we send two photons at different times $t_e$ and $t_e+\delta t_e$ from the surface at $R_e(t_e),$ and consequently a detector located at constant radial coordinate $R_o$  measures their absorption at times $t_o$ and $t_o +\delta t_o.$
This is depicted in fig.~\ref{fig:frequency_photons_dS}.
The corresponding emission and absorption frequencies are
\beq
\omega_e = \frac{2\pi}{\delta \tau_e} \, , \qquad
\omega_o = \frac{2\pi}{\delta \tau_o} \, ,
\eeq
where we denoted with $\delta \tau_e$ and $\delta \tau_o$ the infinitesimal proper times at the positions of the emission and absorption, respectively.
In particular, the redshift factor $z$ at a generic point is defined by
\beq
1 + z = \frac{\omega_e}{\omega_o} =  \frac{\delta \tau_o}{\delta \tau_e} \, .
\label{eq:definition_redshift}
\eeq

\paragraph{Time-independent part of the stretched horizon.}
It is possible to relate the proper time to the time coordinates measured at the positions of emission and absorption of the light rays.
If these positions are located at constant radial coordinate, the relation reads
\beq
\delta \tau_e = \sqrt{f(R_e)} \delta t_e \, , \qquad
\delta \tau_o = \sqrt{f(R_o)} \delta t_o \, ,\label{eq:proper_time}
\eeq
and the redshift factor simplifies to
\beq
1 + z  = \frac{\delta t_o}{\delta t_e} \sqrt{\frac{f(R_o)}{f(R_e)}} \, .
\label{eq:definition_redshift2}
\eeq
In order to determine the relative time difference, we consider the lines at constant $v$ of a light ray and its next peak.
They are parameterized by 
\beq
\begin{aligned}
& v (t_e) = t_e + r^* (R_e) = t_o + r^*(R_o) \, , & \\
& v (t_e+\delta t_e) = t_e+ \delta t_e + r^* (R_e) = t_o + \delta t_o + r^*(R_o) \, ,& 
\label{eq:two_observ}
\end{aligned}
\eeq
where we used the fact that the constant value of the null coordinate can be measured in two different points.
By subtracting the two equations, we get $\delta t_e = \delta t_o,$ which via eq.~\eqref{eq:definition_redshift2} gives a redshift 
\beq
1+z = \sqrt{\frac{f(R_o)}{f(R_e)}} = \frac{\omega_e}{\omega_o} \geq 1 \, .
\label{eq:redshift_timeindep}
\eeq
Now we fix the cosmological redshift in the far past and future, according to the constraints in eq.~\eqref{eq:far_limits_stretched_hor}:
\begin{itemize}
    \item In the far past:
    \beq
1 + z_{\lambda'} = \sqrt{\frac{f_1(R_o)}{f_1 (r^{\rm st}_1)}}  \, .
\label{eq:redshift_st1}
\eeq
\item In the far future:
\beq
1 + z_{\lambda} = \sqrt{\frac{f_2(R_o)}{f_2 (r^{\rm st}_2)}}  \, .
\eeq
\end{itemize}
The requirement $z_{\lambda}=z_{\lambda'}$ implies

\beq
\frac{f_1(R_o)}{f_1(r^{\rm st}_1)}  = \frac{f_2(R_o)}{f_2(r^{\rm st}_2)} \, .
\label{eq:constant_redshift_condition_app}
\eeq
As can be seen from the equation above, when $r_1^{\rm st}\rightarrow r_{C1}$, we also need $r_2^{\rm st}\rightarrow r_{C2}$ for the equation to be satisfied.\footnote{Recall that this was part of our original motivation for this prescription.} 
When the transition occurs between SdS$_3$ black holes, we set $R_o=0$ which gives  $r_1^{\rm st}/a_1 = r_2^{\rm st}/a_2$. 
This means that the values $\rho_i$ associated with the stretched horizons in equation \eqref{eq:generic_location_rst1} are equal $\rho^{\rm st}_1=\rho^{\rm st}_2$. This is no longer the case as soon as  $R_o\neq0$ or for the higher dimensional black hole, where it is not possible to choose $R_0=0$ (because it is the location of the singularity of the black hole).

More generally, for the case with $r_{o,1}\neq r_{o,2}$, the position of the stretched horizon after the shock can be fixed by the condition
\beq
\frac{f_1(r_{o,1})}{f_1(r^{\rm st}_1)}  = \frac{f_2(r_{o,2})}{f_2(r^{\rm st}_2)} \, .
\label{eq:far_past_future_stretched_hor_SdS4}
\eeq
relating the early and late time limits of the stretched horizon \eqref{eq:far_limits_stretched_hor}. Note that also in this case the two stretched horizons approach the cosmological horizons together in order for the equation to be satisfied.
Next, let us look at the time-dependent part of the stretched horizon.

\paragraph{Time-dependent part of the stretched horizon.}
We now consider the case of an outgoing light ray sent at an intermediate time, when it propagates in both parts of the geometry before reaching the observer at $r=R_0$. Such light rays are denoted with $\tilde \lambda$ in fig.~\ref{fig:constant_redshift_SdSall}. 
The constant redshift condition in eq.~\eqref{eq:definition_redshift2} has to be modified taking into account that now the stretched horizon is time dependent,
\begin{align}
1+z_{\tilde \lambda} = \sqrt{\frac{f_2(R_o)}{f_1(R_e)}} \frac{1}{\sqrt{1-\frac{(R'_e)^2}{f_1(R_e)^2}}} \frac{dt_o}{dt_e} 
\label{eq:redshift_SdS4}
\end{align}
where $R'_e \equiv dR_e/dt_e$ and we assumed that light rays are still sent at fixed angular coordinates.

An outgoing light ray in this regime crosses the shockwave. Its parametrization in the region $u<u_s$ reads
\beq
v_1 = t_e + r^*_1 (R_e) = -t_w - r^*_2(r^{\rm st}_2) + 2 r^*_1(r_{\rm sh}) \, , 
\label{eq:v1_SdS4}
\eeq
while after the shockwave we have
\beq
v_2 = t_o + r^*_2(R_o) = -t_w -r^*_2(r^{\rm st}_2) + 2 r^*_2 (r_{\rm sh}) \, . 
\label{eq:v2_SdS4}  
\eeq
In these two expressions we denoted by $r_{\rm sh}$ the radial coordinate where the light ray crosses the shockwave and by $t_o$ the time at which the light ray is received by the observer.
Then we proceed in solving for the time dependent position of the stretched horizon $R_e(t_e)$ as follows:
\begin{enumerate}
    \item Determine numerically (or if possible, analytically) $r_{\rm sh}$ as a function of $t_o$ from eq.~\eqref{eq:v2_SdS4}. 
    \item Take the derivative with respect to $t_o$ of eq.~\eqref{eq:v1_SdS4} and extract the expression for $dt_e/dt_o$. \item Plug the previous expression inside eq.~\eqref{eq:redshift_SdS4} rewriting the unknown function $R_e(t_e)$ as a function of $t_o$.
    The constant redshift condition now gives a differential equation for $R_e(t_o)$ which we solve by using the initial condition \beq
    R_e(\tilde{t}_{o,2}) = r^{\rm st}_1 \, .
    \label{eq:to1_SdS4}
    \eeq
    Here $\tilde{t}_{o,2}$ is the time (defined in the region $u>u_s$) where the shock crosses the observer's worldline, \ie
\begin{equation}
    \tilde{t}_{o,2}=-t_w-r_2^*(r_2^{\rm st})+r_2^*(r_{o,2})  \, .
\end{equation}
\end{enumerate}

\begin{figure}[ht]
    \centering
{\label{subfig:timedep_stretched_SdS4_case1} \includegraphics[scale=0.8]{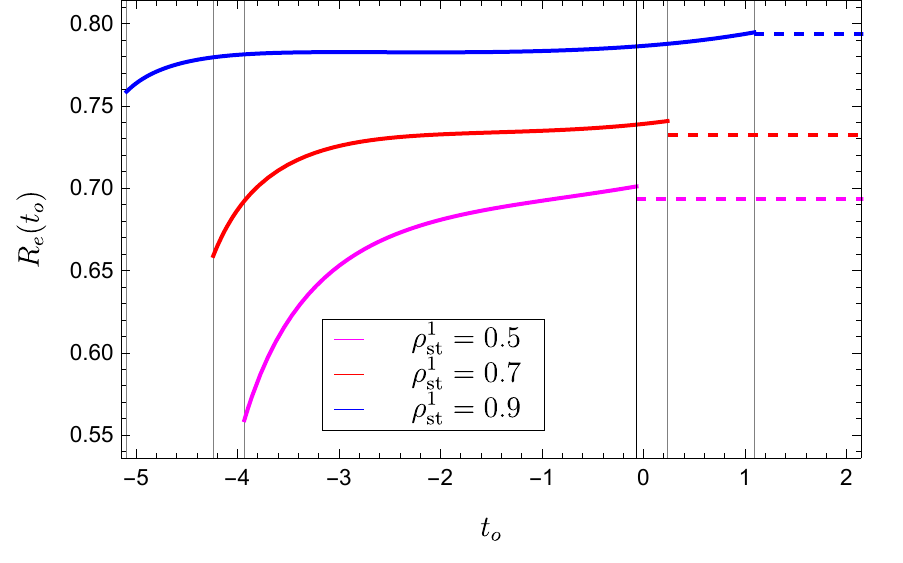}}
\caption{Plots representing the time-dependent solution $R_e(t_o)$ for the stretched horizon in the SdS$_4$ geometry.
     We fix $L=1$, $m_1 = 0.14$, $m_2=0.13, R_o=0.2$  
    and depict the result for various choices of $\rho^1_{\rm st}$ reported in the legend.}
    \label{fig:timedep_stretched_SdS4}
\end{figure}

\noindent
Substituting the solution back into equation \eqref{eq:redshift_SdS4}, we can integrate $t_e(t_o)$ in order to express $R_e(t_e)$. However, in the plots below we have plotted (the curious hybrid) $R_e(t_o)$ for times after the shocks to avoid the jump in the definition of time across the shock.\footnote{The emission time $t_e$ jumps across the shockwave, as a consequence of eq.~\eqref{eq:def_us_appA} with the observation that $R_e(t_{s1})\ne r^{\rm st}_2.$  }
The solution for the time dependent part of the stretched horizon is depicted in fig.~\ref{fig:timedep_stretched_SdS4} for the SdS$_4$ black hole. 
As shown in the figure, the stretched horizon in the presence of a shockwave is not at a constant $r$ coordinate. 
When the first light ray crosses the shock, the stretched horizon becomes time dependent. After the last light ray crosses the shock, the stretched horizon is again at a constant $r$ surface which jumps discontinuously  compared to its previous value.

The discontinuity decreases for $\rho \rightarrow 1,$ for a smaller strength of the shockwave $ \varepsilon \rightarrow 0$ and for an earlier insertion time $t_w \rightarrow \infty $.  
One can check that 
\beq
R_e (t_{s,o2}) \ne r^{\rm st}_2 \, , \qquad
r^*_1 (R_e(t_{s,o2})) \ne r^*_2(r^{\rm st}_2) \, ,
\label{eq:inequalities_timedep_stretched_app}
\eeq 
where $t_{s,o2}$ is defined as the observer's time of the last light ray sent during the time-dependent regime of the stretched horizon.
In other words, it is the time coordinate measured by the observer when a light ray is emitted from the intersection between the stretched horizon $R_e$ and the shockwave.
The latter inequality in \eqref{eq:inequalities_timedep_stretched_app} implies that the coordinate time $t_e$ measured along the stretched horizon is discontinuous when crossing the shockwave.
In this regime the observer's time $t_o$ is instead continuous, and for this reason we used it in the plots. 

Finally, we mention that the constant redshift prescription is not unique: there are many boundary conditions that can be changed such as the location of the observer, the light rays used to compute the redshift and the continuity properties of the stretched horizon.
We discuss these issues and we show that some universal properties of the stretched horizon are preserved by other prescriptions in appendix \ref{app:ping_pong}.

\paragraph{Results in three dimensions.}
The three-dimensional SdS black hole only presents a cosmological horizon, therefore the Penrose diagram reduces to fig.~\ref{fig:constant_redshift} and we can locate the observer at the north pole $R_o=0$.
These simplifications make the problem simpler to study, and now various steps can be performed analytically.

Following the steps outlined below eq.~\eqref{eq:v2_SdS4} with the choice $R_o=0$, we find:
\begin{enumerate}
    \item From eq.~\eqref{eq:v2_SdS4} we determine
\beq
r_{\rm sh}  = a_2 L \tanh \left[ \frac{a_2}{2L} \le t_o+t_w+r^*_2(r^{\rm st}_2) \ri \right] \, ,
\label{eq:rsh_SdS3}
\eeq
where we used the fact that $r_{\rm sh} \leq a_2 L.$
\item Taking the derivative of eq.~\eqref{eq:v1_SdS4} gives
\beq
\frac{dt_e}{dt_o} =\frac{2}{f_1(r_{\rm sh})} \frac{dr_{\rm sh}}{dt_o} \le 1+ \frac{R'(t_e)}{f_1(R_e)}  \ri^{-1}  \, , \qquad
\frac{dr_{\rm sh}}{dt_o} = \frac{a_2^2}{2 \cosh^2 \le \frac{a_2 t_o}{2L} \ri^2} \, ,
\eeq
where the latter identity comes from eq.~\eqref{eq:rsh_SdS3}.
\item Plugging the previous derivative $dt_e/dt_o$ inside eq.~\eqref{eq:redshift_SdS4} leads to the differential equation
\beq
\sqrt{\frac{f_1(r^{\rm st}_1)}{f_1(R_e)} + \frac{\dot{R_e}^2}{f_1(R_e)^2}}
+ \frac{\dot{R_e}}{f_1(R_e)} =  \frac{2}{f_1(r_{\rm sh})} \frac{d r_{\rm sh} }{dt_o} 
\, , 
\eeq
where $\dot{R_e} \equiv dR_e/dt_o$.
After imposing the boundary condition \eqref{eq:to1_SdS4}, we can numerically solve this differential equation for $R_e(t_o)$.
\end{enumerate}

\begin{figure}[ht]
    \centering
 \includegraphics[scale=0.8]{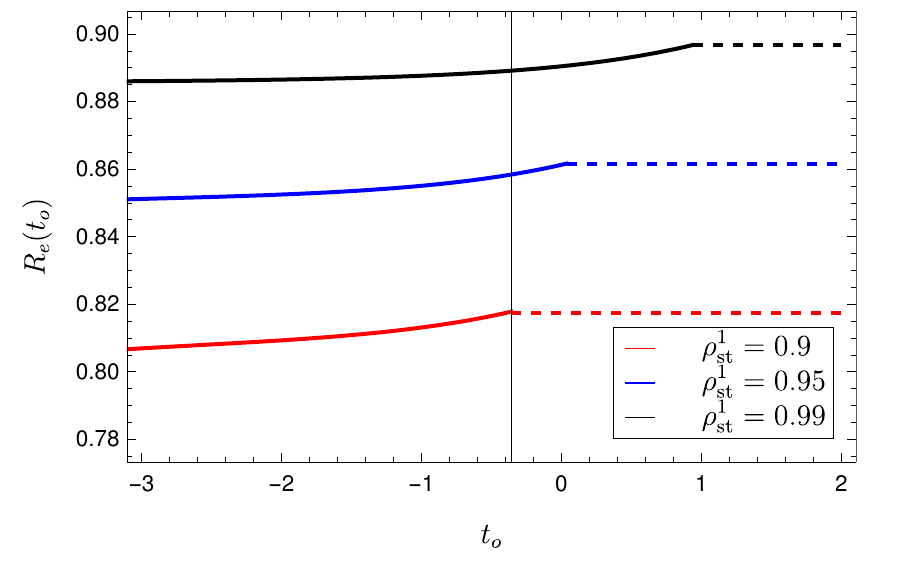}
\caption{Plots representing the time-dependent solution $R_e(t_o)$ for the stretched horizon in the SdS$_3$ geometry.
     We fix $L=1$, $8 G_N \mathcal{E}_1 = 0.2$, $\varepsilon=0.1$ and $R_o=0$  and depict the result for various choices of $\rho^1_{\rm st}$ reported in the legend.}
    \label{fig:timedep_stretched_SdS3}
\end{figure}

\noindent
The solutions for various choices of the parameter $\rho^1_{\rm st}$  are depicted in fig.~\ref{fig:timedep_stretched_SdS3}.
We notice that the stretched horizon jumps after the intersection with the shockwave.
While the jump is initially towards a smaller value of the radial coordinate, for increasing values of $\rho^1_{\rm st}$ the stretched horizon moves towards a bigger value of $r$ instead. 
This is consistent with the requirement that the stretched horizons, either before and after the shockwave, approach the corresponding cosmological horizon together when $\rho^{1}_{\rm st} \rightarrow 1.$
The statement on the jump of the stretched horizon is also valid in higher dimensions, but it is easier to observe in this three-dimensional setting.
The qualitative behaviour is similar to the higher-dimensional case in fig.~\ref{fig:timedep_stretched_SdS4}, therefore the same comments as remarked around the inequalities \eqref{eq:inequalities_timedep_stretched_app} apply here as well.

\subsection{Continuous time prescription}
\label{ssec:details_cont_time}

The constant redshift prescription of section \ref{ssec:details_const_red}
defines the location of the stretched horizon in terms of a physical  experiment, where light rays are sent between a source and a detector in such a way as to keep the cosmological redshift constant.
However, a disadvantage of the previous method is that the de-Sitter time coordinate $t$ along the stretched horizon jumps when crossing the shockwave.
In this section, we examine an alternative to the constant redshift prescription. Instead, we determine the location of the stretched horizon so that the time coordinate is continuous along it.
We further require that the stretched horizon is located at (a different) constant value of the radial coordinate before and after the shockwave insertion.
A welcome outcome of these requirements is that as the stretched horizon before the shockwave approaches the cosmological horizon $r_{C1}$, the stretched horizon after the shockwave will approach the larger cosmological horizon $r_{C2}$. 
This intuitive result was a guiding principles for us to produced this definition of the stretched horizon in a shockwave geometry. This approach is also technically more efficient and, \eg allows us to produce an interesting analytic expression for the scrambling time -- see eq.~\eqref{eq:duration_plateau_SdS4}

Let us assume that the stretched horizon is always located at a constant radial coordinate during all the time evolution, except for a possible jump at the intersection with the shockwave.
We parametrize these constant values according to eq.~\eqref{eq:generic_location_rst1}. 
We now impose that the coordinate time is continuous along the stretched horizon.
By evaluating the constant null direction $u_s$ using the data before and after the shockwave insertion as in eq.~\eqref{eq:def_us_appA}, we find the constraint 
\beq
r^*_1(r_1^{\rm st}) = r^*_2(r_2^{\rm st}) \, .
\label{eq:relation_cont_time}
\eeq
We provide the solution for this identity in some examples of asymptotically dS spacetimes.
In the case of a transition between SdS$_3$ backgrounds, the solutions read
\beq
r_1^{\rm st}= \rho \, a_1 L \, , \qquad
r_2^{\rm st} = a_2 L \, \tanh \left[ \frac{a_2}{2 a_1}  \log \le \frac{1+\rho}{1-\rho}  \ri  \right] \, ,
\label{eq:rst2_SdS}
\eeq
where we denoted with $\rho \equiv \rho_1$ the value of the parameter defined in eq.~\eqref{eq:generic_location_rst1} before the shockwave insertion.
We analytically observe that the stretched horizons before and after the shock approach the corresponding cosmological horizon together when $\rho \rightarrow 1,$ and they both approaches the pole $r=0$ when $\rho \rightarrow 0.$
Similar results can be found for the higher-dimensional SdS black hole, where the solution is numerical.

In section \ref{sec:CV20} we will take a generic and constant value $r^{\rm st}_2$ for the right stretched horizon after the shockwave insertion.
When considering the specific examples in section \ref{sec:examples_CV20}, we will usually specialize to the constant redshift prescription described in section \ref{ssec:details_const_red}.
The requirement of a continuous time coordinate along the stretched horizon discussed here leads to the same qualitative results; minor quantitative differences will be discussed explicitly.

\section{WDW patch}
\label{sec:WDW_patch}

According to the prescription proposed in \cite{Susskind:2021esx,Jorstad:2022mls}, holographic complexity in asymptotically dS spacetime is defined by anchoring extremal surfaces (for the volume) and the WDW patch (for the CV2.0 and action proposals) to the two stretched horizons.
In this section, we study the shape of the WDW patch in preparation for the computation of the CV2.0 conjecture in section \ref{sec:CV20}.

We begin in section \ref{ssec:regularization_WDWpatch} by presenting a regularization prescription that cuts the WDW patch using appropriate cutoff surfaces near timelike infinity.
We study the time dependence of the WDW patch in section \ref{ssec:time_evo_WDW}, identifying in section \ref{ssec:critical_times_dS} a set of critical times parametrizing different regimes of the evolution.
Spacetimes with dS asymptotics present a novel feature compared to the AdS counterpart; they can admit special configurations of the WDW patch where the top and bottom joints move outside the cosmological horizons.
These cases are analyzed in section \ref{ssec:special_configurations_WDW}. 
We finally apply the general techniques to SdS space explicitly in section  \ref{ssec:examples_WDWpatch}.

\subsection{Regularization}
\label{ssec:regularization_WDWpatch}

The hyperfast growth of asymptotically dS spaces is responsible for a divergent complexity, which we regularize by delimiting the integration domain of the WDW patch with cutoff surfaces located near timelike infinity \cite{Jorstad:2022mls}.
Given an asymptotically dS geometry of the form \eqref{eq:dS_metric_shock_wave}, we define the cutoff in the lower region of the Penrose diagram as 
\beq
r_{\rm max,1} = \frac{r_{C1}}{\delta} \, , 
\label{eq:definition_rmax_shocks}
\eeq
where $\delta >0$ is an infinitesimal dimensionless regulator and the cosmological horizon was defined in eq.~\eqref{eq:generic_rc_geometries}.
Unlike the stretched horizon, this cutoff surface near past timelike infinity $\mathcal{I}^-$ does not intersect the shockwave. Therefore there is no need for  prescriptions of the kind discussed in section \ref{sec:stretched_horizon_shocks}. 
We need to independently choose another location for the cutoff surface in the top part of the Penrose diagram (close to $\mathcal{I}^+)$: for simplicity, we set
\beq
r_{\rm max,2} = r_{\rm max,1} = \frac{r_{C1}}{\delta} \, .
\label{eq:rmax2}
\eeq
In the remaining part of the paper, we will be generic and perform the computations by keeping distinct values $r_{\rm max,1}, r_{\rm max,2}$ for the cutoff surfaces; we will use the expressions \eqref{eq:definition_rmax_shocks} and \eqref{eq:rmax2} only in explicit examples.

\begin{figure}[ht]
    \centering
    \subfigure[]{\label{subfig:WDW_dS} \includegraphics[scale=0.48]{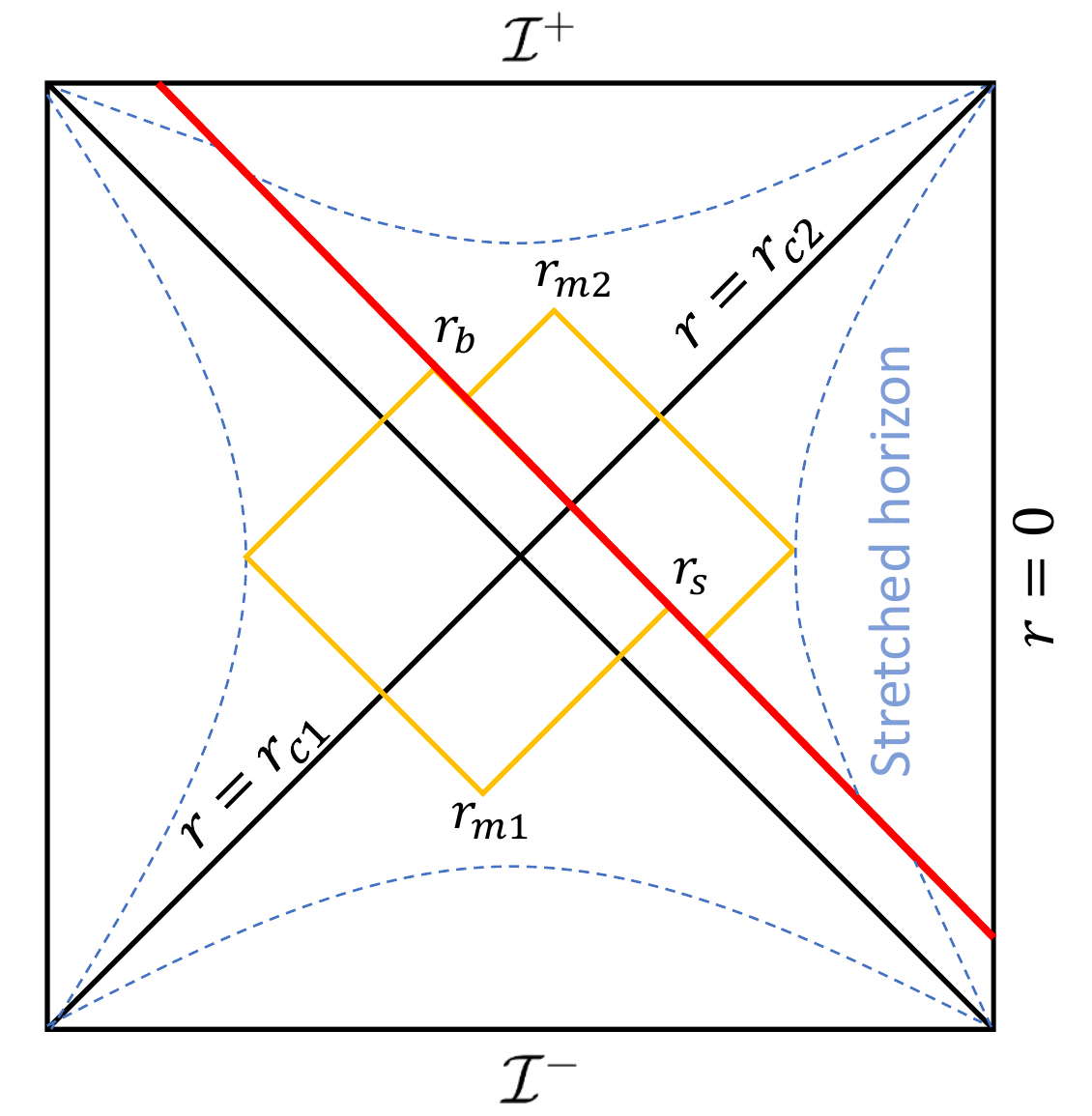} } 
     \subfigure[]{\label{subfig:WDW_SdS} \includegraphics[scale=0.48]{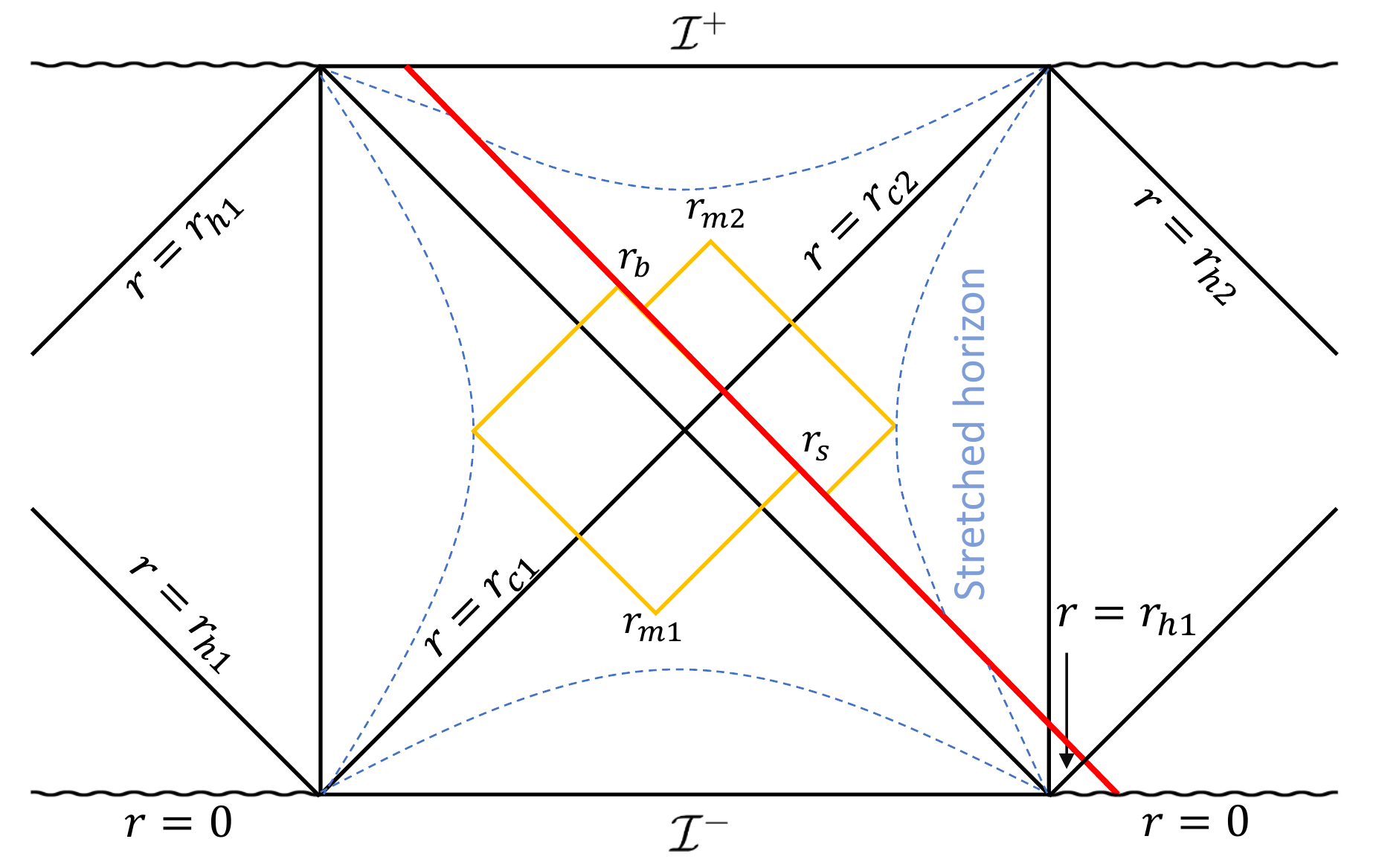} }
    \caption{The WDW patch in SdS$_{d+1}$ perturbed by a shockwave. $r_{c1},r_{h1}$ are the cosmological and black hole horizon before the shockwave. The shockwave can be thought of as throwing away mass from the black hole therefore the mass of the black hole after the shockwave is smaller. $r_{c2},r_{h2}$ are the cosmological and black hole horizon after the shockwave. In the figure $r_s,r_b$ are the intersections between the past and future boundary of the WDW with the shockwave. $r_{m2},r_{m1}$ are the top and bottom joints of the WDW patch.
    (a) WDW patch in SdS$_3$ space perturbed by a shockwave.
   (b) WDW patch in SdS$_{d+1}$ with $d\geq 3$ perturbed by a shockwave. }
   \label{fig:WDW_patches}
  \end{figure}

\subsection{Time evolution of the WDW patch}
\label{ssec:time_evo_WDW}

We will study the time dependence of the WDW patch, referring to the configuration in fig.~\ref{subfig:WDW_dS} for concreteness.
There are certain positions of the WDW patch which play a crucial role in the evolution of the system.
We denote them as follows: 
\begin{itemize}
    \item $r_s$ is the intersection between the shockwave and the past boundary of the WDW patch in the right exterior ($r_s \leq r_{C2}$);
     \item $r_b$ is the intersection between the future boundary of the WDW patch and the shockwave behind the interior of the first cosmological horizon ($r_b \geq r_{C1}$);
    \item $r_{m1}$ is the bottom joint of the WDW patch, obtained as the intersection of its past null boundaries;
     \item $r_{m2}$ is the top joint of the WDW patch, located at the intersection of its future null boundaries.
\end{itemize} 

\noindent
A typical feature of the WDW patch is that the top and bottom joints $r_{m1},r_{m2}$ are located behind the horizon.
This happens for several geometries whose asymptotics is AdS, dS and in many other settings \cite{Brown:2015lvg,Chapman:2016hwi,Carmi:2017jqz,Chapman:2018dem,Chapman:2018lsv,Jorstad:2022mls}. 
In the present case, however, we will find unusual configurations of the WDW patch where the top and bottom vertices of the WDW patch will end up outside the cosmological horizon.

In order to concretely determine the above-mentioned positions, we impose that the appropriate null surfaces intersect.
The idea consists of evaluating the constant null coordinates in two different points: at the stretched horizon where the boundary times are defined, and at the special positions of the WDW patch that we want to determine.
This procedure leads to the identities
\bea
&    t_R+t_w = 2r^*_2(r_s) - 2 r^*_2 (r^{\rm st}_2) \, , &
\label{eq:identity_WDW_times1}\\
&   t_L - t_w =  r^*_1(r^{\rm st}_1) + r^*_2(r^{\rm st}_2) - 2 r^*_1(r_b) \, , &
\label{eq:identity_WDW_times2}\\
&   t_L - t_w = 2 r^*_1 (r_{m1})   - 2 r^*_1 (r_s) - r^*_1(r^{\rm st}_1) + r^*_2(r^{\rm st}_2)  \, , & 
\label{eq:identity_WDW_times3} \\
&   t_R + t_w =  2 r^*_2 (r_{b})  - 2 r^*_2 (r_{m2}) 
\label{eq:identity_WDW_times4} \, .  &
    \eea
Note that the above equations satisfy the symmetry \eqref{eq:time_shift_symmetry} together with the appropriate shift of the shock\footnote{This is not really an isometry of our SdS solution as eq.~\eqref{shifter3} shifts the location of the shockwave. However, our analysis is left invariant with these three shifts of the time parameters.}
\begin{equation}
    t_w \rightarrow t_w+\Delta t\,.
    \label{shifter3}
\end{equation}
By differentiating these relations at fixed boundary time $t_L,$ we find
\beq
\frac{d r_s}{d t_R} =  \frac{f_2 (r_s)}{2} \, , \qquad
\frac{dr_b}{d t_R} = 0 \, , \qquad 
\frac{d r_{m1}}{d t_R} =  \frac{f_1 (r_{m1})}{2} \frac{f_2 (r_s)}{f_1 (r_s)} \, , \qquad
\frac{d r_{m2}}{d t_R} = - \frac{f_2 (r_{m2})}{2}  \, ,
\label{eq:derivatives_tR}
\eeq
while when keeping $t_R$ fixed, we obtain
\beq
\frac{d r_s}{d t_L} = 0 \, , \qquad
\frac{dr_b}{d t_L} = - \frac{f_1 (r_b)}{2}  \, , \qquad 
\frac{d r_{m1}}{d t_L} =   \frac{f_1 (r_{m1})}{2}  \, , \qquad
\frac{d r_{m2}}{d t_L} =  - \frac{f_2 (r_{m2})}{2} \frac{f_1 (r_b)}{f_2 (r_b)} \, .
\label{eq:derivatives_tL}
\eeq

\subsection{Critical times of the WDW patch}
\label{ssec:critical_times_dS}
  
The time dependence of holographic complexity crucially depends on the location of the special positions of the WDW patch, in particular, if they reach the cutoff surfaces near timelike infinity that regularize the spacetime volume.
It turns out that there exist several critical times corresponding to the transition of different shapes of the WDW patch.
While many of the considerations that we will make are not affected by the precise relation between the right and left boundary times, we will mainly focus on the symmetric case \eqref{eq:symmetric_times} for the derivation of the critical times.
We collect in fig.~\ref{fig:full_time_dep_WDW_dS} the various configurations. 
For the sake of simplicity, here and in the following figures, we only show the portion of the Penrose diagram between the two stretched horizons on either side of the cosmological horizon. The full extension of the Penrose diagrams would appear as in fig.~\ref{fig:WDW_patches}.

\begin{figure}[ht]
    \centering
    \subfigure[$t=t_{c1}$.]{\label{subfig:WDW_patch_tc1} \includegraphics[scale=0.6]{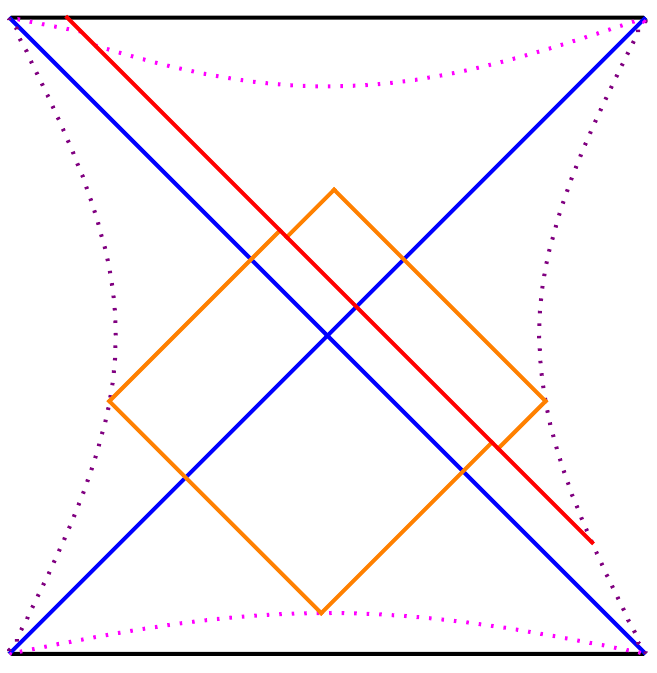}} \quad
      \subfigure[$t=t_{c2}$.]{\label{subfig:WDW_patch_tc2} \includegraphics[scale=0.6]{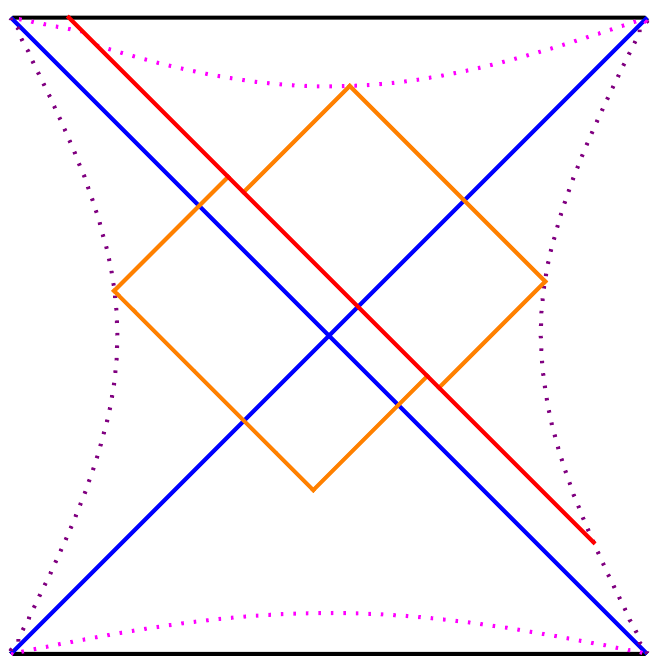}} 
    \quad
     \subfigure[ $t=t_{c3}$.]{\label{subfig:WDW_tc3} \includegraphics[scale=0.383]{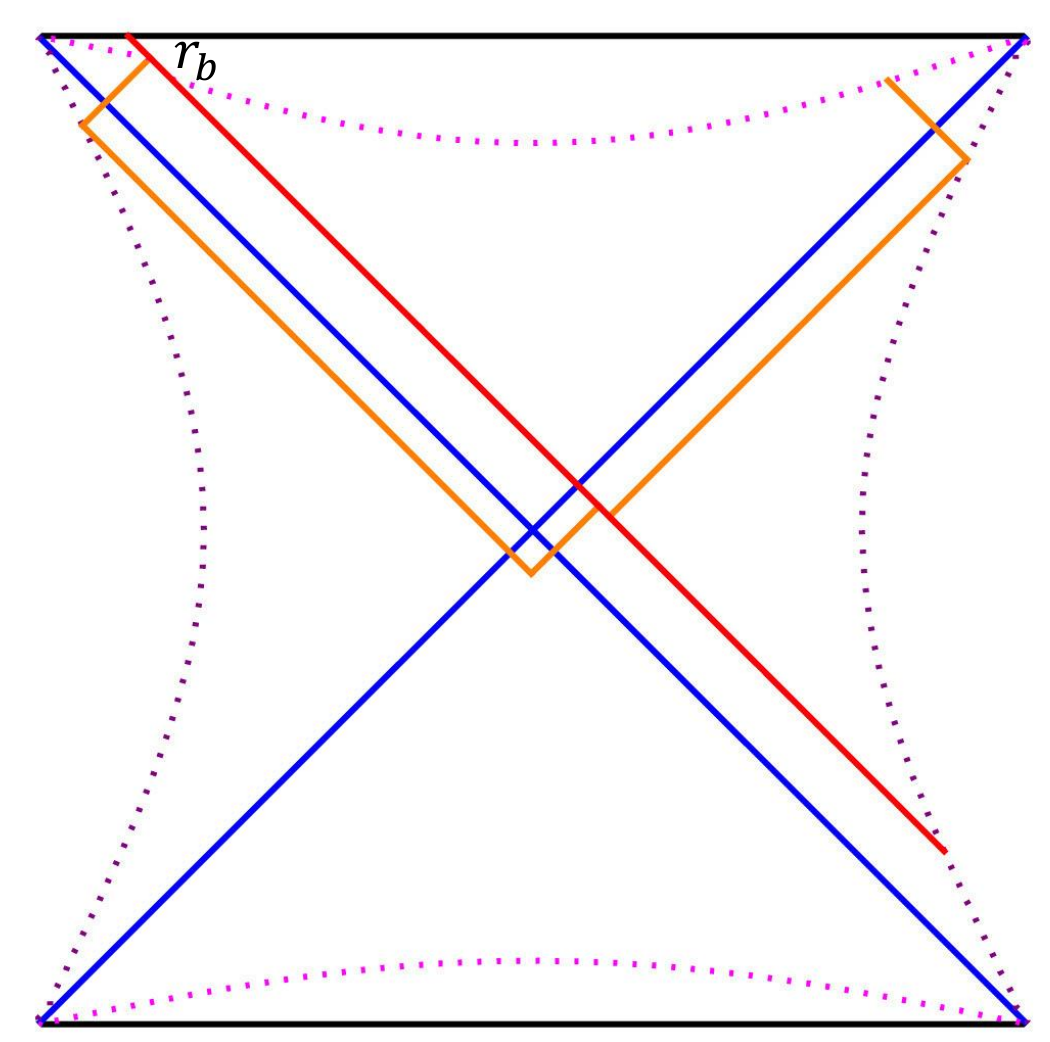}} 
    \caption{Time evolution of the WDW patch in the presence of the shockwave and determination of the critical times.
    (a) Critical time $t_{c1}$ when the bottom vertex of the WDW patch sits on the cutoff surface $r=r_{\rm max,1}$ located close to $\mathcal{I}^-.$
    (b) Critical time $t_{c2}$ when the top vertex of the WDW patch sits on the cutoff surface $r=r_{\rm max,2}$ located close to $\mathcal{I}^+.$
(c) Critical time $t_{c3}$ when $r_b$ intersects the cutoff surface located at $r_{\rm max,2}.$ As indicated in the main text, the Penrose diagrams show only the region between the two stretched horizons. }
    \label{fig:full_time_dep_WDW_dS}
\end{figure}

It is worth mentioning that the cases collected in the above figure always arise for any asymptotically dS metric of the form \eqref{eq:dS_metric_shock_wave}, independently of the value taken by the parameter $\varepsilon$ defined in eq.~\eqref{eq:generic_epsilon_geometries} and of the time $-t_w$ when the shockwave is inserted at the right stretched horizon. 
It turns out that the backgrounds in eq.~\eqref{eq:dS_metric_shock_wave} also allow for configurations which are unusual, \ie the analogous behaviour never appears in asymptotically AdS Vaidya geometries \cite{Chapman:2018dem,Chapman:2018lsv}:
they correspond to the top and bottom vertices of the WDW patch moving outside the cosmological horizon, but the difference is that some of these configurations may or may not occur depending on the choice of parameters describing the shock.
For these reasons, here we only focus on the universal regimes, while we will treat in more detail the special cases in section \ref{ssec:special_configurations_WDW}. 
We will observe that for a shock wave inserted far enough in the past (bigger $t_w$), the special configurations described in section \ref{ssec:special_configurations_WDW} will always appear.

\paragraph{Shockwave insertion.}
The first critical time corresponds to the instant when the shockwave is inserted, at $t_R = - t_w.$
By construction the WDW patch for times $t < - 2t_w$  is completely located in the unperturbed original geometry \eqref{eq:asympt_dS} and the quantities $r_s, r_b$ are not defined simply because the shockwave does not intersect the WDW patch.
In the case of a symmetric time evolution \eqref{eq:symmetric_times}, we will call this critical time
\beq
t_{c0} = - 2 t_w \, .
\eeq

\paragraph{Joints reaching timelike infinities.}
The next critical times $t_{c1}, t_{c2},$ correspond to the instants when the past or future vertices of the WDW patch touch the cutoff surfaces located at $r_{\rm max,1}$ and $ r_{\rm max,2},$ respectively.
The critical time $t_{c1},$ corresponding to the bottom joint touching the cutoff surface $r_{\rm max,1}$ close to past timelike infinity (see fig.~\ref{subfig:WDW_patch_tc1}), is obtained by imposing that $r_{m1}=r_{\rm max,1}.$ 
By manipulating eqs.~\eqref{eq:identity_WDW_times1} and \eqref{eq:identity_WDW_times3} with this requirement, we find 
\bea
& t_w = r^*_1(r_s) +r^*_2(r_s)+ \frac{1}{2} r^*_1(r^{\rm st}_1) - \frac{3}{2}r^*_2(r^{\rm st}_2)  -r^*_1(r_{\rm max,1}) \, ,  &\label{eq:rstc1_identity1} \\
& t_{c1} = 2 t_w - 4 r^*_1(r_s) + 4 r^*_1(r_{\rm max,1}) 
- 2 r^*_1(r^{\rm st}_1) + 2 r^*_2(r^{\rm st}_2) \, . &\label{eq:tc1}
\eea
After fixing the parameter $t_w,$ we obtain 
from eq.~\eqref{eq:rstc1_identity1} the value of $r_s,$ and then we substitute the result into eq.~\eqref{eq:tc1} to obtain the critical time.

One can similarly compute the second critical time when the future vertex touches the cutoff surface $r=r_{\rm max,2},$ corresponding to fig.~\ref{subfig:WDW_patch_tc2}. 
Manipulating 
eqs.~\eqref{eq:identity_WDW_times2} and \eqref{eq:identity_WDW_times4}, we find
\bea
& t_w = r^*_1(r_b) + r^*_2(r_b) - r^*_2(r_{\rm max,2})  - \frac{1}{2} r^*_1(r^{\rm st}_1) - \frac{1}{2} r^*_2(r^{\rm st}_2) \, ,  &\label{eq:rbtc2_identity1} \\
& t_{c2} = - 2 t_w + 4 r^*_2(r_b) - 4 r^*_2(r_{\rm max,2})   \, . &
\label{eq:tc2}
\eea

\paragraph{Special point $r_b$ reaching the cutoff surface.}
The last transition occurs at later times when the joint located at $r=r_b$ touches the future cutoff $r_{\rm max,2}$.
This corresponds to the Penrose diagram in fig.~\ref{subfig:WDW_tc3}.
We impose $ r_b (t_{c3}) = r_{\rm max,2} ,$ which via eq.~\eqref{eq:identity_WDW_times2} gives the analytic solution
\beq
t_{c3} = 2 t_w  + 2 r^*_1(r^{\rm st}_1) + 2r^*_2(r^{\rm st}_2) - 4 r^*_1(r_{\rm max,2}) \, .
\label{eq:critical_time_tc3}
\eeq

\paragraph{Relations between critical times.}
By definition, the critical times satisfy the inequalities $ t_{c1} \leq t_{c2} \leq t_{c3} . $
However, there is in general no definite ordering between $t_{c0}$ and $t_{c1}.$
As summarized in table \ref{tab:critical_times}, when the insertion time is moved closer to $t_w=0,$ there is a transition from the regime where $t_{c0} < t_{c1}$ to the opposite case where $t_{c0} < t_{c1}$. 

\begin{table}[ht]   
\begin{center}    
\begin{tabular}  {|c|c|c|c|c|} \hline  \textbf{Choice of parameters} & \multicolumn{4}{c|}{\textbf{Time ordering}} \\ \hline
\rule{0pt}{3ex} $t_w \gg L $  & $t_{c0}$ &  $t_{c1}$ &  $t_{c2}$ &  $t_{c3}$   \\[0.1cm]
\hline
\rule{0pt}{3ex}
$t_w \ll L $  & $t_{c1}$ &  $t_{c0}$ &  $t_{c2}$ &  $t_{c3}$  \\[0.2cm]
\hline
\end{tabular}   
\caption{Hierarchy between the critical times in the regimes with either $t_w \gg L$ or $t_w\ll L.$ When the shockwave is sent at times closer to 0, there is a change in the ordering of $t_{c0}$ and $t_{c1}.$
The other critical times always have a definite order.} 
\label{tab:critical_times}
\end{center}
\end{table}

We remind the reader that the shockwave is always inserted in the past, \ie at the time $t_R=-t_w$ with $t_w \geq 0.$ We will focus mostly on the case $t_{c0}<t_{c1}$. As noted in the table, this occurs when the insertion time of the shock is very early, $t_w\gg L$. Further, as mentioned already, our analysis focuses on the time dependence of the volume of the WDW patch after it crosses the shock, for $t>t_{c0}$. Earlier than that, the time dependence is influenced by the time-dependent part of the stretched horizon. This analysis is more complicated and we leave it for future investigations.

\subsection{Special configurations of the WDW patch}
\label{ssec:special_configurations_WDW}

\begin{figure}[ht]
    \centering
   \subfigure[]{\label{fig:alternative1_WDWpatch} \includegraphics[scale=0.64]{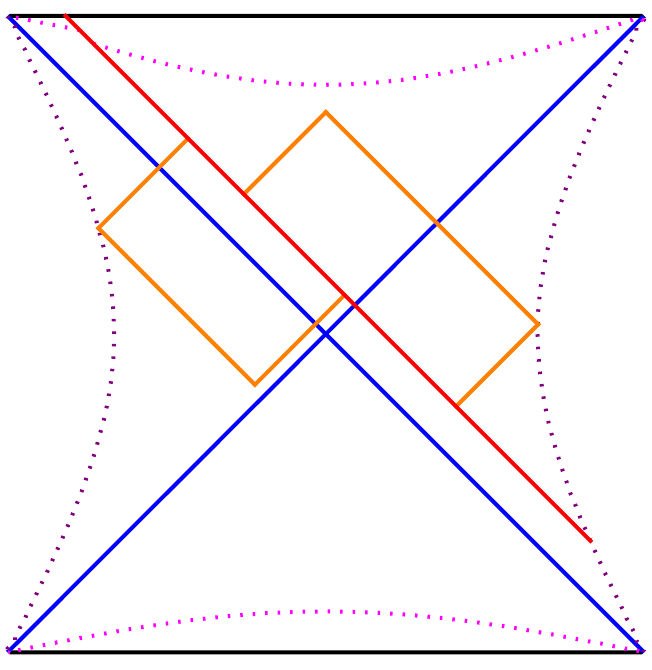}} \quad
    \subfigure[]{\label{fig:alternative2_WDWpatch} \includegraphics[scale=0.64]{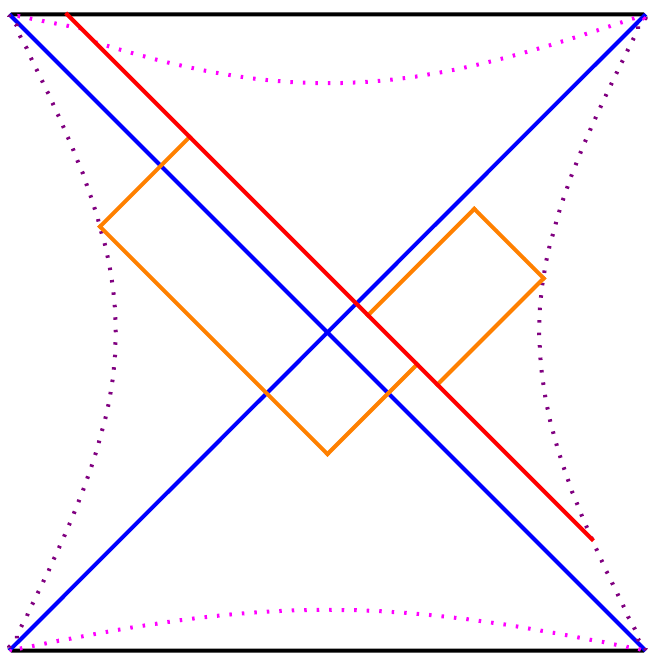}} \quad
    \subfigure[]{\label{fig:alternative3_WDWpatch} \includegraphics[scale=0.64]{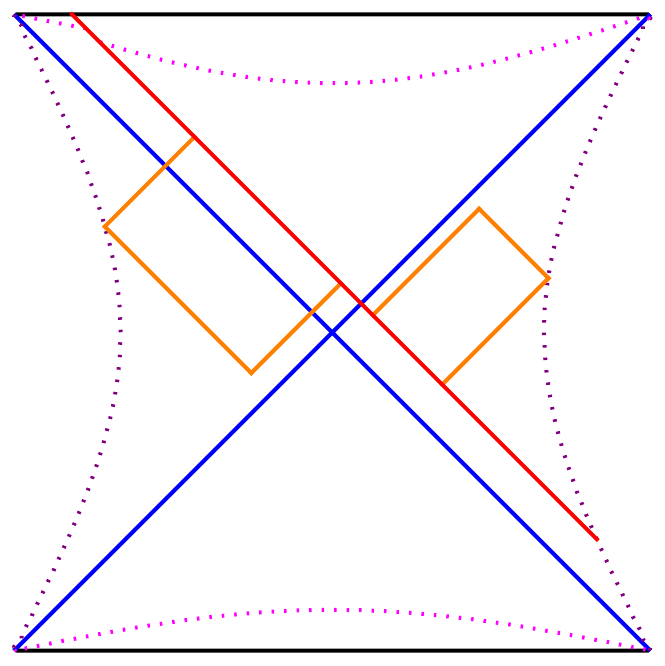}}
    \caption{Exceptional shapes of the WDW patch, corresponding to the case (a) when the special point $r_s$ of the WDW patch sits in the range $ [r_{C1}, r_{C2}],$ (b) or when the special point $r_b$ sits in the same interval.
    Case (c) corresponds to both $r_s, r_b$ being located in the range $[r_{C1}, r_{C2}].$}
    \label{fig:alternative_WDWpatch}
\end{figure}

We remark on a special feature of the special positions of the WDW patch discussed in section \ref{ssec:time_evo_WDW}, which does not manifest in the AdS counterpart: the null energy condition fixes $r_{C1} \leq r_{C2}$ for all the geometries in asymptotically dS spacetimes.
For this reason, there exist finite times $t_{c,s}$ and $t_{c,b}$ such that the special positions of the WDW patch reach the values $r_s=r_{C1}$ and $r_b=r_{C2},$ respectively.
By plugging these results inside the identities defining the position of the top and bottom joints of the WDW patch, see eqs.~\eqref{eq:identity_WDW_times3} and \eqref{eq:identity_WDW_times4}, we find
\beq
\begin{aligned}
& r_s (t_{c,s}) = r_{C1}   \, \Rightarrow \, 
r_{m1} = r_{C1} \, , & \\
& r_b (t_{c,b}) = r_{C2}  \, \Rightarrow \, r_{m2} = r_{C2} \, . &
\end{aligned}
\label{eq:special_limit_rsrb}
\eeq
Therefore, we conclude that the critical time $t_{c,s}$ can be equivalently identified either as the special position $r_s$ or the bottom vertex $r_{m1}$ reaching the cosmological horizon $r_{C1}.$
In a similar way, the critical time $t_{c,b}$ corresponds to $r_b$ and the top joint of the WDW patch to reach the cosmological horizon $r_{C2}$.
These configurations change the shape of the WDW patch, as depicted in figs.~\ref{fig:alternative1_WDWpatch} and \ref{fig:alternative2_WDWpatch}.\footnote{
 It is important to observe that these configurations arise due to the time advance experienced by light rays crossing the shockwave. As a result of the latter, the two stretched horizons are in causal contact, \eg as can be seen by extending the null rays defining the top left boundary of the WDW patch in fig.~\ref{fig:alternative2_WDWpatch}. We discuss this feature further in section \ref{sec:discussion}. 
\label{footy178}  }

Indeed, one can show for all the geometries described by eq.~\eqref{eq:dS_metric_shock_wave} that $r_s$ is a monotonically increasing function of time which runs along the range $[0,r_{C2}].$ 
Once it sits in the interval $[r_{C1}, r_{C2}],$ its location in the Penrose diagram jumps from the right to the top quadrant of the Penrose diagram (this refers to the intersection point before the shockwave), changing the shape of the WDW patch as depicted in fig.~\ref{fig:alternative1_WDWpatch}.
The critical time corresponding to the beginning of this regime is obtained from eq.~\eqref{eq:identity_WDW_times1} by imposing $r_s=r_{C1}.$
For symmetric boundary times as in eq.~\eqref{eq:symmetric_times}, this gives 
\beq
t_{c,s} = -2 t_w + 4 r^*_2 (r_{C1})  -4 r^*_2 (r^{\rm st}_2)  \, .
\label{eq:definition_tcs}
\eeq
A consequence of this peculiar regime is that the bottom joint of the WDW patch stops being located outside of the cosmological horizon $r_{C1}$ (\ie with $r_{m1}>r_{C1}$) rather moves to the interior region corresponding to the left quadrant of the Penrose diagram.
In terms of the radial coordinate, $r_s$ reaches the cosmological horizon $r_{C1}$ from below, while the joint $r_{m1}$ reaches the same location from above.

Similar considerations apply to $r_b,$ which is generically a monotonically increasing function of time along the range $[r_{C1}, r_{\rm max,1}]$.
When $r_b$ sits in the interval $[r_{C1}, r_{C2}],$ it 
is located in the right quadrant of the Penrose diagram (this refers to the intersection point after the shockwave),
as depicted in  fig.~\ref{fig:alternative2_WDWpatch}.
The critical time $t_{c,b}$ such that this configuration is achieved corresponds to imposing  $r_b=r_{C2}$ in eq.~\eqref{eq:identity_WDW_times2}, giving in the symmetric setting \eqref{eq:symmetric_times} the expression
\beq
t_{c,b} = 2 t_w - 4 r^*_1 (r_{C2})  + 2 r^*_1 (r^{\rm st}_1) + 2 r^*_2 (r^{\rm st}_2) \, . 
\label{eq:definition_tcb}
\eeq
In this regime, the top joint of the WDW patch moves towards the exterior of the cosmological horizon $r_{C2}.$

Depending on the choice of the parameters $(\rho, \varepsilon, t_w)$ describing the shockwave geometry \eqref{eq:dS_metric_shock_wave}, it can happen that $t_{c,s}  \leq t_{c,b}$.
In the list of parameters, we used $\rho$ to denote $\rho_1$ from equation \eqref{eq:generic_location_rst1} fixing the location of the stretched horizon in the far past, before the shockwave. We will refer to this parameter as simply $\rho$ from now on.
In that case, there exists a time regime $t_{c,s}\leq t \leq t_{c,b}$ such that both the special points of the WDW patch are located in the interval $[r_{C1}, r_{C2}]$ and the corresponding Penrose diagram is represented in fig.~\ref{fig:alternative3_WDWpatch}.

\subsubsection{Joints behind the stretched horizon} \label{special99}

\begin{figure}[ht]
    \centering
    \includegraphics[scale=0.5]{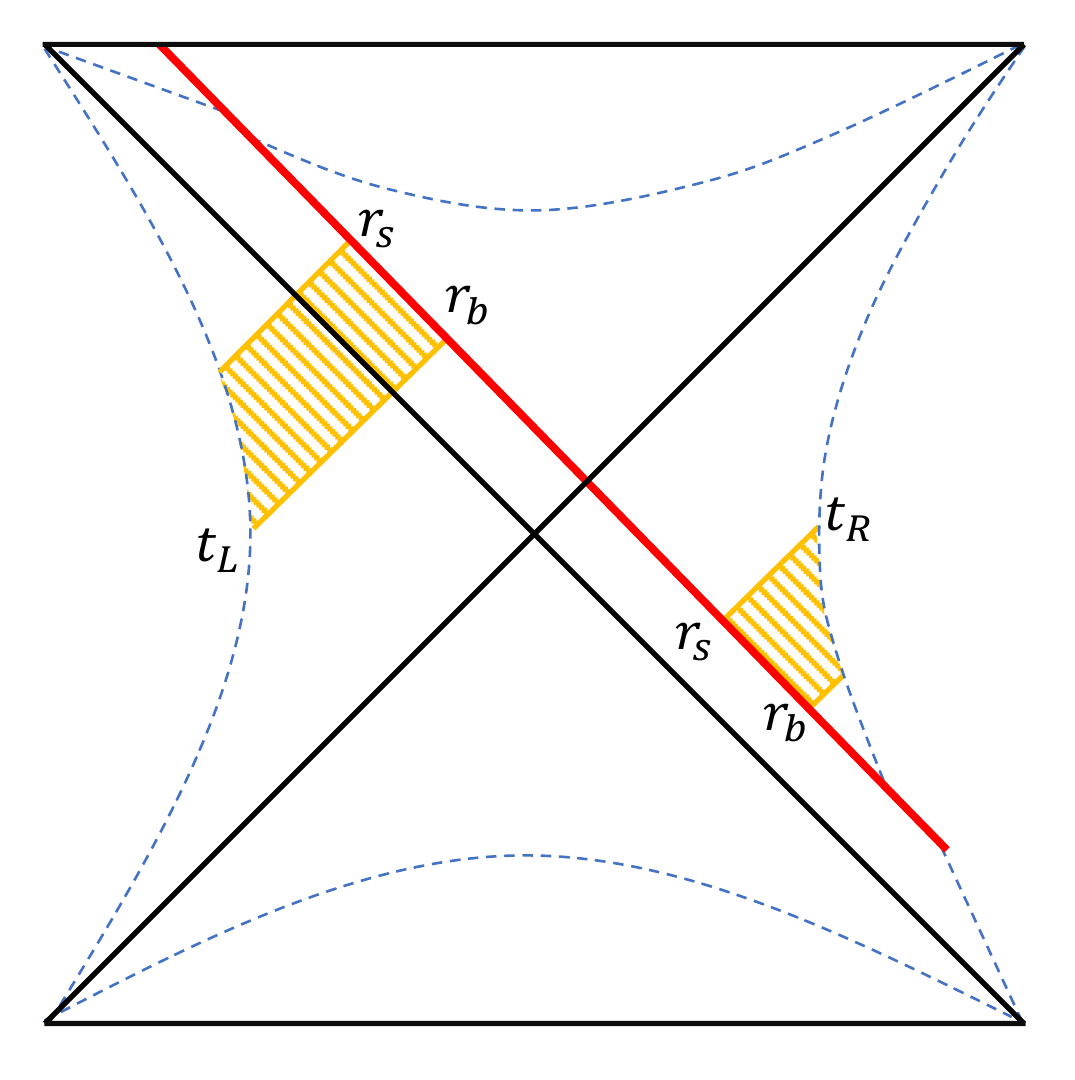}
    \caption{Plot of the WDW patch in the regime when the top and bottom joints move behind the stretched horizons. }
    \label{fig:alternative4_WDWpatch}
\end{figure}

Since the special configurations studied here allow for the joints $r_{m1}$ and $r_{m2}$ to move outside from the cosmological horizon, it is also possible for them to reach the location of the stretched horizons, thus changing even further the form of the WDW patch and introducing two other critical times into the game.
For a symmetric time evolution \eqref{eq:symmetric_times}, this only happens for a part of the time range $t_{c,s}  \leq t\leq t_{c,b}$, corresponding to the configuration depicted in fig.~\ref{fig:alternative3_WDWpatch}. In this case, the WDW patch looks as in figure  \ref{fig:alternative4_WDWpatch}. There are two critical times defined by the following (equivalent) conditions defining the beginning and ending times of these special configurations:
\begin{itemize}
    \item The left future and right past boundaries of the WDW patch meet the shockwave with $r_s=r_b$. 
    \item The bottom joint reaches the stretched horizon before the shockwave insertion, \ie $r_{m1}=r^{\rm st}_1.$ 
    \item The top joint reaches the stretched horizon after the shockwave insertion, \ie $r_{m2}=r^{\rm st}_2.$
\end{itemize}

\noindent
It is easy to show the equivalence of these statements by using the identities which describe the time evolution of the WDW patch.
If we impose $r_s=r_b$ and subtract the eqs.~\eqref{eq:identity_WDW_times2} and \eqref{eq:identity_WDW_times3}, we obtain
\beq
2 r^*_1 (r^{\rm st}_1) - 2 r^*_1(r_{m1}) = 0 \quad \Rightarrow \quad
r_{m1} = r^{\rm st}_1 \, .
\eeq
If we subtract the identities \eqref{eq:identity_WDW_times1} and \eqref{eq:identity_WDW_times4} instead, we find
\beq
2 r^*_2 (r^{\rm st}_2) - 2 r^*_2(r_{m2}) = 0 \quad \Rightarrow \quad
r_{m2} = r^{\rm st}_2 \, ,
\eeq
which shows that $r_s=r_b$ implies that the vertices of the WDW patch reach the corresponding stretched horizon.
By performing the same manipulations, one can also show the opposite direction, thus proving the equivalence of the previous bullets.

We denote with $t_{c,\mathrm{st}}$ the critical times (there can be either two or none of those) corresponding to the instant when the joints reach (or leave) the stretched horizons.
By manipulating eqs.~\eqref{eq:identity_WDW_times1} and \eqref{eq:identity_WDW_times2} under the assumptions that $r_s=r_b,$  $r_{m1}=r^{\rm st}_{1}$ and $r_{m2}=r^{\rm st}_{2}$, we find
\bea
& t_w = r^*_1(r_s) + r^*_2(r_s) - \frac{1}{2} r^*_1(r^{\rm st}_1) - \frac{3}{2} r^*_2 (r^{\rm st}_2) \, ,  & 
\label{eq:identity1_critical_time_behind_stretched} \\
& t_{c,\mathrm{st}} = 2 t_w - 4 r^*_1(r_s) + 2 r^*_1(r^{\rm st}_1)  + 2 r^*_2(r^{\rm st}_2) \, . &  
\label{eq:critical_time_behind_stretched}
\eea
We numerically solve the first equation for $r_s$ and then plug the result into the second one to obtain the corresponding critical times.
In the computation of CV2.0 conjecture, we need to take into account the possibility that the WDW patch assumes a new shape after the joints moved behind the stretched horizons, corresponding to the Penrose diagram depicted in fig.~\ref{fig:alternative4_WDWpatch}.
In this configuration, the stretched horizons cut the WDW patch.

\paragraph{Hierarchy between critical times.}
It is important to observe that the identity \eqref{eq:identity1_critical_time_behind_stretched} may or may not admit solutions for fixed values of the parameters $(\rho,\varepsilon,t_w)$ and generic time.
If the solutions exist, they must be two because the WDW patch will enter and then exit the configuration in fig.~\ref{fig:alternative4_WDWpatch} with the time scales satisfying

\beq
 t_{c,s} \leq t_{c,\mathrm{st1}} \leq 
 t_{c,\mathrm{st2}} \leq t_{c,b}  \, .
\eeq
The reason for this hierarchy is evident from the corresponding definitions: the critical times $t_{c,\mathrm{st}}$ can only exist when $r_s$ and $r_b$ are both located inside the interval $[r_{C1}, r_{C2}],$ which is determined by the critical times $t_{c,s}$ and $t_{c,b}.$
Furthermore, it is also clear that the following inequalities hold
\beq
t_{c1} \leq t_{c,\mathrm{st1}} \leq t_{c,\mathrm{st2}} \leq t_{c2} \, ,
\label{eq:hierarchy_tcst_tc1}
\eeq
because by definition the critical times $t_{c1}, t_{c2}$ require the joints of the WDW patch to reach timelike infinity and to be located inside the cosmological horizon, while $t_{c,\mathrm{st}}$ assume the joints to move outside from the horizon.
These statements are summarized in table \ref{tab:special_critical_times}.

\begin{table}[ht]   
\begin{center}    
\begin{tabular}  {|c|c|c|c|c|} \hline  \textbf{Comparison} & \multicolumn{4}{c|}{\textbf{Time ordering}} \\ \hline
\rule{0pt}{3ex} Hierarchy with fig.~\ref{fig:alternative1_WDWpatch} and \ref{fig:alternative2_WDWpatch} & $t_{c,s}$ &  $t_{c,\mathrm{st1}}$ &  $t_{c,\mathrm{st2}}$ &  $t_{c,b}$   \\[0.1cm]
\hline
\rule{0pt}{3ex} Hierarchy with fig.~\ref{subfig:WDW_patch_tc1} and \ref{subfig:WDW_patch_tc2}
 & $t_{c1}$ &  $t_{c,\mathrm{st1}}$ &  $t_{c,\mathrm{st2}}$ &  $t_{c2}$  \\[0.2cm]
\hline
\end{tabular}   
\caption{Hierarchies involving the critical times $t_{c,\mathrm{st1}},t_{c,\mathrm{st2}}$ characterized by the shape in fig.~\ref{fig:alternative4_WDWpatch} of the WDW patch. In the first row, the comparison is made with respect to the critical times $t_{c,s}$ and $t_{c,b}$ defining the beginning of the regimes in fig.~\ref{fig:alternative1_WDWpatch} and \ref{fig:alternative2_WDWpatch}, respectively. In the second row, the comparison is done with the standard shapes in fig.~\ref{subfig:WDW_patch_tc1} and \ref{subfig:WDW_patch_tc2}.} 
\label{tab:special_critical_times}
\end{center}
\end{table}

Before proceeding, we remark that for large enough $t_w$, the special configuration in fig.~\ref{fig:alternative4_WDWpatch} exists.
In general, there is no definite relation between $t_{c1}$ and $t_{c,s}$ or between $t_{c2}$ and $t_{c,b}$. 
Even further, there is no definite relation between $t_{c,s}$ and the insertion time of the shockwave $t_{c0}=-2t_w.$
This is clear from eqs.~\eqref{eq:definition_tcs}, which implies
\beq
t_{c,s} - t_{c,0} = 4 r^*_2 (r_{C1}) - 4 r^*_2(r^{\rm st}_2) \, .
\eeq
Depending on the choice of the parameters $(\rho,\varepsilon),$ one can make the second term subleading or dominant, thus allowing for two different orderings between these times.
One can similarly take eq.~\eqref{eq:definition_tcb} and compute
\beq
t_{c,b} - t_{c,0} = 4 t_w - 4 r^*_1 (r_{C2}) +2 r^*_1(r^{\rm st}_1) +2 r^*_2(r^{\rm st}_2) \, .
\eeq
In most of the cases, one finds $t_{c0} < t_{c,b},$ but one can get the opposite ordering when the parameters satisfy
\beq
t_w \ll L \, , \qquad
\varepsilon \gg 1 \, , \qquad
\rho \sim 1 \, .
\eeq
This shows that several hierarchies are possible, which are relevant for the time evolution of CV2.0.

\subsection{Explicit cases}
\label{ssec:examples_WDWpatch}

In three bulk dimensions, we can analytically compute the time dependence of the special positions of the WDW patch.
In the following analysis, we will assume to work at $t\geq t_{c0}$. We leave the treatment of earlier times for future studies.

\paragraph{Three dimensions.}
The results for the three-dimensional SdS black hole can be determined analytically.
Starting from eqs.~\eqref{eq:identity_WDW_times1} and \eqref{eq:identity_WDW_times2}, we find
\beq
r_s = La_2 \tanh \left[ \frac{a_2}{2L}(t_R + t_w +2 r^*_2(r^{\rm st}_2))  \right] \, ,
\label{eq:rs_analytic_SdS3}
\eeq
\beq
r_b = La_1 \coth \left[ \frac{a_1}{2L}(-t_L+t_w+r^*_1(r^{\rm st}_1) +r^*_2(r^{\rm st}_2)) \right] \, .
\label{eq:rb_analytic_SdS3}
\eeq
The lower and top joints of the WDW patch are defined by cases, depending on the range of $r_s$ and $r_b.$

\begin{figure}[ht]
    \centering
  \subfigure[]{\label{fig:plotrs_rb1} \includegraphics[scale=0.6]{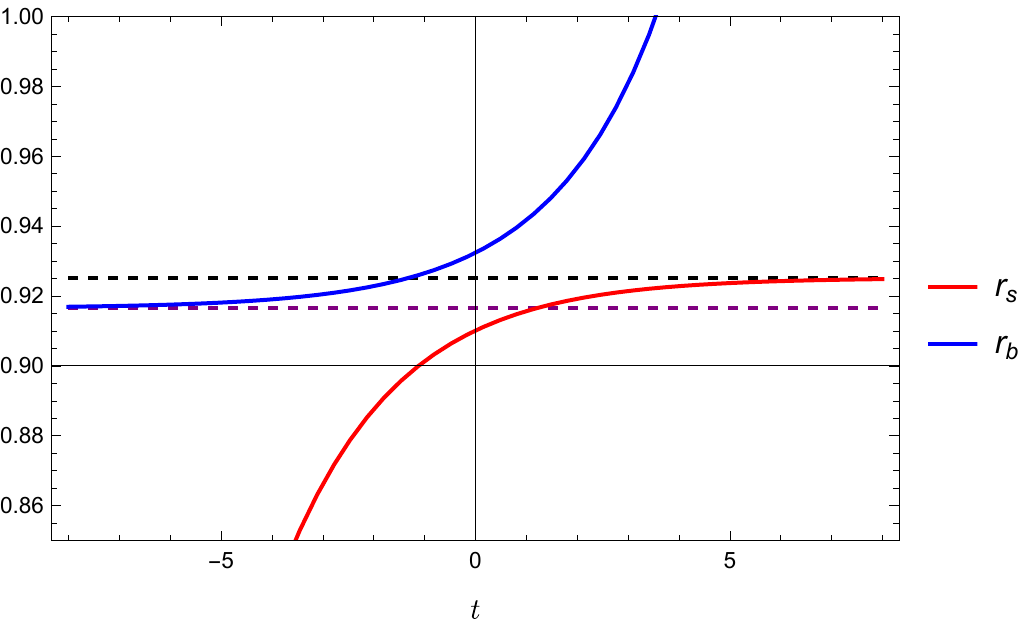}}
  \subfigure[]{\label{fig:plotrs_rb2} \includegraphics[scale=0.6]{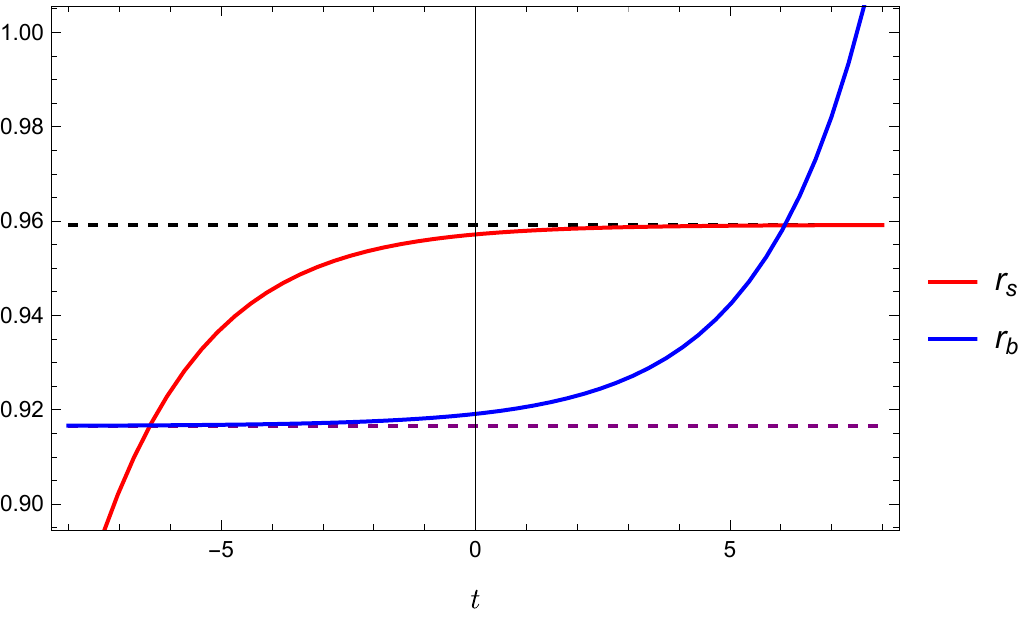}}
    \caption{Plot of the quantities $r_s$ and $r_b$ in eqs.~\eqref{eq:rs_analytic_SdS3} and \eqref{eq:rb_analytic_SdS3}.
    The horizontal dashed lines correspond to the cosmological horizons $a_1L$ and $a_2L$. 
    We fix the values $L=1, \rho=0.5, G_N\mathcal{E}_1 = 0.02 $ and we consider the symmetric time configuration \eqref{eq:symmetric_times}.
    The stretched horizons are selected according to the prescriptions in eq.~\eqref{eq:generic_location_rst1} (and \eqref{eq:constant_redshift_condition_app} with $R_0=0$).
    Panel (a)  $\varepsilon=0.1, t_w=4$.
    Panel (b) $\varepsilon=0.5, t_w=6$. In this case, there is a regime when both the solutions are located inside the interval $[r_{C1}, r_{C2}],$ and the WDW patch takes the shape in fig.~\ref{fig:alternative3_WDWpatch} and \eqref{fig:alternative4_WDWpatch}.}    \label{fig:plot_rsrb_dS}
\end{figure}

The time dependence of these functions in the symmetric configuration \eqref{eq:symmetric_times} is depicted in fig.~\ref{fig:plotrs_rb1} and \ref{fig:plotrs_rb2}.
The plots represent the evolution for fixed values of $(\rho, t_w)$ and two different choices of $\varepsilon$.
As it is clear from the analytic expressions, after the insertion of the shockwave at $t=-2t_w$, both the positions are monotonically increasing functions of the boundary time. 
In particular, $r_s$ moves from regions closer to the north pole towards the cosmological horizon $r_{C2},$ while $r_b$ moves from the horizon $r_{C1}$ towards future timelike infinity $\mathcal{I}^+.$

Substituting the previous results into eqs.~\eqref{eq:identity_WDW_times3} and \eqref{eq:identity_WDW_times4}, we obtain
\beq
r_{m1} = \begin{cases}
La_1 \coth \left[ \frac{ a_1}{2L}(t_L-t_w + 2 r^*_1(r_s)+r^*_1(r^{\rm st}_1)-r^*_2(r^{\rm st}_2)) \right] & \mathrm{if} \,\, r_s \leq r_{C1} \\
La_1 \tanh \left[ \frac{ a_1}{2L}(t_L-t_w + 2 r^*_1(r_s)+r^*_1(r^{\rm st}_1)-r^*_2(r^{\rm st}_2)) \right] & \mathrm{if} \,\, r_{C1} < r_s \leq r_{C2} ,\\
\end{cases}
\label{eq:rm1_analytic_SdS3}
\eeq
\beq
r_{m2} = \begin{cases}
La_2 \coth \left[ \frac{a_2}{2 L} (-t_R - t_w + 2 r^*_2(r_b))\right] & \mathrm{if} \,\, r_b \geq r_{C2} \\
La_2 \tanh \left[ \frac{a_2}{2 L} (-t_R - t_w + 2 r^*_2(r_b))\right] & \mathrm{if} \,\, r_{C1} \leq r_b < r_{C2}. \\
\end{cases}
\label{eq:rm2_analytic_SdS3}
\eeq
Notice that these functions are defined by cases because we need to pick a different solution depending on the evolution of the WDW patch.
When either of the positions $r_s,r_b$ enter the interval $[r_{C1}, r_{C2}],$ then the joints $r_{m1}, r_{m2}$ of the WDW patch move outside the corresponding cosmological horizon.

\paragraph{Higher dimensions.}
In higher dimensions, it is not possible to analytically solve the full-time evolution of the special positions of the WDW patch.
However, one can check numerically that the time dependence of the special points of the WDW patch is qualitatively similar to the previous plots.

\section{Complexity=Volume 2.0}
\label{sec:CV20}

\begin{figure}[ht]
    \centering
     \subfigure[]{\label{subfig:first_regime_ds} \includegraphics[scale=0.63]{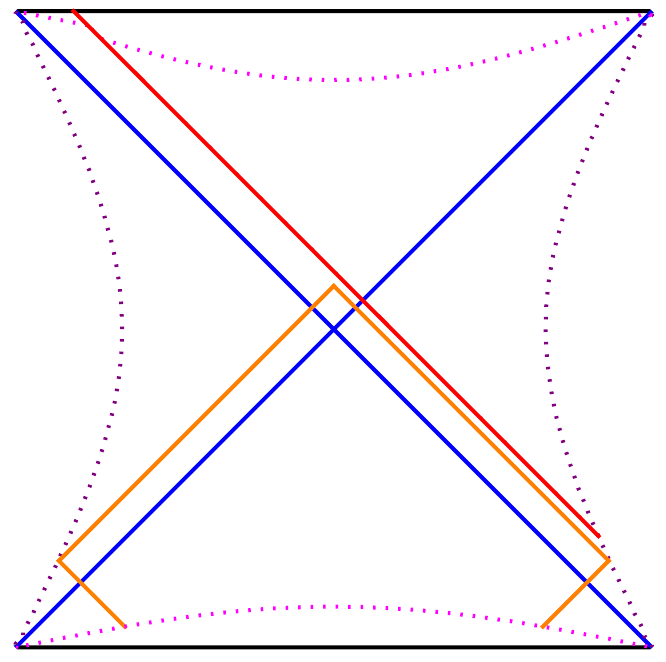}} \quad
     \subfigure[]{\includegraphics[scale=0.63]{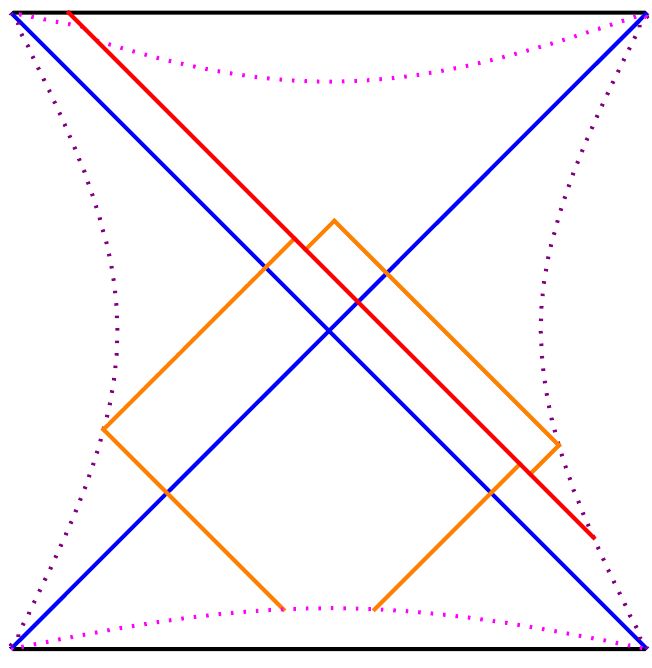}} \quad
      \subfigure[]{\label{subfig:intermediate_regime_dS}  \includegraphics[scale=0.63]{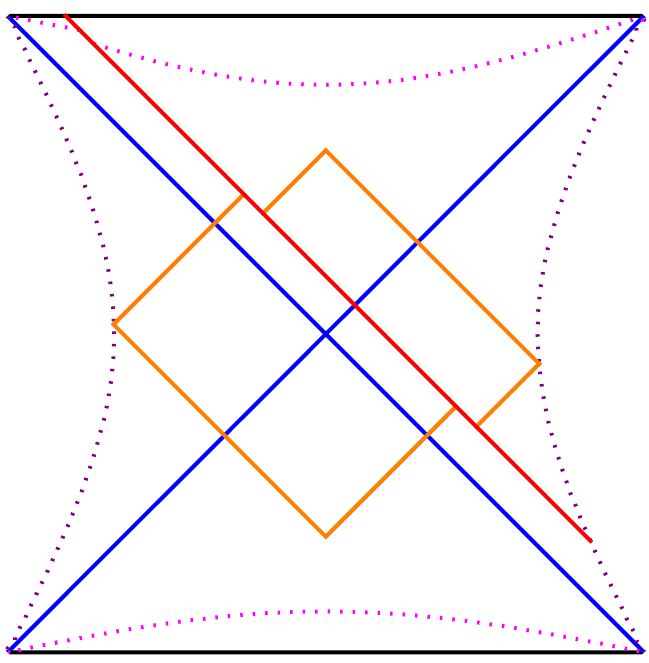}} \\
      \subfigure[]{\label{fig:WDW3reg}\includegraphics[scale=0.63]{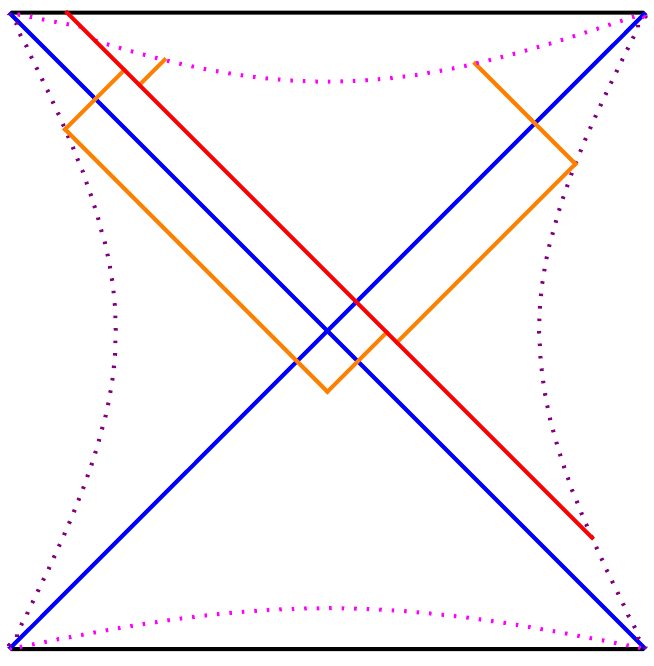}} \quad
        \subfigure[]{\label{fig:WDW4reg}\includegraphics[scale=0.63]{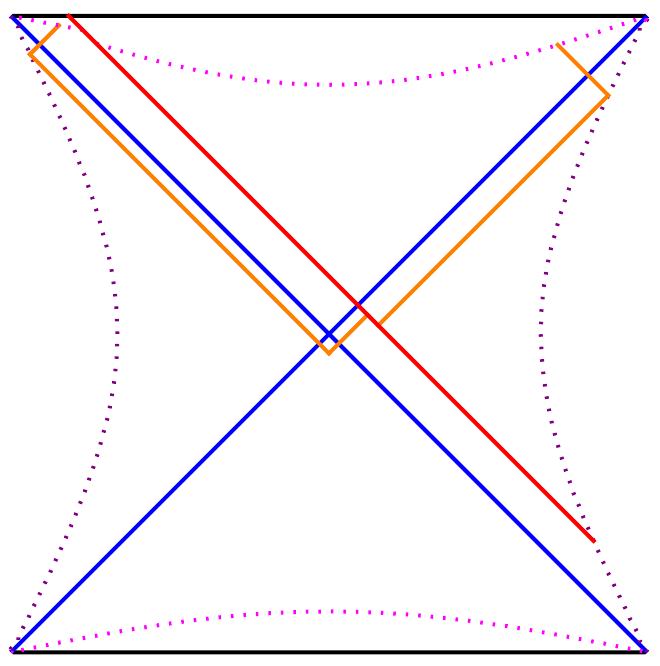}}
    \caption{Time evolution of the WDW patch in the presence of the shockwave.
    (a) Early times $t<t_{c0}$ when the WDW patch completely sits in the dS geometry before the shockwave insertion.
    (b) Times $t_{c0} \leq t < t_{c1}$ after the shockwave insertion, when the past vertex of the WDW patch sits behind $\mathcal{I}^-.$ 
    (c) Intermediate times  $t_{c1} \leq t < t_{c2}$ when both the vertices of the WDW patch are located inside the cosmological horizons.
    (d) Times  $t_{c2} \leq t < t_{c3}$ when the future joint sits  behind the cutoff surface at $r=r_{\rm max,2}.$
    (e) Late times $t \geq t_{c3}$ when the future vertex sits behind the cutoff surface at $r=r_{\rm max,2}.$  }
    \label{fig:time_evolution_WDW_reg}
\end{figure}

We apply the CV2.0 conjecture \cite{Couch:2016exn} to compute the holographic complexity in the asymptotically dS geometries introduced in eq.~\eqref{eq:dS_metric_shock_wave}.
To evaluate of the spacetime volume of the WDW patch, we use 
\beq
\mathcal{C}_{V2.0} \equiv \frac{V_{\rm WDW}}{G_N L^2} = 
\frac{\Omega_{d-1}}{G_N L^2} \int dr \, r^{d-1} \int dt \, \mathcal{F} (r) \, , \qquad
\Omega_{d-1} = \frac{2 \pi^{\frac{d}{2}}}{\Gamma\! \le \frac{d}{2} \ri} \, ,
\label{eq:generic_prescription_CV20}
\eeq
where we have written the measure explicitly and collected the volume of the $(d-1)-$dimensional unit sphere.
The integrand $\mathcal{F}(r)$ is a formal way to denote that we will split the integration region into several terms, which will only depend on the radial coordinate.
This is similar to the analysis performed in the Vaidya case \cite{Chapman:2018lsv}; we will specify below the explicit form of the integrands.
The generic time evolution of the WDW patch (in the absence of the special regimes discussed in section \ref{ssec:special_configurations_WDW}) is depicted in fig.~\ref{fig:time_evolution_WDW_reg}.

We will start by presenting in section \ref{ssec:CV20_general_strategy} the general computation of CV2.0 in the regime $t_{c1} \leq t \leq t_{c2},$ which corresponds to the WDW patch not reaching the cutoff surfaces located at timelike infinities.
Besides being an important regime because it presents a plateau of complexity, it is also convenient because we can obtain the other cases from limits of this calculation.

One can repeat the process described in section \ref{ssec:CV20_general_strategy} for the special configuration depicted in figs.~\ref{fig:alternative1_WDWpatch} and \ref{fig:alternative2_WDWpatch} and find that the functional for the complexity remains the same. 
The case when the joints of the WDW patch move behind the stretched horizon is different and plays an important role.
For this reason, the explicit computation will be reported in section \ref{ssec:CV20_joints_behind_stretched}. 
We then list the expressions for the integrated CV2.0 in all the regimes in section \ref{ssec:integrated_CV20_dS} and we derive the rate of change of complexity in section \ref{ssec:rate_CV20}.

As in the previous section, the following Penrose diagrams depict  only the region between the two stretched horizons on either side of the cosmological horizons. We will perform the general computations in terms of generic cosmological and stretched horizons, without referring to a specific dimension of SdS. We will focus on the specific cases of three and four bulk dimensions in section \ref{sec:examples_CV20}.
Furthermore, we will only consider times after the shockwave insertion,\footnote{\label{footnote:CV20_before_shock}  If we decide to impose fixed cosmological redshift, then the stretched horizon becomes time-dependent for a certain regime before the shockwave insertion. Thus the computations become much more complicated. For this reason, we leave investigating these earlier times for future work.} thus excluding the regime at $t<-2 t_w$ depicted in fig.~\ref{subfig:first_regime_ds}.

\subsection{General strategy}
\label{ssec:CV20_general_strategy}

We perform the general computation of CV2.0 conjecture in eq.~\eqref{eq:generic_prescription_CV20} for the regime depicted in fig.~\ref{subfig:intermediate_regime_dS}; the other cases can be obtained with small changes in the endpoints of integration.

\begin{figure}[ht]
    \centering
    \includegraphics[scale=0.5]{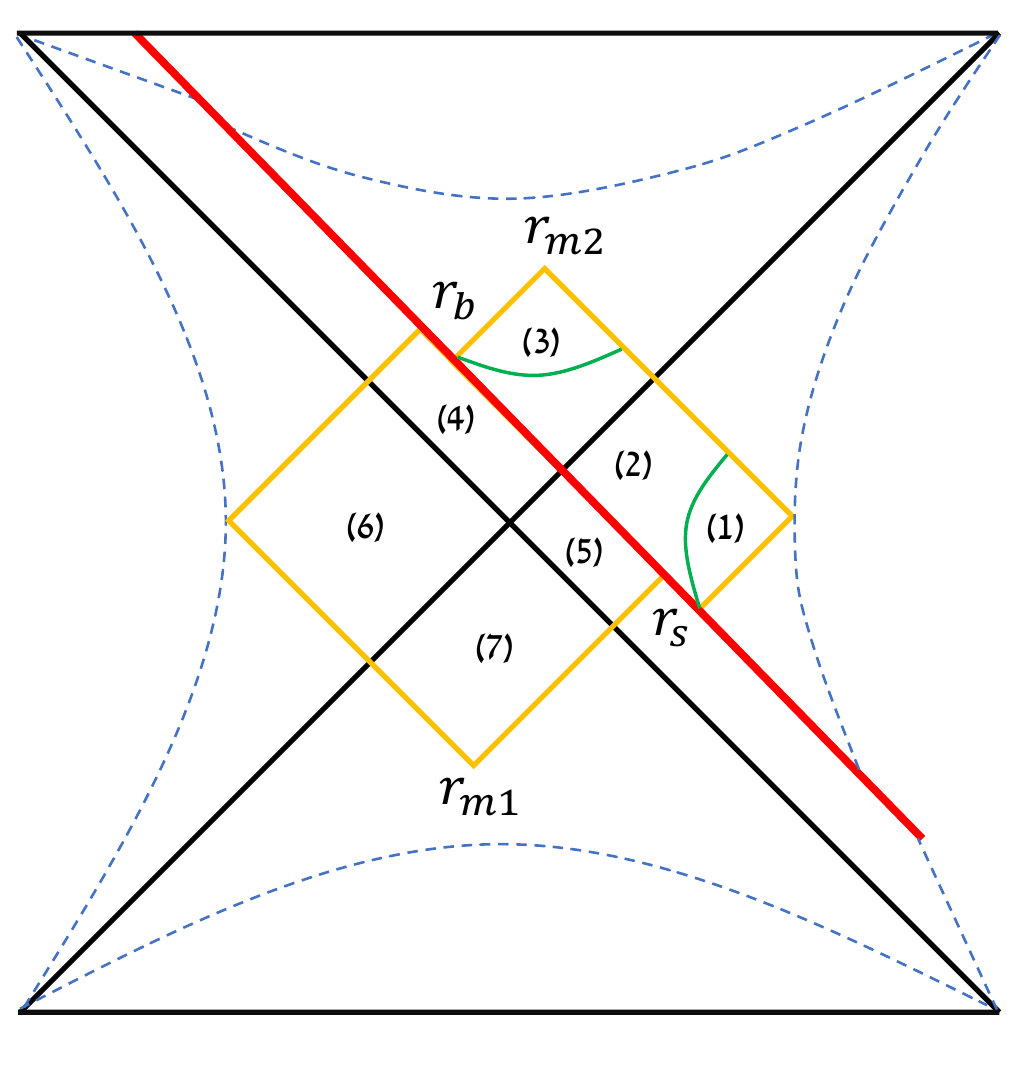}
    \caption{Decomposition of the WDW patch into seven regions for the evaluation of CV2.0.}
    \label{fig:decomposition_integral_CV20}
\end{figure}

\noindent
The total CV2.0 complexity is given by
\beq
\mathcal{C}_{V2.0} = \sum_{k=1}^7 \mathcal{I}_k \, ,
\label{eq:total_CV20_constred}
\eeq
where each of the following seven terms refers to a corresponding part in the splitting depicted in fig.~\ref{fig:decomposition_integral_CV20}:
\begin{subequations}
\beq
\mathcal{I}_1 \equiv \frac{\Omega_{d-1}}{G_N L^2} \int_{r^{\rm st}_2}^{r_s} dr \, r^{d-1} \le 2 r^*_2 (r) - 2 r_2^* (r^{\rm st}_2)   \ri \, ,
\eeq
\beq
\mathcal{I}_2 \equiv \frac{\Omega_{d-1}}{G_N L^2} \int_{r_s}^{r_b} dr \, r^{d-1} \le t_R + t_w  \ri \, ,
\eeq
\beq
\mathcal{I}_3 \equiv \frac{\Omega_{d-1}}{G_N L^2} \int_{r_b}^{r_{ m2}} dr \, r^{d-1} \le  t_R + t_w +  2 r_2^* (r) - 2 r^*_2 (r_b) \ri  \, ,
\eeq
\beq
\mathcal{I}_4 \equiv \frac{\Omega_{d-1}}{G_N L^2} \int_{r_{C1}}^{r_b} dr \, r^{d-1} \le t_L - t_w + 2 r^*_1(r) - 
     r^*_1 (r^{\rm st}_1) - r^*_2 (r^{\rm st}_2)   \ri  \, ,
\eeq
\beq
\mathcal{I}_5 \equiv \frac{\Omega_{d-1}}{G_N L^2}  \int_{r_s}^{r_{C1}} dr \, r^{d-1} \le  2 r^*_1(r) -2  r^*_1 (r_s)   \ri \, ,
\eeq
\beq
\mathcal{I}_6 \equiv \frac{\Omega_{d-1}}{G_N L^2} \int_{r^{\rm st}_1}^{r_{C1}} dr \, r^{d-1} \le 2 r^*_1 (r) - 2 r^*_1 (r^{\rm st}_1) \ri \, ,
\eeq
\beq
\mathcal{I}_7 \equiv \frac{\Omega_{d-1}}{G_N L^2} \int_{r_{C1}}^{r_{m1}} dr \, r^{d-1} \le t_w - t_L + 2 r^*_1 (r) - 2 r^*_1 (r_{s}) - r^*_1(r^{\rm st}_1) + r^*_2 (r^{\rm st}_2) \ri \, .
\eeq
\end{subequations}
By applying eqs.~\eqref{eq:derivatives_tR} and \eqref{eq:derivatives_tL}, we obtain the rate of change of holographic complexity. The derivatives with respect to the right and left boundary times are 
\beq
\frac{d\mathcal{C}_{V2.0}}{dt_R} = \frac{\Omega_{d-1}}{G_N L^2} \frac{1}{d} 
\left[ r_{m2}^{d} - r_s^d - \frac{f_2(r_s)}{f_1(r_s)} \le r_{m1}^d - r_s^d \ri \right] \, ,
\label{eq:derCV20_tR}
\eeq
\beq
\frac{d\mathcal{C}_{V2.0}}{dt_L} = \frac{\Omega_{d-1}}{G_N L^2} \frac{1}{d} 
\left[ r_b^d - r_{m1}^{d}  + \frac{f_1(r_b)}{f_2(r_b)} \le r_{m2}^d - r_b^d \ri \right] \, .
\label{eq:derCV20_tL}
\eeq
In order to find this result, we used the fundamental theorem of integral calculus, the chain rule, and eqs.~\eqref{eq:identity_WDW_times1}--\eqref{eq:identity_WDW_times4}.
Focusing on the symmetric case \eqref{eq:symmetric_times} by using the chain rule and taking the average of the previous results, we find
\beq
\begin{aligned}
\frac{d \mathcal{C}_{V2.0}}{dt} = & \frac{\Omega_{d-1}}{2 G_N L^2} \frac{1}{d} 
\left[ r_{m2}^{d} \le 1 + \frac{f_1(r_b)}{f_2(r_b)}  \ri
- r_{m1}^d \le  1 + \frac{f_2(r_s)}{f_1(r_s)} \ri \right. \\
& \left. + r_b^d \le 1 - \frac{f_1(r_b)}{f_2(r_b)} \ri
- r_s^d \le 1 - \frac{f_2(r_s)}{f_1(r_s)}   \ri
\right] \, .
\label{eq:rate_CV20}
\end{aligned}
\eeq
It is clear that in any regime, this rate diverges when the bottom or top vertices of the WDW patch reach past or future timelike infinity, respectively.
This is the reason for introducing the cutoff surfaces to act as regulators.
Naively, it looks like the rate of growth of complexity diverges at $r_s=r_{C1},$ when the factors $f_1(r_s)$ appear at the denominator vanish.
However, one can check that when $r_s=r_{C1},$ we also have $r_{m1}=r_{C1}.$
As a result, this limit in the complexity rate \eqref{eq:rate_CV20} remains finite.

\subsection{CV2.0 with the joints behind the stretched horizons}
\label{ssec:CV20_joints_behind_stretched}

\begin{figure}[ht]
    \centering
    \includegraphics[scale=0.5]{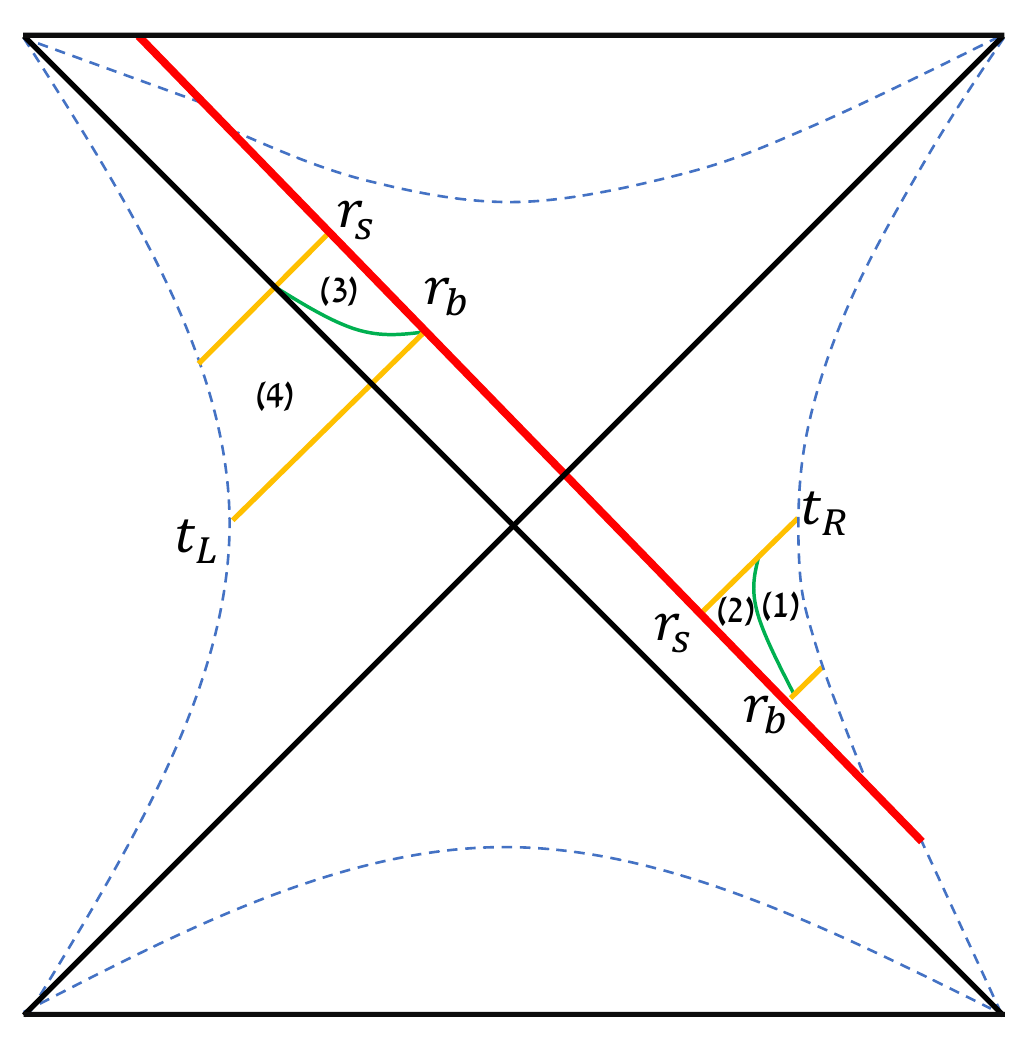}
    \caption{Decomposition of the WDW patch into four regions for the evaluation of CV2.0 when the special points of the WDW patch satisfy $r_s \geq r_b,$ thus implying that the top and bottom joints lie behind the stretched horizons. }
    \label{fig:CV20_alternativeWDWpatch4}
\end{figure}

The analysis performed in section \ref{ssec:special_configurations_WDW} revealed that it is possible to achieve a time regime such that the top and bottom joints of the WDW patch move behind the stretched horizon, corresponding to the Penrose diagram in fig.~\ref{fig:alternative4_WDWpatch}.
We perform the corresponding computation of the CV2.0 conjecture, which is based on a different splitting of the integration into the regions depicted in fig.~\ref{fig:CV20_alternativeWDWpatch4}.
We find 
\beq
\mathcal{C}_{V2.0} = \sum_{k=1}^4 \mathcal{J}_k \, ,
\label{eq:total_CV20_alt5}
\eeq
where the spacetime volume corresponding to each subregion is given by
\begin{subequations}
\beq
\mathcal{J}_1 \equiv \frac{\Omega_{d-1}}{G_N L^2} \int_{r^{\rm st}_2}^{r_b} dr \, r^{d-1} \, \le t_R + t_w + 2 r^*_2(r^{\rm st}_2)  - 2 r^*_2 (r_b) \ri \, ,
\eeq
\beq
\mathcal{J}_2 \equiv \frac{\Omega_{d-1}}{G_N L^2} \int_{r_b}^{r_s} dr \, r^{d-1} \, \le t_R + t_w + 2 r^*_2(r^{\rm st}_2)  - 2 r^*_2 (r)  \ri  \, ,
\eeq
\beq
\mathcal{J}_3 \equiv \frac{\Omega_{d-1}}{G_N L^2} \int_{r_b}^{r_s} dr \, r^{d-1} \, \le  2 r^*_1(r) - 2 r^*_1 (r_s) \ri   \, ,
\eeq
\beq
\mathcal{J}_4 \equiv \frac{\Omega_{d-1}}{G_N L^2}  \int_{r^{\rm st}_1}^{r_b} dr \, r^{d-1} \, \le -t_L + t_w+ r^*_1(r^{\rm st}_1)  + r^*_2(r^{\rm st}_2)  - 2 r^*_1 (r_s) \ri  \, ,
\eeq
\end{subequations}
It turns out that this expression can be obtained by a careful limiting procedure of the result \eqref{eq:total_CV20_constred}.
This is more evident by considering the growth rate, which requires to use of the identities \eqref{eq:derivatives_tR} and \eqref{eq:derivatives_tL}. 
In the case with symmetric boundary times \eqref{eq:symmetric_times}, we get
\beq
\begin{aligned}
\frac{d \mathcal{C}_{V2.0}}{dt} = & - \frac{\Omega_{d-1}}{2 G_N L^2} \frac{1}{d} 
\left[ (r^{\rm st}_{2})^{d} \le 1 + \frac{f_1(r_b)}{f_2(r_b)}  \ri
- (r^{\rm st}_{1})^d \le  1 + \frac{f_2(r_s)}{f_1(r_s)} \ri \right. \\
& \left. + r_b^d \le 1 - \frac{f_1(r_b)}{f_2(r_b)} \ri
- r_s^d \le 1 - \frac{f_2(r_s)}{f_1(r_s)}   \ri
\right] \, .
\label{eq:rate_CV20_alt5}
\end{aligned}
\eeq
By comparing with eq.~\eqref{eq:rate_CV20}, we notice that this result is obtained by setting $r_{m1} \rightarrow r^{\rm st}_1 $ and $r_{m2} \rightarrow r^{\rm st}_2 $, but in addition, we also need to reverse the overall sign.
This change of sign corresponds to the fact that $r_s$ and $r_b$ exchange their location with respect to the other configurations of the WDW patch.
Therefore, null rays which were previously delimiting bottom integration endpoints, now end up becoming top integration endpoints and vice versa.
Indeed, one can check that this prescription is correct, because it gives a positive-definite expression for CV2.0, as expected from a spacetime volume.

\subsection{Integrated Complexity=Volume 2.0}
\label{ssec:integrated_CV20_dS}

As we mentioned earlier, the other regimes in fig.~\ref{fig:time_evolution_WDW_reg} are very similar to the one studied above. Therefore, we only list the regularized CV2.0 for the different regimes after the shockwave insertion depicted in fig.~\ref{fig:time_evolution_WDW_reg}, rather than presenting the full derivation.
In the following computations, we regularize the spacetime volume using the cutoff surfaces introduced in eqs.~\eqref{eq:definition_rmax_shocks} and \eqref{eq:rmax2}.
We locate the stretched horizon on the left side as summarized in eq.~\eqref{eq:generic_location_rst1}, and on the right side as determined by the condition in eq.~\eqref{eq:constant_redshift_condition_app}. 
The different regimes are limits of the intermediate regime described in section \ref{ssec:CV20_general_strategy}. These are the results:  

\begin{itemize}
\item The regime  $t_{c0} \leq t < t_{c1}$,
corresponds to setting $r_{m1} \rightarrow r_{\rm max,1}$ in eq.~\eqref{eq:total_CV20_constred}.
\item The regime $t_{c1} \leq t < t_{c2}$ (the intermediate regime): the result was found in eq.~\eqref{eq:total_CV20_constred}. 
If the choice of parameters allows for the configuration in fig.~\ref{fig:alternative4_WDWpatch}, during the time period delimited by the critical times $t_{c,\mathrm{st}}$ defined in eq.~\eqref{eq:critical_time_behind_stretched}, we use the expression \eqref{eq:total_CV20_alt5} instead.
\item The regime $t_{c2} \leq t < t_{c3}$,
is equivalent to setting $r_{m2} \rightarrow r_{\rm max,2}$ in eq.~\eqref{eq:total_CV20_constred}.
\item The regime $ t_{c3} \leq t$,
In this case, we effectively put $r_b=r_{\rm max,1}$ in the geometry before the shockwave, corresponding to region 4 in fig.~\ref{fig:decomposition_integral_CV20}. We effectively put $r_b = r_{\rm max,2}$ in the geometry after the shock, which also corresponds to the region 2,3 in fig.~\ref{fig:decomposition_integral_CV20}. 
In particular, the term corresponding to region 3 in fig.~\ref{fig:decomposition_integral_CV20} now vanishes.
\end{itemize}

\noindent
One may wonder if the special configurations of the WDW patch depicted in fig.~\ref{fig:alternative_WDWpatch} provide a different result.
The outcome of the analysis is that the formal functional dependence of CV2.0 for the exceptional configurations of the WDW patch (introduced in section \ref{ssec:time_evo_WDW}) remains unchanged, and the only difference in the computation is that one has to pick the correct solution for the joints $r_{m1}, r_{m2}$ (\ie inside or outside the cosmological horizon).
For this reason, it is not restrictive to use the expressions listed in the previous bullets.

\subsection{Complexity rate of change}
\label{ssec:rate_CV20}

We list the rates of complexity change in the symmetric time configuration \eqref{eq:symmetric_times} for the different regimes mentioned above:
\begin{itemize}
    \item Regime $t_{c0} \leq t <  t_{c1}$:
    \beq
    \begin{aligned}
\frac{d \mathcal{C}_{V2.0}}{dt} = & \frac{\Omega_{d-1}}{2 G_N L^2} \frac{1}{d} 
\left[ r_{m2}^{d} \le 1 + \frac{f_1(r_b)}{f_2(r_b)}  \ri
- \le r_{\rm max,1} \ri^d \le  1 + \frac{f_2(r_s)}{f_1(r_s)} \ri \right. \\
& \left. + r_b^d \le 1 - \frac{f_1(r_b)}{f_2(r_b)} \ri
- r_s^d \le 1 - \frac{f_2(r_s)}{f_1(r_s)}   \ri
\right] \, .
\label{eq:reg_rate_CV20_tc1}
\end{aligned}
\eeq
\item Regime $t_{c1} \leq t < t_{c2}$: the rate is given in eq.~\eqref{eq:rate_CV20}.
When the parameters allow for the existence of a configuration where the joints move behind the stretched horizon, we need to use the rate \eqref{eq:rate_CV20_alt5} instead.
\item Regime $t_{c2} \leq t < t_{c3}$:
    \beq
    \begin{aligned}
\frac{d \mathcal{C}_{V2.0}}{dt} = & \frac{\Omega_{d-1}}{2 G_N L^2} \frac{1}{d} 
\left[ \le r_{\rm max,2} \ri^d  
\le 1 + \frac{f_1(r_b)}{f_2(r_b)}  \ri
- r_{m1}^d \le  1 + \frac{f_2(r_s)}{f_1(r_s)} \ri \right. \\
& \left. + r_b^d \le 1 - \frac{f_1(r_b)}{f_2(r_b)} \ri
- r_s^d \le 1 - \frac{f_2(r_s)}{f_1(r_s)}   \ri
\right] \, .
\label{eq:reg_rate_CV20_tc2}
\end{aligned}
\eeq
\item Regime $ t \geq t_{c3}$: 
 \beq
    \begin{aligned}
\frac{d \mathcal{C}_{V2.0}}{dt} = & \frac{\Omega_{d-1}}{2 G_N L^2} \frac{1}{d} 
\left[ \le r_{\rm max,1} \ri^d  + \le r_{\rm max,2} \ri^d  
- r_{m1}^d \le  1 + \frac{f_2(r_s)}{f_1(r_s)} \ri 
- r_s^d \le 1 - \frac{f_2(r_s)}{f_1(r_s)}   \ri
\right] \, .
\label{eq:reg_rate_CV20_tc4}
\end{aligned}
\eeq
\end{itemize}

\noindent
Next, we will analytically evaluate the late time limit of the complexity rate.

\paragraph{Late times.}
The late time limit $t \gg t_{c3} $ corresponds to the case when the geometric data go to 
\beq
r_{m1} \rightarrow r_{C1} \, , \quad
r_s \rightarrow r_{C2} \quad  \Rightarrow \quad
f_2(r_s) \rightarrow 0 \, ,
\eeq
as can be understood geometrically and checked in specific examples such as the eqs.~\eqref{eq:rs_analytic_SdS3} and \eqref{eq:rm1_analytic_SdS3}.
In this case the CV2.0 rate \eqref{eq:reg_rate_CV20_tc4} becomes 
\beq
\begin{aligned}
\frac{d \mathcal{C}_{V2.0}}{dt} (t \gg t_{c3}) & = \frac{\Omega_{d-1}}{2 G_N L^2} \frac{1}{d} 
\left[ \le  r_{\rm max,1} \ri^d  + \le r_{\rm max,2} \ri^d  
- (r_{C2})^d - (r_{C1})^d 
\right] \approx \\
& \approx \frac{\Omega_{d-1}}{2 G_N L^2} \frac{1}{d} 
\left[ \le  r_{\rm max,1} \ri^d  + \le r_{\rm max,2} \ri^d  \right] =
\frac{\Omega_{d-1}}{ G_N L^2} \frac{1}{d} 
\le  \frac{r_{C1}}{\delta} \ri^d  \, ,
\end{aligned}
\label{eq:late_time_rate_dS}
\eeq
where we approximated the results by assuming $\delta \ll 1$ and in the last step we used the definitions \eqref{eq:definition_rmax_shocks} and \eqref{eq:rmax2}. 
We find that the result is the same as in empty dS without a shockwave for late times \cite{Jorstad:2022mls}.

In the strict limit $\delta \rightarrow 0,$ this expression becomes positively divergent, reproducing the original hyperfast growth of complexity \cite{Susskind:2021esx}.
As soon as we keep the $\delta$ regulator finite, this is a constant rate, corresponding to the usual linear behavior of complexity for late times.
In particular, the growth rate is given by the sum of the rates in the two geometries, related to the location of the regulator surfaces close to timelike infinities.
This behavior parallels the statement, valid in the AdS-Vaidya case, that the complexity rate at late times is the sum of the masses of the black holes between which the geometry jumps \cite{Chapman:2018lsv}.

\section{Explicit results for the Complexity=Volume 2.0}
\label{sec:examples_CV20}

We specialize the general computation of CV2.0 and the corresponding growth rate, performed in section \ref{sec:CV20}, to the cases of shockwaves geometries describing a transition between SdS black holes in three dimensions (section \ref{ssec:CV20_SdS3}) and in four dimensions (section \ref{ssec:CV20_SdS4}).
One of the main results is that there is always a time interval during the evolution where a plateau appears in the complexity (\ie the growth rate essentially vanishes).
We will investigate various features of this regime in terms of the parameters describing the shockwave perturbation.

\subsection{Three-dimensional SdS space}
\label{ssec:CV20_SdS3}

We consider the transition between two SdS$_3$ geometries with blackening factor \eqref{eq:blackening_factor_shock_SdS}. 
We explicitly evaluate the complexity according to CV2.0 conjecture following the procedure presented in section \ref{ssec:integrated_CV20_dS}, where the special positions of the WDW patch were given in eqs.~\eqref{eq:rs_analytic_SdS3}--\eqref{eq:rm2_analytic_SdS3}.

\begin{figure}[ht]
    \centering
  \subfigure[]{ \label{fig:integratedCV20SdS3} \includegraphics[scale=0.6]{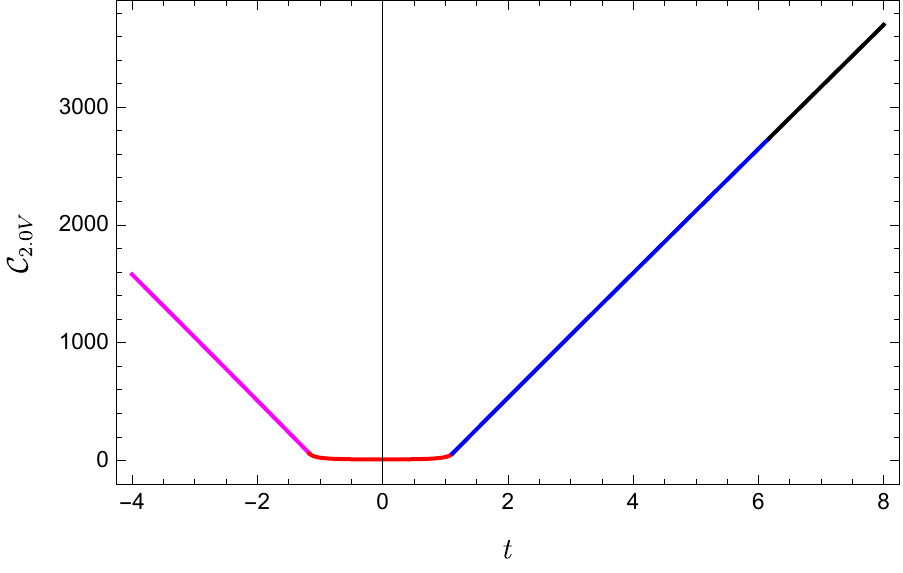}}
 \subfigure[]{\label{subfig:focus_CV20_SdS3_plateau}  \includegraphics[scale=0.575]{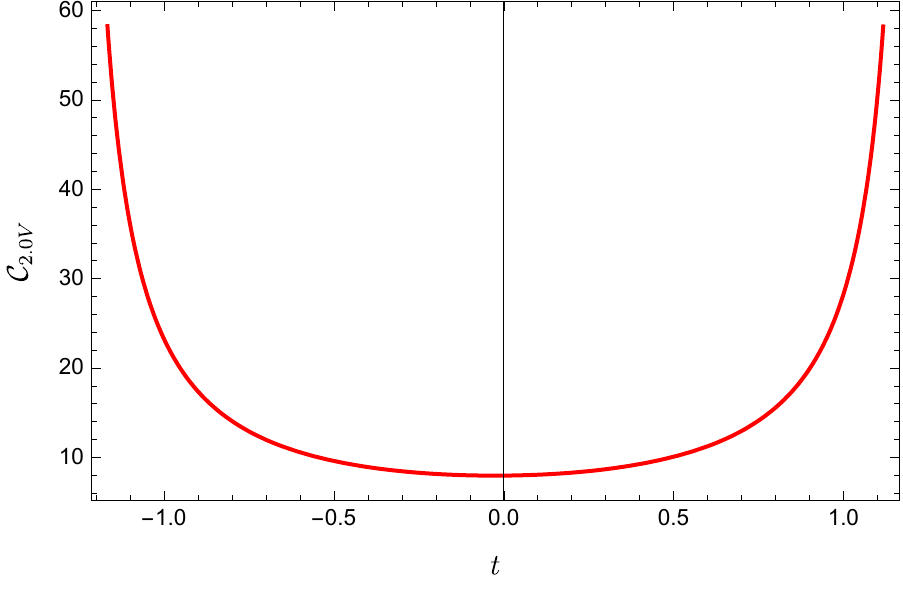}} \\
  \subfigure[]{\includegraphics[scale=0.52]{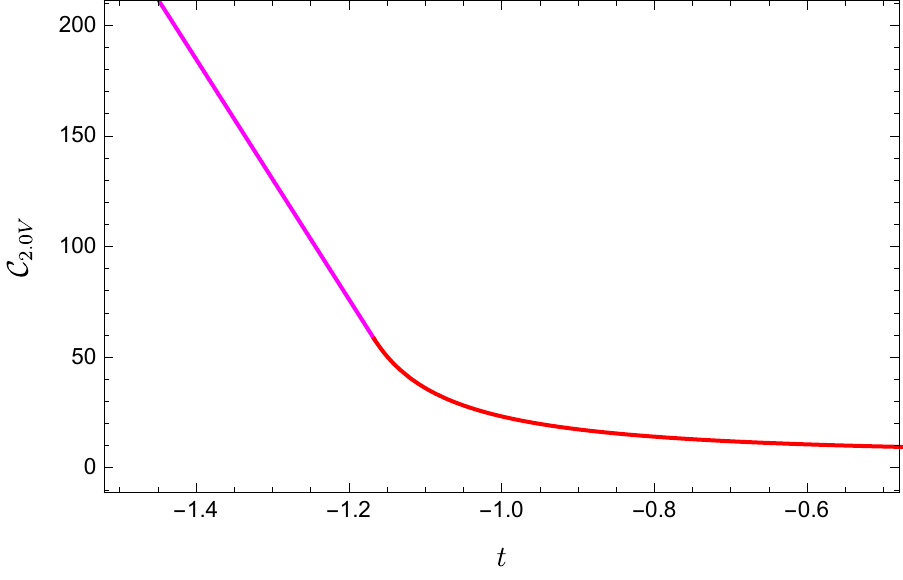}}
   \subfigure[]{ \includegraphics[scale=0.52]{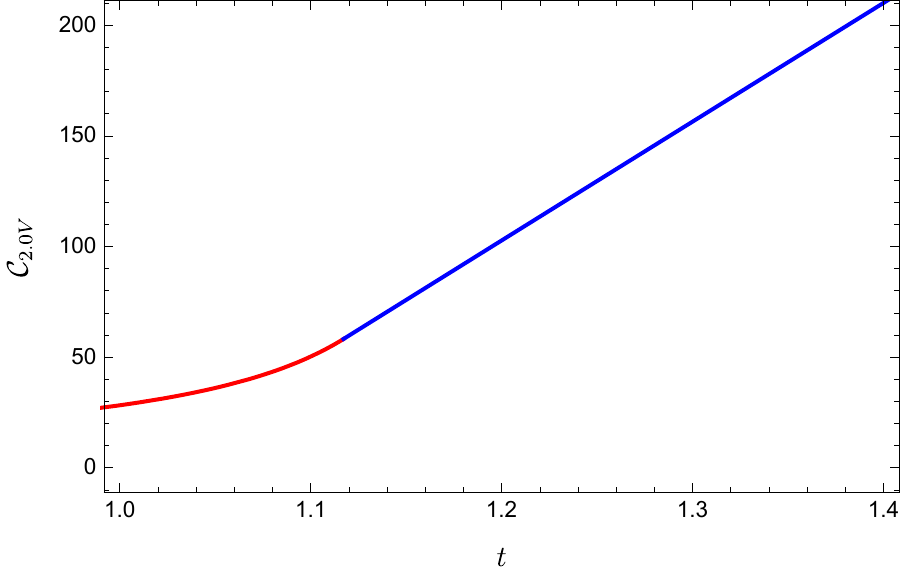}}
   \subfigure[]{\includegraphics[scale=0.52]{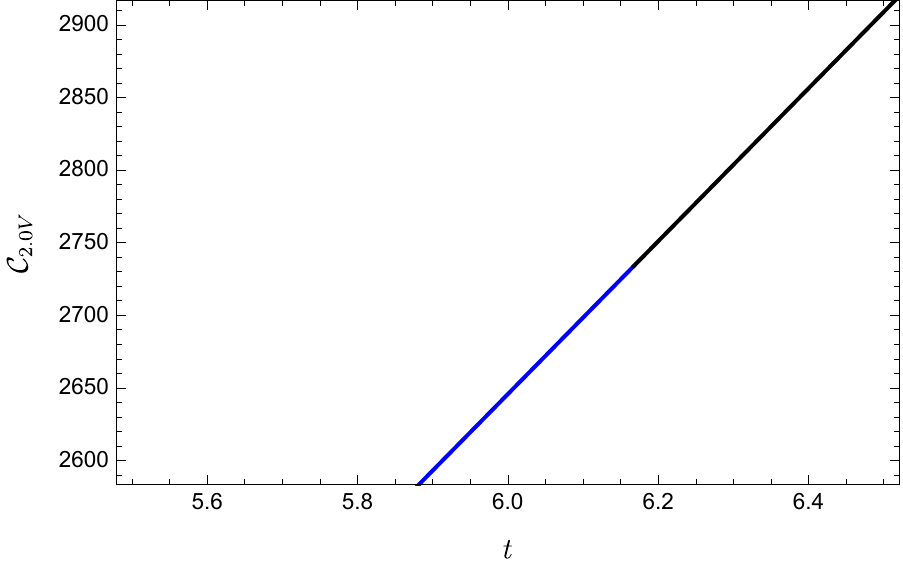}}
    \caption{(a) Complexity computed according to the CV2.0 proposal as a function of time. We fix $L=1, \rho=0.5, t_w=2, \delta=0.05, G_N\mathcal{E}_1=0.02$  and $\varepsilon=0.1$, according to the definition \eqref{eq:generic_epsilon_geometries}.
    (b) Focus on the plateau regime during the interval $t \in [t_{c1}, t_{c2}].$ In the lower panels, we focus on the junctions at critical times
    (a) $t_{c1}$, (b) $t_{c2}$, (c)  $t_{c3}$.  }\label{fig:integratedCV20SdS3case1}
\end{figure}

\paragraph{Integrated complexity.}
The time dependence of the integrated CV2.0 complexity is plotted in fig.~\ref{fig:integratedCV20SdS3}. 
We observe that the complexity increases linearly at late times $t\gtrsim t_{c2}$, where the critical time $t_{c2}$ was defined in eq.~\eqref{eq:tc2}. The rate of linear increase is fixed by the cutoff near future infinity, as we have seen in equation \eqref{eq:late_time_rate_dS}. The linear evolution is preceded by a \emph{plateau} region in the time interval $t \in [t_{c1}, t_{c2}]$, where the critical times $t_{c1}$ and $t_{c2}$ were defined in eqs.~\eqref{eq:tc1} and \eqref{eq:tc2}. 
We call this part of the time evolution a \emph{plateau} region, referring to the fact that complexity is approximately constant and much smaller than its value at later times because the WDW patch does not reach timelike infinity, and therefore we do not need to regularize its spacetime volume. 
We plot the plateau region in more detail in fig.~\ref{subfig:focus_CV20_SdS3_plateau}. Note that its behavior is similar to the one observed in the absence of a shockwave, see fig.~5 in reference \cite{Jorstad:2022mls}.
A focus on the transition between various regimes of the complexity evolution is shown in the lower panels demonstrating that the complexity is continuous at all times after the shockwave insertion. As we will see in a moment, for different parameters of the shockwaves, the size of the plateau region can become arbitrarily long, especially for very early shocks.

\begin{figure}[ht]
    \centering
  \subfigure[]{ \label{fig:integratedCV20SdS3_new_st_long_pl} \includegraphics[scale=0.6]{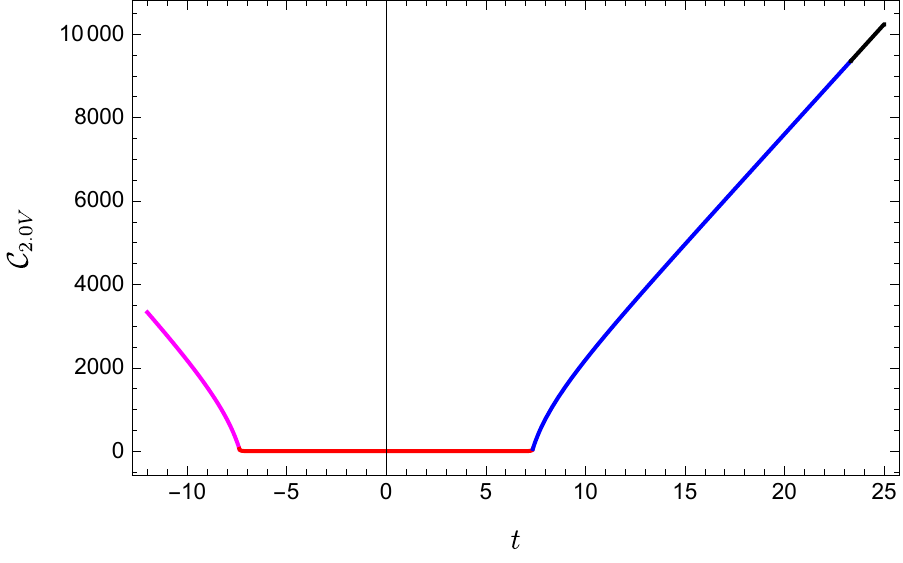}}
 \subfigure[]{\label{subfig:focus_CV20_SdS3_plateau_new_st_long_pl}  \includegraphics[scale=0.57]{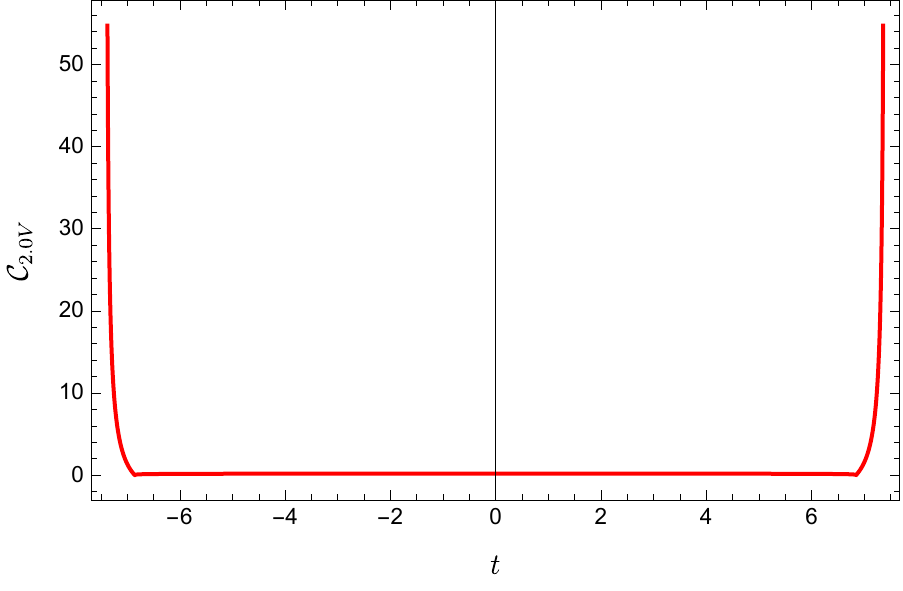}}
    \caption{(a) Complexity computed according to the CV2.0 proposal as a function of time in $d=2$. We fix $L=1, \rho=0.99, t_w=6, \delta=0.05, G_N\mathcal{E}_1=0.02$ and the quantity $\varepsilon=0.01$ defined in eq.~\eqref{eq:generic_epsilon_geometries}. 
    The parameters are chosen such that the plateau is in the regime where the length of the plateau is growing linearly as a function of $t_w$ and the special configurations described in section \ref{ssec:special_configurations_WDW} appear. (b) Focus on the time interval $t \in [t_{c1}, t_{c2}].$}
    \label{fig:integrated_CV20_SdS3_new_st_long_pl}
\end{figure}

The choice of parameters presented in the plot \ref{fig:integratedCV20SdS3}, where the insertion time of the shock which is not too large, corresponds to a time evolution of the positions where the WDW patch intersect the shockwave (\ie $r_s$ and $r_b$) is similar to that plotted in fig.~\ref{fig:plotrs_rb1}.
In particular, the special configurations depicted in fig.~\ref{fig:alternative4_WDWpatch} are not realized.
Now we consider instead a choice of parameters with a bigger value of the insertion time $t_w$ of the shock, which leads to the configuration in fig.~\ref{fig:alternative3_WDWpatch}.
The integrated complexity is represented in fig.~\ref{fig:integrated_CV20_SdS3_new_st_long_pl}.
First of all, we notice that the duration of the plateau region is much longer than in the previous case.
Secondly, the focus on the plateau regime (plotted in  fig.~\ref{subfig:focus_CV20_SdS3_plateau_new_st_long_pl}) shows a kink along the time evolution.
This precisely corresponds to the transition to the regime where the joints of the WDW patch move behind the stretched horizon, as can be understood by the fact that the complexity vanishes at the location of the kink.
This happens because at the critical time defined in eq.~\eqref{eq:critical_time_behind_stretched}, we have $r_s=r_b$ and the spacetime volume of the WDW patch vanishes.
One can check that complexity is continuous at all the critical times in this case, too.

According to the hierarchies of critical times shown in table~\ref{tab:critical_times}, it can happen that $t_{c1} < t_{c0}.$
When this situation occurs, we find yet another different scenario for the evolution of complexity, where CV2.0 directly starts at the shockwave insertion in the plateau regime. This scenario is plotted in fig.~\ref{fig:integratedCV20_case3}. 

Finally, in appendix \ref{app:formation}, we also analyze  the complexity of formation of shockwaves in SdS$_3$, defined as the excess complexity at $t_L=t_R=0$ with a shockwave compared to the case of empty dS space as a function of the parameters of the problem. 
In the appendix, we find that the complexity of formation shows a plateau regime where it is approximately constant, and slightly decreases until it meets a kink, which corresponds to the transition to a special configuration \ref{fig:alternative3_WDWpatch} of the WDW patch.
After the kink, the complexity of formation increases and it can be approximated by a linear function proportional to $(t_w-t_*)$, see eq.~\eqref{eq:Cf_largetw_SdS3}.
The linear growth is delayed by a scrambling time $t_*$, in a similar fashion as what happens for the time evolution of complexity.
We will analyze in detail the features of the scrambling time for the plateau region of complexity in section \ref{ssec:Plateau_of_complexity}.

\begin{figure}[ht]
    \centering
  \subfigure[]{\label{subfig:general_integratedCV20_case3} \includegraphics[scale=0.6]{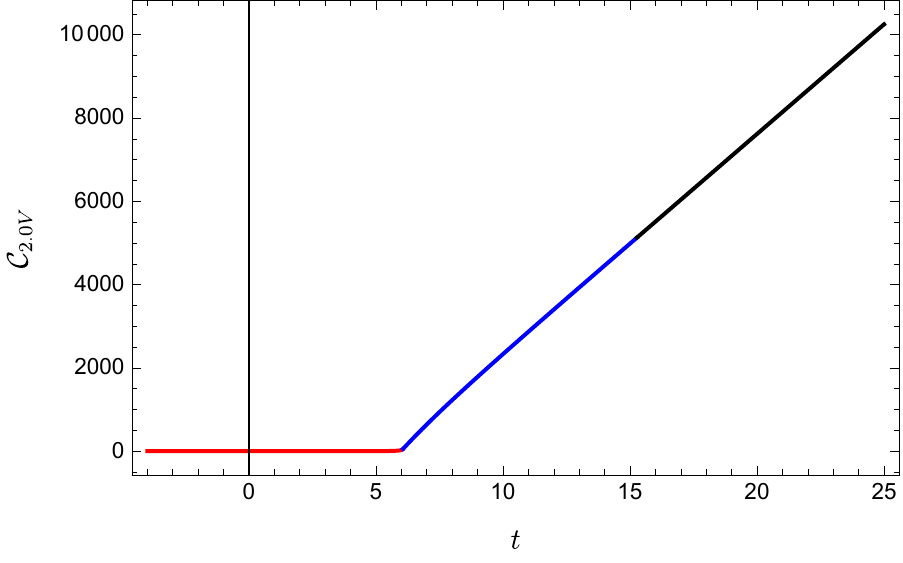}}
 \subfigure[]{\label{subfig:focus_CV20_dS_plateau_case3}  \includegraphics[scale=0.57]{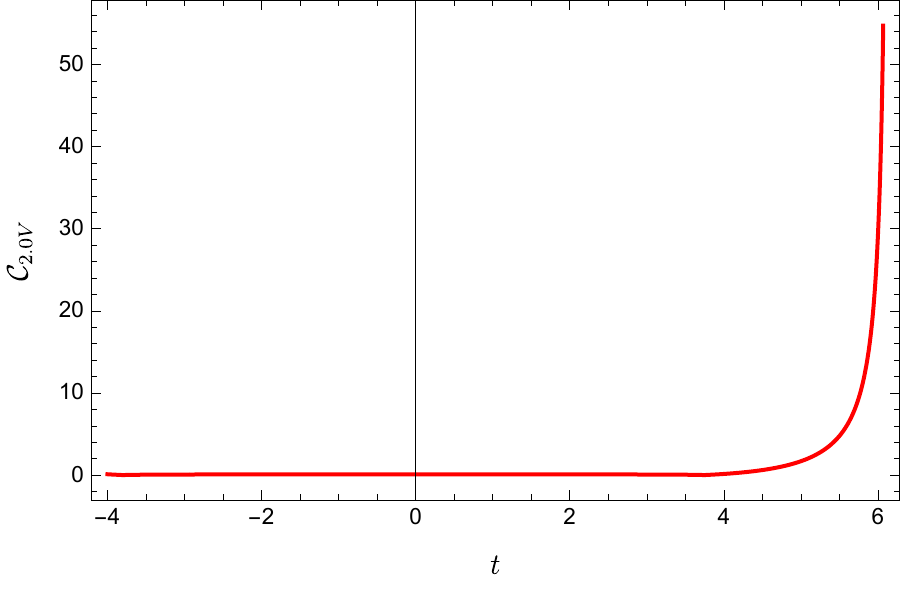}}
    \caption{(a) Complexity computed according to the CV2.0 proposal as a function of time in $d=2$. We fix $L=1, \rho=0.99, t_w=2, \delta=0.05, G_N\mathcal{E}_1=0.02$  and the quantity $\varepsilon=0.1$ defined in eq.~\eqref{eq:generic_epsilon_geometries}. 
    (b) Focus on the time interval $t \in [t_{c1}, t_{c2}].$}
     \label{fig:integratedCV20_case3}
\end{figure}

\begin{figure}[ht]
    \centering
\subfigure[]{\label{subfig:plotCV20rateSdS_case1} \includegraphics[scale=0.69]{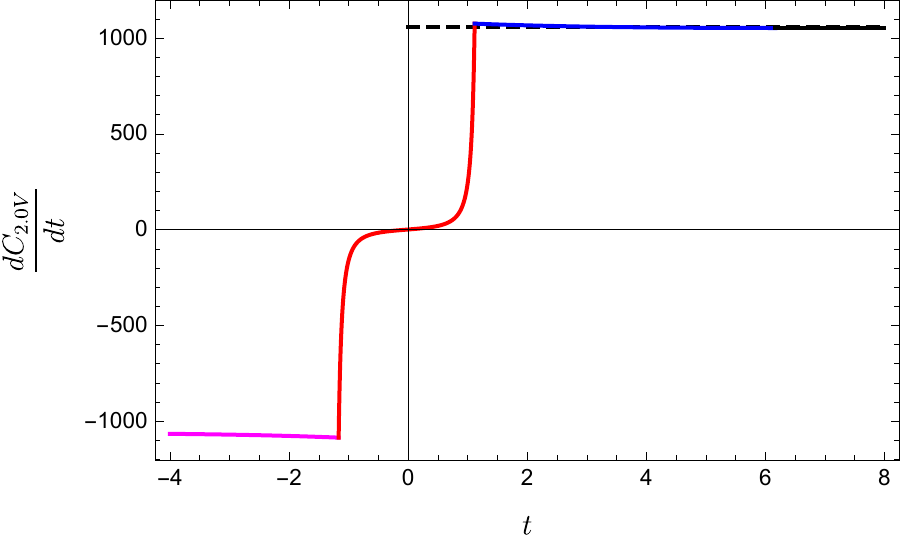} }
\subfigure[]{\label{subfig:plotCV20rateSdS_case2} \includegraphics[scale=0.69]{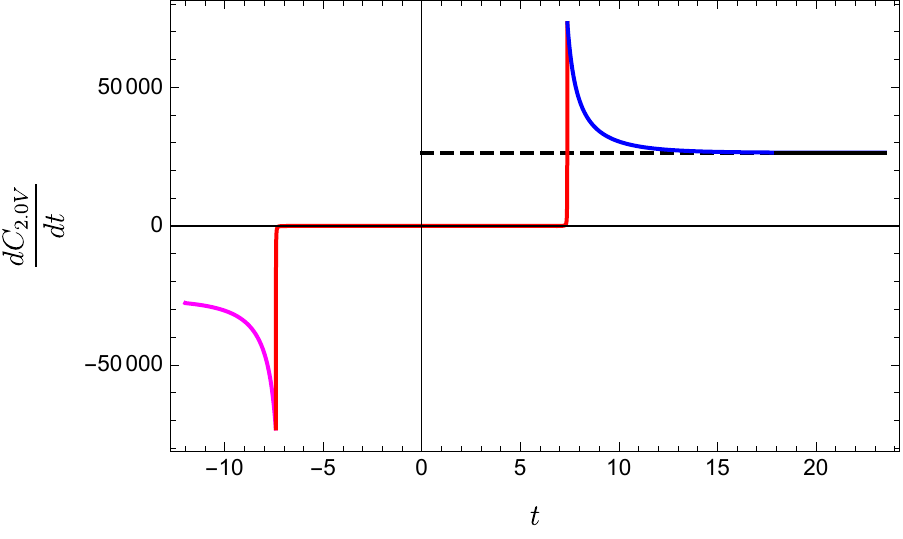}  }
\subfigure[]{\label{subfig:plotCV20rateSdS_case3} \includegraphics[scale=0.69]{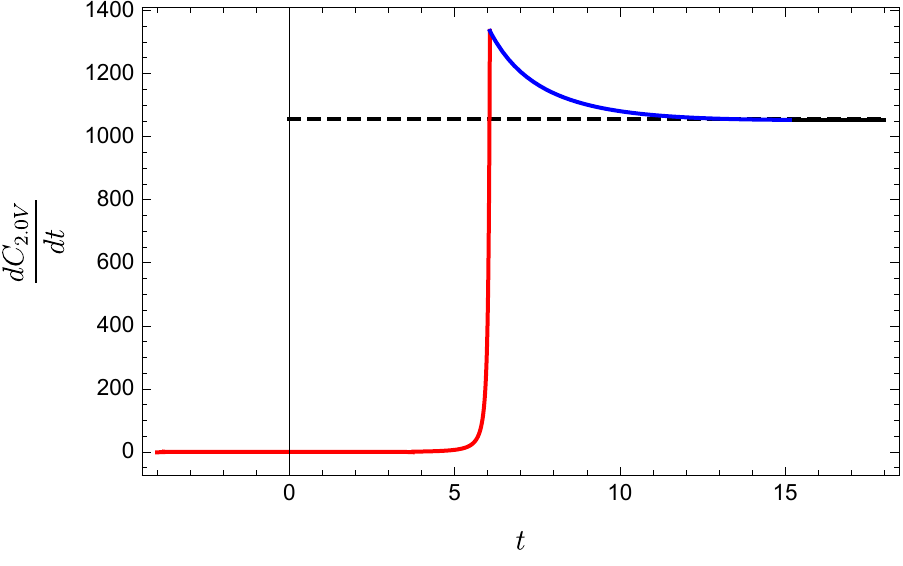} }
    \caption{(a) Plot of the rate of change of CV2.0 for the geometry described by the blackening factor \eqref{eq:blackening_factor_shock_SdS} with $d=2.$
    We set $L=1, t_w=2, \rho=0.5, G_N\mathcal{E}_1=0.02, \varepsilon=0.1, \delta=0.05.$ Different colors correspond to different regimes delimited by the corresponding critical times. 
    (b) Plot of the rate, but choosing the parameters
     $ 
    G_N\mathcal{E}_1=0.02, \varepsilon=0.01, \rho=0.99, t_w =6, \delta=0.01$ instead.
     (c) Plot of the rate, but choosing the parameters
     $ G_N\mathcal{E}_1=0.02, \varepsilon=0.1, \rho=0.99, t_w =2, \delta=0.01$ instead.}
     \label{fig:plotCV20rateSdS3}
\end{figure}

\paragraph{Complexity rate of change.}
The time dependence of the rate of change of the holographic complexity is plotted in fig.~\ref{fig:plotCV20rateSdS3} for the three cases illustrated in figures \ref{fig:integratedCV20SdS3case1}, \ref{fig:integrated_CV20_SdS3_new_st_long_pl} and \ref{fig:integratedCV20_case3}. These rates are determined according to the expressions derived in section \ref{ssec:rate_CV20}.
Note that the rate is continuous across the transitions between various regimes (after the shockwave insertion).
The red region corresponds to the interval when CV2.0 admits the plateau, and after that, we notice that the rate decreases until it reaches the constant value \eqref{eq:late_time_rate_dS} for late times, represented by the horizontal dotted line. 
This is analogous to the behavior seen for AdS black holes \cite{Carmi:2017jqz}. 

One may also consider the limiting case when the shockwave perturbation produces a transition from a black hole to empty dS space: the qualitative behavior does not change with respect to the plots reported here.

\subsubsection{Plateau of complexity}
\label{ssec:Plateau_of_complexity}

A peculiar aspect of the complexity rate of change in the presence of shockwaves, which was observed in \cite{Chapman:2018lsv} for the asymptotically AdS case, is that there exists a time interval such that the complexity rate of change is approximately vanishing. In the case of AdS-Vaidya geometries, this regime appears when both the top and bottom joints of the WDW patch touch future (past) timelike infinity. For SdS this interval starts at the critical time $t_{c1}$ and ends at $t_{c2},$ corresponding to the bottom (top) joint of the WDW patch touching past (future) timelike infinity  (if $t_{c0}>t_{c1}$ the plateau will start when the WDW patch crosses the shock as shown in fig.~\ref{fig:integratedCV20_case3}). The origin of this behavior for each of the geometries is different. For AdS black holes, the complexity rate of change is approximately vanishing due to the cancellation between the increase of the complexity from the top quadrant and decrease from the bottom region. While in SdS, it arises because of the stretched horizon, the top and bottom joints of the WDW patch do not reach the 
future and past singularity at the same time. The cancellation appears only when the WDW patch is not touching the singularities. It is possible that for small enough values of the stretched horizon parameter $\rho$
(\ie the stretch horizon approaches the black hole horizon), the reasoning for the cancellation might be similar to that for the AdS black holes. We leave the exploration of this question for the future.

\begin{figure}[ht]
    \centering
     \subfigure[]
   { \includegraphics[scale=0.75]{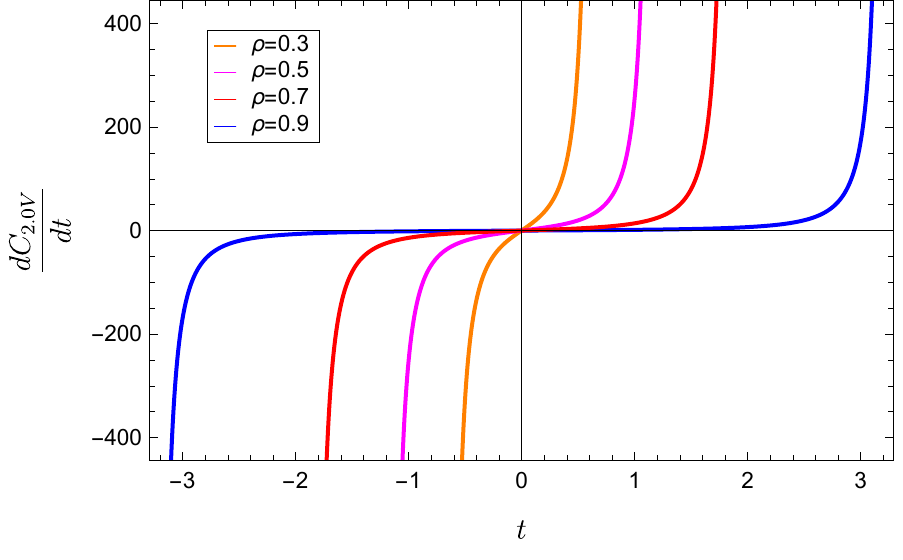}} \,
    \subfigure[]{ \includegraphics[scale=0.75]{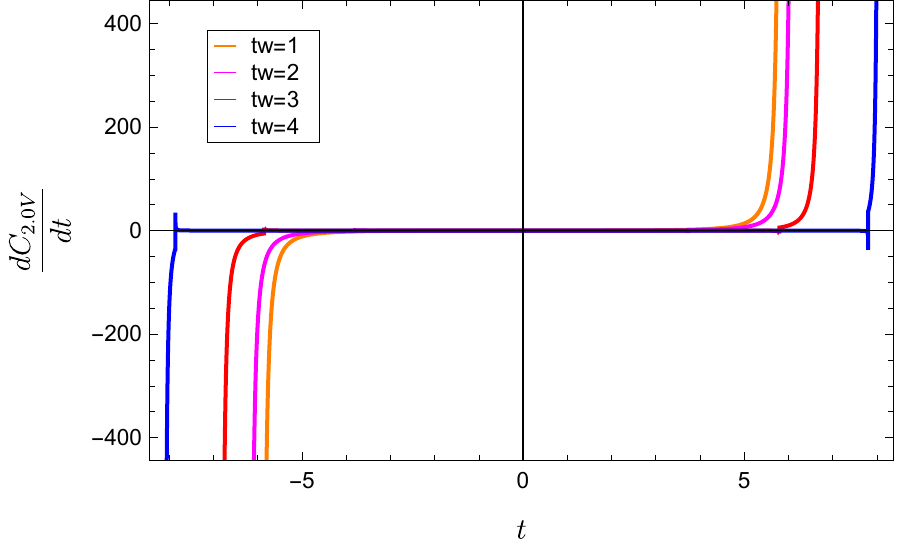}}
    \subfigure[]{\includegraphics[scale=0.75]{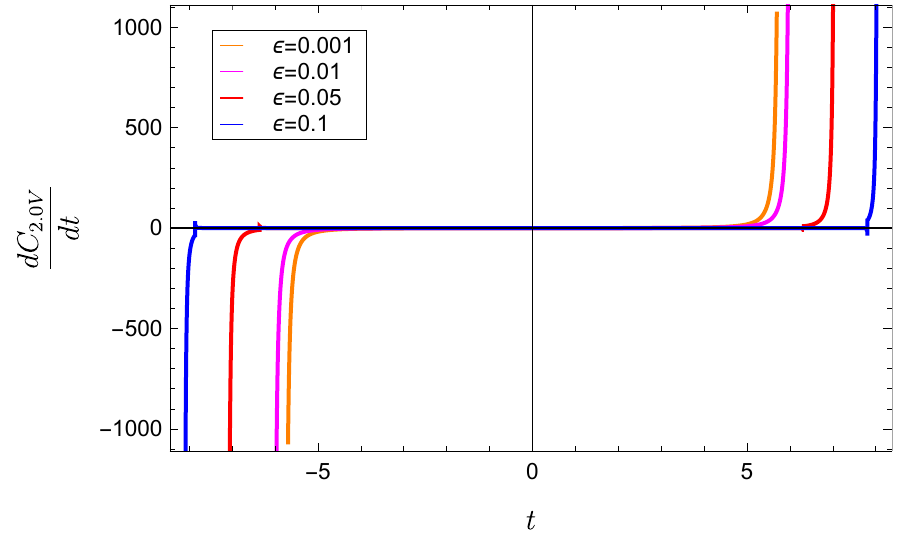}}
    \caption{Plot of the rate of evolution of CV2.0 conjecture as determined by eq.~\eqref{eq:rate_CV20}. We put $L=1,\delta=0.05,G_N\mathcal{E}_1=0.02 $ for all the cases, then we fix two out of three parameters $\rho, t_w, \varepsilon$ and we let one of them vary.
    (a) We vary the parameter $\rho$ at the stretched horizon while setting $\varepsilon=0.001,t_w=6$ .
    (b) We vary the time $t_w$ when the shockwave is inserted while setting $\varepsilon=0.1,\rho=0.99$.
    (c) We vary the strength $\varepsilon$ of the perturbation while setting $t_w=4,\rho=0.99$.  
    \label{fig:CV20_3d4d}}
\end{figure}

In fig.~\ref{fig:CV20_3d4d}, we plot  the numerical dependence of the complexity rate determined in eq.~\eqref{eq:rate_CV20} for intermediate times, referring to the choice of parameters used in the plot \ref{subfig:plotCV20rateSdS_case1}.
While the graphs have a dependence on the geometric data similar to the unperturbed case studied in fig.~5 of reference \cite{Jorstad:2022mls}, the main difference is that there are several cases that show a kink in the evolution of the rate.
This happens when the joints of the WDW patch move behind the stretched horizon, and it is a peculiar property of the asymptotically dS spacetimes.
We notice that the plateau regime lasts longer in the following circumstances: when the stretched horizon is closer to the cosmological one ($\rho \rightarrow 1$), when the shockwave is inserted earlier (increasing $t_w$), and when the shock is heavier (increasing $\varepsilon$).
The same qualitative dependence on the parameters $(t_w, \varepsilon)$ was observed in the AdS-Vaidya case \cite{Chapman:2018lsv}. 

Before diving into a detailed analysis of the length of  the plateau region, we specify that the presence of an additional regime where the top and bottom joints of the WDW patch move behind the stretched horizons does not affect the discussion presented here.
Indeed, complexity stays very small even in such regime, and furthermore the critical time $t_{c,\mathrm{st}}$ when this occurs always satisfies  eq.~\eqref{eq:hierarchy_tcst_tc1},  as commented in section \ref{ssec:special_configurations_WDW}. 

\paragraph{Duration of the plateau.}
We estimate the time duration of the plateau according to
\beq
t_{\rm pl} \equiv t_{c2} - t_{c1} \, .
\label{eq:def_plateau_time}
\eeq
Using eqs.~\eqref{eq:tc1} and \eqref{eq:tc2}, we find
\beq
t_{\rm pl} = -4 t_w - 4 \le r^*_2(r_{\rm max,2}) + r^*_1 (r_{\rm max,1})  \ri + 4 \le r^*_2(r_b) + r^*_1 (r_s) \ri+ 2 \le r^*_1(r_1^{\rm st}) - r^*_2 (r_2^{\rm st}) \ri \, .
\label{eq:tpl_SdS3}
\eeq
We will use the definitions for $r_s$ and $r_b$ given at the critical times in eqs.~\eqref{eq:rstc1_identity1} and \eqref{eq:rbtc2_identity1}, respectively. 
The previous identity has an explicit linear dependence on $t_w,$ plus another implicit dependence on $t_w$ encoded by the positions $r_s$ and $r_b$ determined at the corresponding critical times.
The critical times approach a linear dependence on the time $t_w$ when $t_w \gg L,$ and such regime begins after a non-vanishing delay that we identify as the scrambling time.

\begin{figure}[ht]
\centering
\subfigure[]{ \includegraphics[scale=0.64]{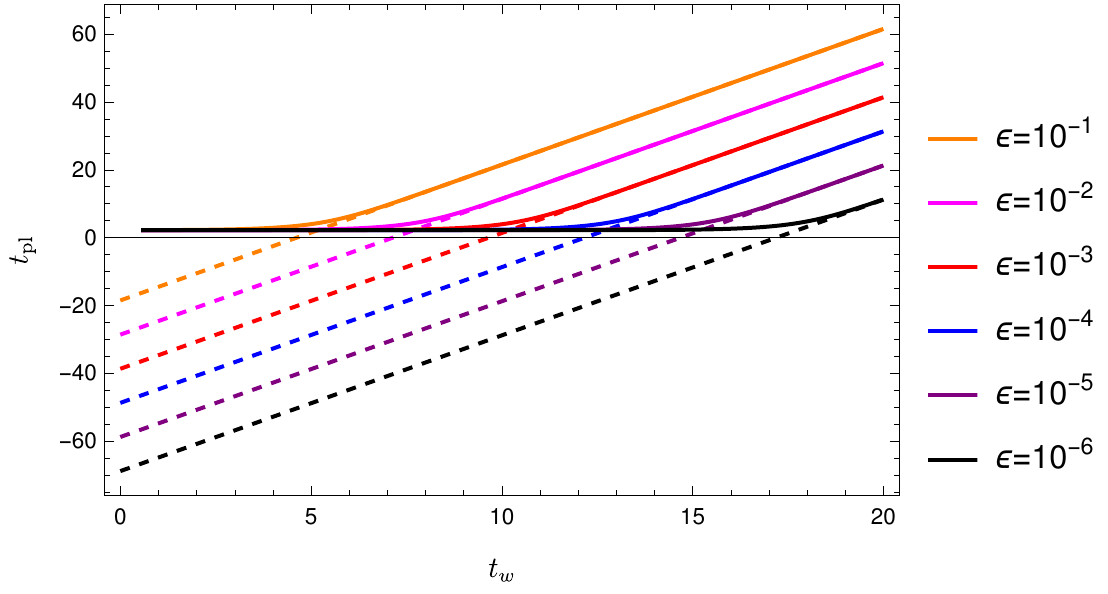}}
\subfigure[]{\label{subfig:plateau_SdS_rho099} \includegraphics[scale=0.64]{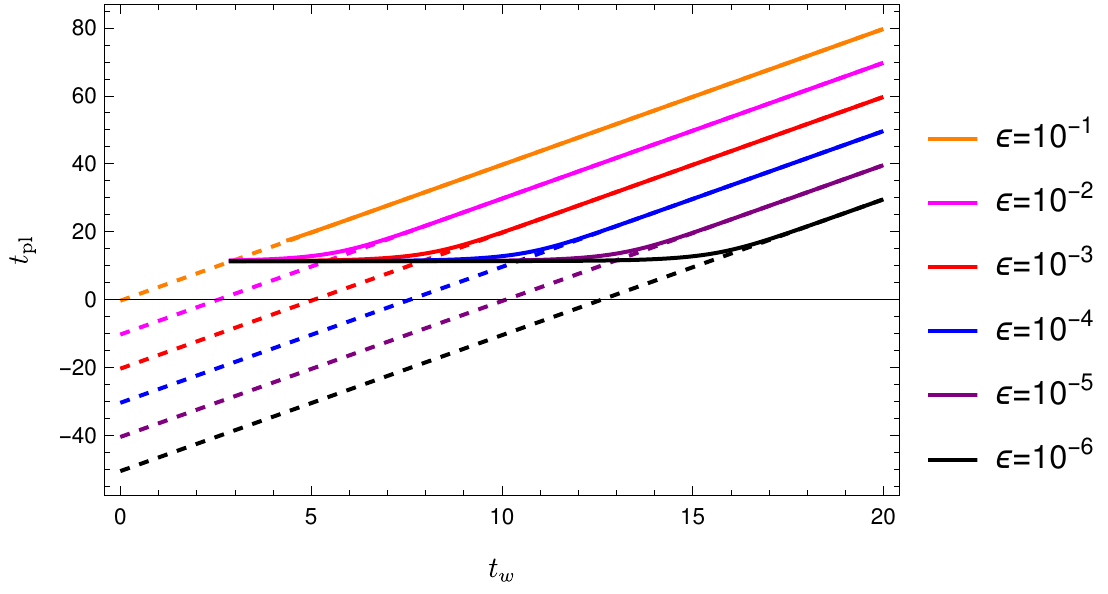}}
\caption{Dependence of the duration of the plateau in terms of the insertion time of the shockwave, for various choices of $\varepsilon$ defined in eq.~\eqref{eq:generic_epsilon_geometries} in $d=2$. (a) We fix $L=1,\delta =0.05, G_N\mathcal{E}_1 = 0.02$ and $\rho=0.5$ . The dotted lines corresponds to the curves $t_{\rm pl} = 4 (t_w - t_*)$ with scrambling time in eq.~\eqref{eq:duration_plateau_SdS3}.
In panel (b), we only change the location of the stretched horizons by choosing $\rho=0.99.$ }
\label{fig:plateau_rho099_SdS3}
\end{figure}

We plot the numerical time duration of the plateau \eqref{eq:def_plateau_time} as a function of the time insertion of the shockwave $t_w$ in fig.~\ref{fig:plateau_rho099_SdS3}, for various choices of the strength $\varepsilon$ of the perturbation.
The scrambling time becomes smaller for bigger values of the deformation parameter $\varepsilon,$ as can be expected by the fact that the transition between the geometries is sharper. This is also an expected feature of the behavior of complexity, where the smearing of a smaller perturbation throughout the system takes longer.

The numerical computation shows that the duration of the plateau always starts from a value independent of $\varepsilon.$
This constant can be analytically determined in the limiting case without shockwave, \ie when $\varepsilon=0.$
Indeed, the duration of the plateau in SdS$_3$ reads\footnote{This result can be obtained by considering the three-dimensional SdS solution without shockwaves, and computing analytically the critical time in a fashion similar to the case of empty dS space, studied in \cite{Jorstad:2022mls}.}
\beq
t_{\rm pl}^{\rm dS} = \frac{2L}{a_1} \log \left[ \frac{(1+\rho)(1-\delta)}{(1-\rho)(1+\delta)}  \right] \, .
\label{eq:tpl_empty_dS}
\eeq
This regime does not exist (\emph{i.e.}, we find $t_{\rm pl}^{\rm dS}=0$) when we do not regularize the geometry (\ie $\delta=0$) and at the same time we attach the stretched horizon at the pole (\ie $\rho=0$).
Instead, this regime becomes longer (and approaches infinity) when we move the stretched horizon closer to the cosmological one (\ie $\rho\to1$). 

When $t_{c1} < t_{c0}$, the plateau will start before the shockwave insertion, and this involves a computation of complexity during the time-dependent regime of the stretched horizon.
We leave this analysis for future investigations.
For this reason, the numerical results represented by the continuous lines in fig.~\ref{fig:plateau_rho099_SdS3} are only depicted after a certain value of $t_w,$ which depends on $\varepsilon,$ such that we meet the requirement $t_{c0}<t_{c1}.$

\paragraph{Scrambling time.}
In the case of shockwaves in AdS space, we can extract from the duration of the plateau the time it takes the system to scramble a perturbation \cite{Stanford:2014jda,Chapman:2018lsv}. This time is known as the scrambling time and it determines the beginning of the linear regime in the evolution of $t_{\rm pl}$.
Indeed, there is a linear regime for $t_w \gg L$ such that
\beq
t_{\rm pl} = 4 (t_w - t_*) \, .
\label{eq:linear_approx_plateau}
\eeq
In the case of asymptotically dS space, we will work in the same way, to find an analogous notion of scrambling time for the shockwave perturbation. This scrambling time can be found analytically in an expansion for sufficiently large $t_w$.

We parameterize the position $r_s$ at the critical time $t_{c1}$ in this limit as
\beq
r_s = a_1 L ( 1 - e^{-x_s}) \, .
\label{eq:assumptions1_tc_SdS3}
\eeq
A numerical analysis shows that $e^{-x_s} \ll \varepsilon$ and  $e^{-x_s} \ll 1$.
Therefore, we will work under the assumption that the exponential $e^{-x_s}$ is small compared to any other (dimensionless) parameters characterizing the solution.
One can similarly show that a good ansatz for the special position $r_b$ of the WDW patch at the critical time $t_{c2}$ is determined by
\beq
r_b = a_2 L ( 1 + e^{-x_b}) \, , \qquad
e^{-x_b}  \ll \varepsilon \, , \qquad
e^{-x_b} \ll 1 \, ,
\label{eq:assumptions2_tc_dS}
\eeq
where again we do not assume any condition on the strength of the shockwave insertion.

One can show that the total duration of the plateau in the regime $t_w \gg L$ is well approximated by eq.~\eqref{eq:linear_approx_plateau}, with scrambling time
\beq
 t_* = \le\frac{L}{4a_1}+\frac{3L}{4a_2} \ri\log \le \frac{1-\rho}{1+\rho} \ri
+ \frac{L}{2} \le \frac{1}{a_1} + \frac{1}{a_2} \ri \log \le \frac{a_1+a_2}{a_2-a_1} \ri \, .
\label{eq:duration_plateau_SdS3}
\eeq
Here we assumed that the stretched horizon after the shockwave insertion is determined according to eq.~\eqref{eq:constant_redshift_condition_app} by requiring a constant cosmological redshift.
One can notice from the plot in fig.~\ref{fig:plateau_rho099_SdS3} that this approximation (marked by the dotted lines) works very well in the regime $t_w \gg L$. 
It is important to stress that this prediction is valid in any regime where the assumptions \eqref{eq:assumptions1_tc_SdS3} and \eqref{eq:assumptions2_tc_dS} hold, independently of the values of $\rho, \varepsilon $.

If we further assume that $\varepsilon \ll 1,$ such that the factors $a_i$ defined in eq.~\eqref{eq:ai_SdS3} are related by
\beq
a_2 \approx
a_1 \le 1  + \varepsilon \, \frac{4 G_N \mathcal{E}_1}{a_1^2} \ri \, ,
\eeq
then the corresponding approximation for the scrambling time reads
\beq
t_* \underset{\varepsilon \rightarrow 0}{\approx} \frac{L}{a_1} \log \le \frac{a_1^2}{2 G_N \mathcal{E}_1} \frac{1}{\varepsilon} \ri = 
 \frac{1}{2 \pi T_1} \log \le \frac{a_1^2}{2 G_N \mathcal{E}_1} \frac{1}{\varepsilon} \ri  \, ,
 \label{eq:limit_scrambling_time_SdS3}
\eeq
where we recognized that the prefactor of the logarithmic divergence is proportional to the Hawking temperature $T_1= a_1/(2 \pi L)$ given in eq.~\eqref{eq:Hawking_temperature_SdS3}. For comparison, in the three-dimensional AdS-Vaidya geometry the expression for the duration of the plateau $t_{\rm pl}$ for small deformations of the shockwave ($\varepsilon \ll 1$) is given by \cite{Chapman:2018lsv} 
\beq
t_{\rm pl}^ {\rm Vaidya} = 4 (t_w - t_{*}) \, , \qquad
t_*^{\rm Vaidya} \underset{\varepsilon \ll 1}{\approx} \frac{1}{2 \pi T_1} \log \le \frac{2}{\varepsilon} \ri \, ,
\label{eq:scrambling_smalleps_Vaidya}
\eeq
where $T_1$ is the temperature \eqref{eq:Hawking_temperature_dS} of the black hole before the shockwave insertion and $\varepsilon$ is related to the jump in the black hole masses $m_2/m_1=(1+\varepsilon)^2.$

In this three-dimensional case, we also find an interpretation for the argument of the logarithm in terms of dual quantities defined on the would-be boundary dual theory.
We assume that the energy of the shockwave, defined as the difference between the energy of the black hole solution before and after the perturbation, is proportional to a few thermal quanta: $\mathcal{E}_1 - \mathcal{E}_2 = 2 T_1 $.
According to eq.~\eqref{eq:generic_epsilon_geometries}, this implies
\beq
\frac{a_1^2}{2 G_N \mathcal{E}_1} \frac{1}{\varepsilon} = 
\frac{2T_1 S_1}{\mathcal{E}_1 - \mathcal{E}_2} = S_1 \, ,
\label{help}
\eeq
such that the scrambling time \eqref{eq:limit_scrambling_time_SdS3} reads
\beq
t_* \underset{\varepsilon \ll 1}{\approx}
\frac{1}{2 \pi T_1} \log S_1 \, . \label{smart5}
\eeq
Now, we note that this result precisely matches with the one obtained for three-dimensional AdS-Vaidya geometry connecting two BTZ black hole backgrounds, since $2/\varepsilon=S_1$ in eq.~\eqref{eq:scrambling_smalleps_Vaidya} once we require the energy of the shockwave to be $\Delta E=2T_1.$
Remarkably, these very different settings reproduce the same leading behavior for the scrambling time in three dimensions.\footnote{Recall that in eq.~\eqref{smart5}, $S_1$ corresponds to the entropy of the cosmological horizon (there is no black hole horizon for SdS$_3$) while for the AdS-Vaidya background, $S_1$ is the black hole entropy. In both cases, this can be regarded as the entropy of the state in the dual theory.}

The opposite limiting case corresponds to the maximal change of the energy of the black hole, which happens when $\mathcal{E}_2=0,$ or analogously $\varepsilon=1.$
In such a framework, we have a transition from SdS$_3$ to empty dS space, and the position of the cosmological horizon simply reduces to the value of the curvature radius $L.$ This is equivalent to setting $a_2=1$, which gives
\beq
t_* \underset{\varepsilon =1}{=} \le\frac{L}{4a_1}+\frac{3L}{4} \ri\log \le \frac{1-\rho}{1+\rho} \ri
+ \frac{L}{2} \le \frac{1}{a_1} + 1 \ri \log \le \frac{1+a_1}{1-a_1} \ri \, .
\eeq
We might note here that the prefactors in both terms, which set the scale of $t_*$, involve some average of the inverse temperatures of the initial and final cosmological horizons. When the critical times discussed in section \ref{ssec:critical_times_dS} satisfy $t_{c1}<t_{c0}$, the duration of the plateau changes because it is measured starting from the shockwave insertion.
Nonetheless, one can check that there is still a linear regime when $t_w \gg L$, and that the scrambling time defined in eq.~\eqref{eq:duration_plateau_SdS3} is still the same.

One may wonder what is the duration of the regime delimited by the critical times $t_{c, \rm st1} \leq t \leq t_{c, \rm st2}$ defined in eq.~\eqref{eq:critical_time_behind_stretched}, corresponding to the joints of the WDW patch in fig.~\ref{fig:alternative3_WDWpatch} moving behind the stretched horizons.
One can show that in the limit $t_w \gg L,$ the duration of such regime is approximated by the same analytic function obtained in eqs.~\eqref{eq:linear_approx_plateau} and \eqref{eq:duration_plateau_SdS3}.
Indeed, the kinks in fig.~\ref{subfig:focus_CV20_SdS3_plateau_new_st_long_pl} are very close to the beginning and ending of the plateau regime, and we find that $t_{c, \rm st 1} \rightarrow t_{c1}, t_{c, \rm st2} \rightarrow t_{c2}$ when we increase $t_w.$

The results discussed so far apply when the stretched horizon is determined by requiring a constant cosmological redshift, as discussed in section \ref{ssec:details_const_red}.
We report for completeness the duration of the plateau and the scrambling time when the stretched horizons are determined by requiring a continuous flux of time instead, see section \ref{ssec:details_cont_time}:
\beq
t_{\rm pl} = 4 (t_w - t_*) \, , \quad
t_* = \frac{L}{a_1} \log \le \frac{1-\rho}{1+\rho} \ri 
+ \frac{L}{2 a_1} \le \frac{1}{a_1} + \frac{1}{a_2} \ri  \log \le \frac{a_1+a_2}{a_2-a_1} \ri \, .
\label{eq:duration_plateau_SdS3_conflux}
\eeq
Comparing to eq.~\eqref{eq:duration_plateau_SdS3}, we notice that the difference is only contained in the prefactor of the first term, and it becomes negligible when $\varepsilon \rightarrow 0$.

\subsection{Four-dimensional SdS space}
\label{ssec:CV20_SdS4}

\begin{figure}[ht]
    \centering
\includegraphics[scale=0.72]{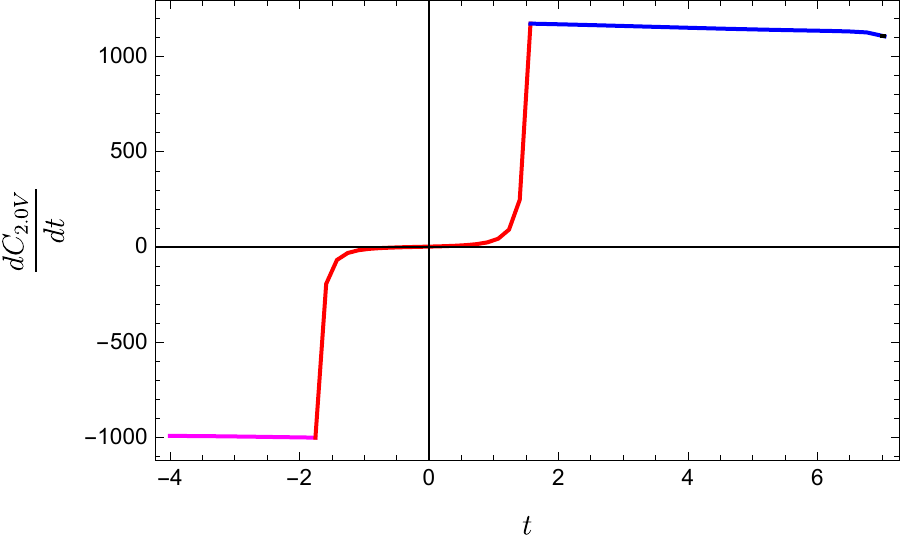} 
    \caption{Plot of the rate of CV2.0 for the geometry described by the blackening factor \eqref{eq:blackening_factor_shock_SdS} in $d=3.$
    We set $ t_w=2, \rho=0.5, m_1=0.14, m_2=0.13, \delta=0.1.$ Different colors correspond to different regimes delimited by the corresponding critical times.}
\label{fig:plotCV20rateSdS4} 
\end{figure}

In the case of general higher-dimensional black holes in dS space, the treatment becomes numerical and we do not have a closed form for the special positions of the WDW patch -- see however the next subsection.
While the integrations required to compute CV2.0 are not known analytically, the complexity rate is still determined by the analysis given in section \ref{ssec:rate_CV20}.
We focus on the four-dimensional case, which corresponds to the blackening factor \eqref{eq:blackening_factor_shock_SdS} with $d=3.$
The resulting plot of the rate of change of the holographic complexity is plotted in fig.~\ref{fig:plotCV20rateSdS4}.
While the Penrose diagram is different from the lower-dimensional counterpart, the prescription to attach the stretched horizon between the black hole and cosmological horizon is responsible for a similar time evolution compared to the other cases.
By numerical integration, we can show that just as in SdS$_3$, there is also a plateau region at intermediate times in SdS$_4$.

\subsubsection{Plateau of complexity}

As mentioned above, higher-dimensional SdS black holes still admit a plateau regime delimited by the critical times defined in eqs.~\eqref{eq:tc1} and \eqref{eq:tc2}. 
When $t_w \gg L,$ the position $r_s$ of the WDW patch can be parametrized as
\beq
r_s = r_{C1} \le 1- e^{-x_s} \ri  \qquad {\rm with}\quad
e^{-x_s} \ll \frac{\varepsilon}{L} \, , \quad
e^{-x_s} \ll 1 \, ,
\label{eq:assumptions1_SdS4}
\eeq
and a similar ansatz applies to the position $r_b$
\beq
r_b = r_{C2} (1+ e^{-x_b}) \qquad {\rm with}\quad
e^{-x_b} \ll \frac{\varepsilon}{L} \, , \quad
e^{-x_b} \ll 1 \, .
\label{eq:assumptions2_SdS4}
\eeq
As in SdS$_3$, we will only consider the complexity after the shockwave in order to not be affected by the time-dependent part of the stretched horizon.

\begin{figure}[ht]
\centering
\subfigure[]{\includegraphics[scale=0.66]{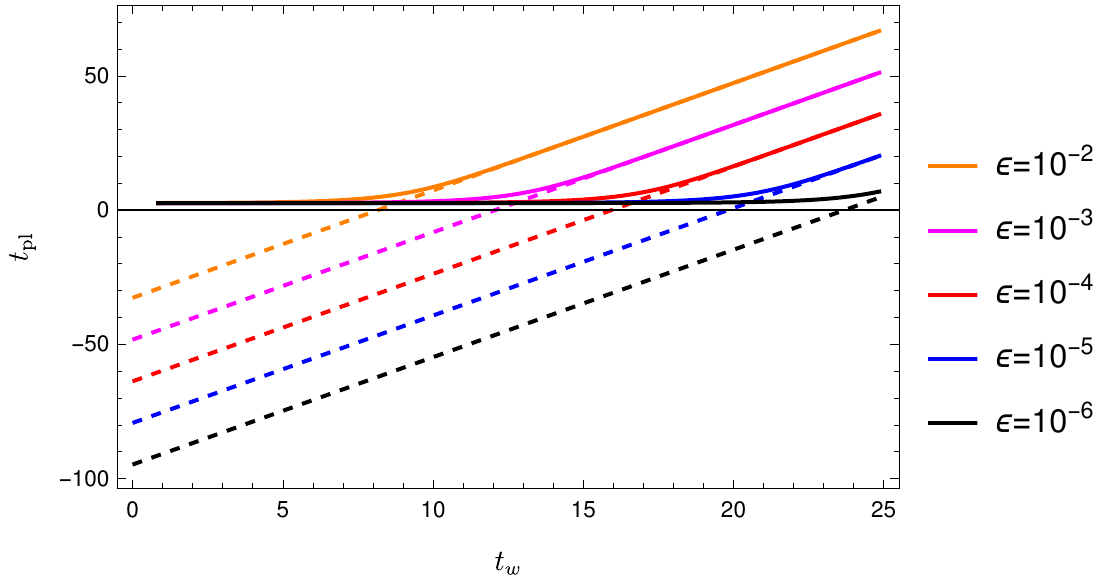}} \,
\subfigure[]{\includegraphics[scale=0.66]{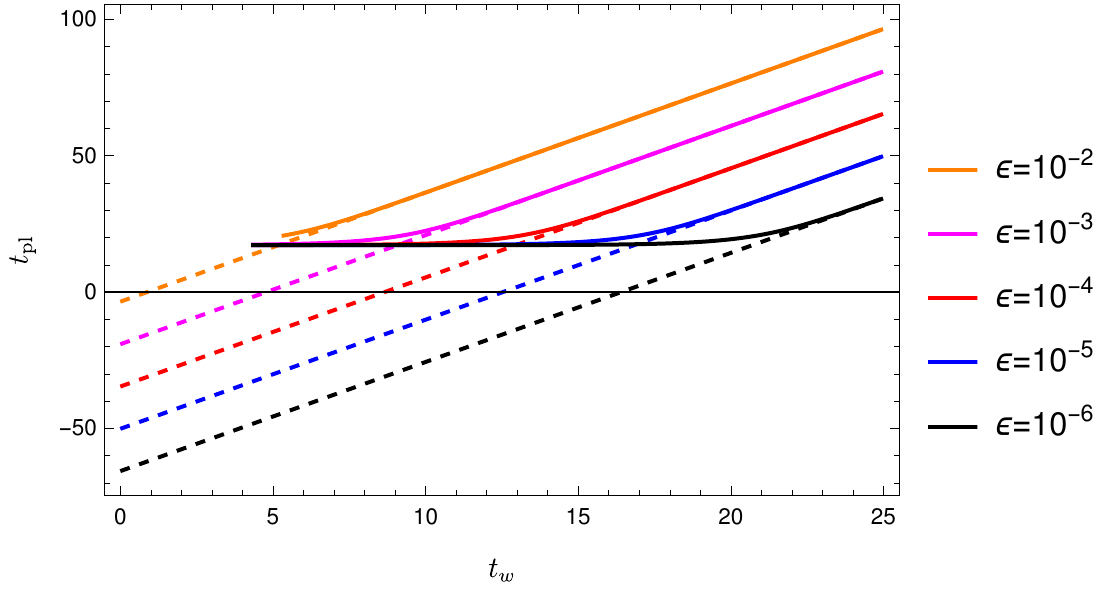}}
\caption{Dependence of the duration of the plateau on the insertion time of the shockwave, for various choices of $\varepsilon$ defined in eq.~\eqref{eq:generic_epsilon_geometries} in $d=3$. We fix $\delta =0.1, m_1 = 0.14$ and $\rho=0.5$ in the left panel, $\rho=0.99$ in the right one. The dotted lines corresponds to the curves $t_{\rm pl} = 4 (t_w - t_*)$ in eq.~\eqref{eq:duration_plateau_SdS4}.}
\label{fig:plateau_rho099_SdS4}
\end{figure}

Using the ansatz \eqref{eq:assumptions1_SdS4} and \eqref{eq:assumptions2_SdS4}, one can derive that
the linear behaviour for insertion times satisfying $t_w \gg L$  is well approximated by the interpolating function 
\beq
t_{\rm pl} = 4 (t_w - t_*) \, , 
\eeq
with scrambling time
\beq
\begin{aligned}
t_* & =  -  \frac{ r_{C2}^2 + r_{h2} r_{C2} + r_{h2}^2 }{(r_{C2}-r_{h2})(2r_{C2}+r_{h2}) (r_{C2}+2r_{h2})} 
\left[ r_{C2}^2 \log \left| \frac{r_{C1}-r_{C2}}{r_{C1}+r_{C2}+r_{h2}} \right| \right. \\
& \left. + 2r_{C2} r_{h2} \log \left| \frac{r_{C1}-r_{C2}}{r_{C1}-r_{h2} } \right| 
- r_{h2}^2 \log \left| \frac{r_{C1}-r_{h2}}{r_{C1} + r_{C2} +r_{h2}} \right|
\right] \\
& - \frac{ r_{C1}^2 + r_{h1} r_{C1} + r_{h1}^2 }{(r_{C1}-r_{h1})(2r_{C1}+r_{h1}) (r_{C1}+2r_{h1})} 
\left[ r_{C1}^2 \log \left| \frac{(r_{C1}-r_{C2}) \le (\rho+1)r_{C1}+(2-\rho)r_{h1} \ri}{(\rho-1)(r_{C1}+r_{C2}+r_{h1})(r_{C1}-r_{h1})}  \right|  \right. \\
& \left.   
+ 2 r_{C1} r_{h1} \log \left| \frac{(r_{C1}-r_{C2})\rho}{(r_{C2}-r_{h1})(1-\rho)} \right|
- r_{h1}^2 \log \left| \frac{(r_{C2}-r_{h1})\le (\rho+1)r_{C1}+(2-\rho)r_{h1} \ri}{\rho (r_{C1}+r_{C2}+r_{h1})(r_{C1}-r_{h1})} \right|  
\right] 
\, .
\end{aligned}
\label{eq:duration_plateau_SdS4}
\eeq
Indeed, we can notice from fig.~\ref{fig:plateau_rho099_SdS4} that the interpolation is precise.\footnote{The scrambling time \eqref{eq:duration_plateau_SdS4} is determined using the continuous time prescription described in section \ref{ssec:details_cont_time} for the stretched horizon. A similar computation using the constant redshift prescription of section \ref{ssec:details_const_red} is technically harder. However, we observe that in three dimensions the two prescriptions give the same result in the relevant limit $\varepsilon \rightarrow 0,$ see eq.~\eqref{eq:duration_plateau_SdS3_conflux}. We believe that a similar phenomenon happens in higher dimensions.  }
We find a qualitatively similar behavior to the one found for SdS$_3$.
As discussed below eq.~\eqref{eq:tpl_empty_dS}, the equation for the duration of the plateau only holds when $t_{c0} \leq t_{c1}$.
Therefore, the continuous curves in the plot refer to this regime only.

We expand the scrambling time for small values of the $\varepsilon$ parameter defined in eq.~\eqref{eq:generic_epsilon_geometries}.
To this aim, we first expand at linear order the event horizons as
\beq
 r_{h2} = r_{h1} - \alpha r_{\rm cr} \, \varepsilon + \mathcal{O}(\varepsilon^2) \, , \qquad
r_{C2} = r_{C1} + \beta r_{\rm cr} \, \varepsilon + \mathcal{O}(\varepsilon^2) \, .
\label{eq:expansion_radii_SdSd} 
\eeq
Exploiting the fact that the blackening factor of the black hole vanishes at the horizons and expanding at first order around $\varepsilon=0,$ we find the explicit expressions for $d=3$ 
\beq
\alpha \, r_{\rm cr} = \frac{r_{h1} r_{C1} (r_{h1}+r_{C1})}{r_{C1}^2 + r_{h1} r_{C1} - 2 r_{h1}^2} \, , \qquad
\beta \, r_{\rm cr} = \frac{r_{h1} r_{C1} (r_{h1}+r_{C1})}{2 r_{C1}^2 - r_{h1} r_{C1} - r_{h1}^2} \, .
\label{eq:expansion_radii_SdS4}
\eeq
Substituting these expressions inside the scrambling time \eqref{eq:duration_plateau_SdS4} and keeping only the leading logarithmic divergence, we find
\beq
t_* \underset{\varepsilon \ll 1}{\approx}  
\frac{2 r_{C1} L^2}{(r_{C1}-r_{h1})(2r_{C1}+r_{h1})}
\log \le  \frac{1}{\varepsilon} \ri = 
 \frac{1}{2 \pi T_{C1}} \log \le \frac{1}{\varepsilon} \ri  \, ,
 \label{eq:approx_scrambling_plateau_SdS4}
\eeq
where we recognized the Hawking temperature at the cosmological horizon of the SdS$_4$ black hole, see eq.~\eqref{eq:Hawking_temperature_SdS4}.
Therefore, we notice that this limit for the scrambling time shows a universal pattern common to all asymptotically dS spacetimes. Note that, in the limit $\varepsilon\ll 1$ we have  $T_{c1}\approx T_{c2}$ for first order corrections in $\varepsilon$. Finally, we can also anticipate that the argument of the logarithm in eq.~\eqref{eq:approx_scrambling_plateau_SdS4} will also acquire a factor of $1-\rho$ in the double scaling limit considered below.

\subsection{Higher dimensions}
\label{ssec:higherd}
For higher dimensions, we are able to find a general description of the scrambling time. The description is more cumbersome than in eq.~\eqref{eq:duration_plateau_SdS4}, but it greatly simplifies in the double scaling limit defined by
\beq
\varepsilon \rightarrow 0 \, , \qquad
\rho \rightarrow 1 \, , \qquad
\frac{1-\rho}{\varepsilon} \quad \mathrm{fixed} \, .
\label{eq:double_scaling_limit}
\eeq
The result reads
\beq
t^{\rm SdS_{d+1}}_* =  \frac{1}{2 \pi T_{C1}} 
\log \le   \frac{1-\rho}{\beta r_{\rm cr} \varepsilon} \le r_{C1}- r_{h1} \ri  \ri
+ \mathcal{O} (1-\rho, \varepsilon) \, 
\label{rocker4}
\eeq
This formula is valid in any number of spacetime dimensions, with the values of the critical radius given in eq.~\eqref{eq:critical_mass_SdS} and the constant $\beta$ defined by the leading order in $\varepsilon$,
\beq
r_{\rm cr} \equiv L \sqrt{\frac{d-2}{d}}
 \, , \qquad
r_{C2} = r_{C1} + \beta r_{\rm cr} \, \varepsilon + \mathcal{O}(\varepsilon^2) \, .
\label{rocker3}
\eeq
To better understand the scrambling time for SdS$_{d+1}$, we consider the limit of small black holes $r_{h1}/r_{C1} \ll 1$. Then using eq.~\eqref{rocker3}, we can recast the above result \eqref{rocker4} as
\beq
t^{\rm SdS_{d+1}}_* \approx  \frac{1}{2 \pi T_{C1}} 
\log \le  (1-\rho) \frac{(d-1)\, S_{C1}}{\Delta S_{C1}}   \ri\,.
\label{smarter8}
\eeq
Here, $S_{C1}$ is the initial entropy of the cosmological horizon while $\Delta S_{C1}=S_{C2}-S_{C1}$ is the increase in the cosmological entropy.

In the regime $r_{h1}/r_{C1} \ll 1$, we can also write the scrambling time as
\beq
t^{\rm SdS_{d+1}}_* \approx  \frac{1}{2 \pi T_{C1}} 
\log \left[   \le \frac{r_{C1}}{r_{h1}} \ri^{d-2} \frac{2(1-\rho)}{\varepsilon}  \right]
+ \mathcal{O} (1-\rho, \varepsilon) \, .
\eeq
One can always exploit the definitions of Hawking temperature to express the scrambling times in terms of $T_{C1}, T_{h1}.$
The precise dictionary depends on the number of dimensions; to avoid cumbersome expressions, we do not report them explicitly. 
However, some simplifications occur in the limit of small black holes $r_{h1}/r_{C1} \ll 1$.
In this limit, we send $r_{h1} \rightarrow 0$ and $r_{C1} \rightarrow L,$ so that the expressions \eqref{eq:hor_cosm_temp_SdS} for the Hawking temperatures become
\beq
T_{h1} \approx \frac{d-2}{4 \pi r_{h1}} \, , \qquad
T_{C1} \approx \frac{1}{2 \pi r_{C1}} \, \quad  \Rightarrow \quad 
\frac{r_{h1}}{r_{C1}} \approx \frac{d-2}{2} \frac{T_{C1}}{T_{h1}} \, .
\eeq
Therefore we can express the scrambling time for a light black hole in SdS as
\beq
t^{\rm SdS_{d+1}}_* \approx  \frac{1}{2 \pi T_{C1}} 
\log \left[   \le \frac{T_{h1}}{T_{C1}} \ri^{d-2} \le \frac{2}{d-2} \ri^{d-2} \frac{2(1-\rho)}{\varepsilon}  \right]
 \, .
\label{rocker9}
\eeq
For $T_{h1}\gg 1$ the stretch horizon can approach the cosmological horizon without $t_*$ becoming negative. The condition for positive $t_*$ is $T_{h1}^{d-2}(1-\rho)>\varepsilon$. We note that in this expression \eqref{rocker9}, the black hole seems to play a central role in defining the scrambling time in contrast to the formula in eq.~\eqref{smarter8} only refers to the cosmological horizon. Of course, these two expressions are entirely equivalent even though their physical interpretation is somewhat disparate. We would like to understand if either of these conditions have a good geometric interpretation in future studies.

\section{Discussion} 
\label{sec:discussion}

\subsection{Summary of results}

In order to better understand dS holography, we studied shockwaves in SdS geometries and how they affect the spacetime volume of the WDW patch as a function of the time along the stretched horizon. We found several interesting properties. 
For times around $t_L=t_R=0$, we found that there is a plateau regime where the spacetime volume of the WDW patch is approximately constant. 
Similar behavior was found for shockwaves of AdS black holes \cite{Chapman:2018lsv}, but the origin of the phenomena in the two settings is different. In AdS black holes the plateau appears when the top and bottom joints of the WDW patch are both behind the singularities of the black hole, while in SdS background the plateau appears in the regime where the future and past joints both do not yet reach future and past timelike infinities. 
The difference between the two frameworks comes from the inclusion of the stretched horizon in SdS geometry.
In both cases, there is a scrambling time that determines the length of the plateau, after which complexity starts to increase linearly. That is, $t_{\rm pl} = 4 (t_w-t_*)$.

We considered a double scaling limit \eqref{eq:double_scaling_limit} where the strength of the shockwave is small and the stretched horizon approaches the cosmological horizon for general $d+1$ dimensions. Further in the regime of small black holes (\ie $r_{h1} \ll r_{C1}$), we found two interesting expressions for the scrambling time. The first of these \eqref{smarter8},
\beq
t^{\rm SdS_{d+1}}_* \approx  \frac{1}{2 \pi T_{C1}} 
\log \le  (1-\rho) \frac{(d-1)\, S_{C1}}{\Delta S_{C1}}   \ri\,,
\label{smarter8a}
\eeq
refers only to properties of the cosmological and stretched horizons. Of course, this may seem natural as the scrambling time should be a property of the dual boundary theory residing on the stretched horizon. However, the second expression \eqref{rocker9}, which is mathematically equivalent to the first,
\beq
t^{\rm SdS_{d+1}}_* \approx  \frac{1}{2 \pi T_{C1}} 
\log \left[   \le \frac{T_{h1}}{T_{C1}} \ri^{d-2} \le \frac{2}{d-2} \ri^{d-2} \frac{2(1-\rho)}{\varepsilon}  \right]
 \, ,
\label{eq:tpl_conclusions}
\eeq
seems to give an important role to the black hole horizon as well. Of course, in both expressions, the scrambling time is proportional to the inverse temperature of the cosmological horizon, and it is logarithmic in the parameter $(1-\rho)$ describing the location of the stretched horizon. It may be useful to note that in this regime ($r_{h1}\ll r_{C1} $), the black hole makes a small entropic contribution (\ie $S_{h1}\ll S_{C1})$ but it is very hot (\ie $T_{h1}\gg T_{C1})$. Hence the system is far from an equilibrium configuration by some measure.

In appendix \ref{app:formation},
we provided an alternative derivation of the scrambling time in terms for the complexity of formation \cite{Chapman:2016hwi}  . There, in working with the double scaling limit \eqref{eq:double_scaling_limit}, we found a precise match between the expressions for the scrambling time in eqs.~\eqref{rocker4} and \eqref{eq:scrambling_formation} when we also consider the simplifying limit of small black holes, as was considered above. Hence for any number of dimensions (\ie $d\ge3$), both approaches yield precisely the same expression for the scrambling time, as given in eq.~\eqref{smarter8a} or eq.~\eqref{eq:tpl_conclusions}.

It is interesting to compare eq.~\eqref{smarter8a} for general $d$ with the expression the scrambling time in eq.~\eqref{smart5} for $d=2$, since both apply for small shocks (\ie $\varepsilon\ll 1$). First, however, we observe that it is straightforward to show that beginning with eq.~\eqref{eq:duration_plateau_SdS3} and applying the same double scaling limit as in eq.~\eqref{eq:double_scaling_limit}, one recovers the scrambling time in eq.~\eqref{smarter8a} with $d=2$.

There are two key differences between the present analysis (\ie the double scaling limit) and our derivation of the expression in eq.~\eqref{smart5} (or eq.~\eqref{eq:simplified_scrambling_time_Cf_SdS3}). The first is that for the former, we assumed $(1-\rho)\sim O(1)$ and so did not keep track of the contribution proportional to $\log(1-\rho)$ which appears in eq.~\eqref{eq:duration_plateau_SdS3} (or eq.~\eqref{eq:scrambling_time_Cf_SdS3}). 
The second difference is that we made a specific choice for the strength of the shockwave $\varepsilon$ so that the change in energy corresponds to a few thermal quanta. That is, $\varepsilon\mathcal{E}_1=\mathcal{E}_1 - \mathcal{E}_2 = 2 T_1 $ -- see eq.~\eqref{help} and the preceding discussion. In the context of eq.~\eqref{smarter8a}, it is interesting to consider what this choice implies for the change in the entropy of the cosmological horizon. 
Applying eqs.~\eqref{eq:generic_rc_geometries} and \eqref{eq:ai_SdS3}, we find
\beq
\Delta S_{C1}=\frac{\pi L }{2G_N}\,(a_2-a_1)\approx \frac{\mathcal{E}_1 - \mathcal{E}_2}{T_{C1}}=2\,,
\label{changeS2}
\eeq
where we have used the ``small black hole" approximation that $8G_N\mathcal{E}_i\ll 1$.
A natural generalization to higher dimensions of this result for $d=2$ would be to choose $\varepsilon$ so that $\Delta S_{C1}\sim d$ in which case the scrambling time \eqref{smarter8a} reduces to
\beq
t^{\rm SdS_{d+1}}_* \approx  \frac{1}{2 \pi T_{C1}} 
\log \le\frac{d-1}{d}\,(1-\rho) \, S_{C1}   \ri\,.
\label{smarter8b}
\eeq
Of course, this expression is very reminiscent of those found in holographic studies of AdS black holes, in that the scrambling time is proportional to the log of the horizon entropy.

The primary difference between the formula for AdS scrambling times and those found here is the appearance of the factor of $1-\rho$ in the argument of the logarithm in eq.~\eqref{smarter8a} or eq.~\eqref{smarter8b}. We are considering a limit where this factor is small and so the latter is simply related to the redshift factor experienced by observers on the stretched horizon, \ie $f(r_{\rm st})\simeq 2(1-\rho)$ -- see further discussion below. 
It is curious that in, \eg eq.~\eqref{smarter8a}, one can tune this factor to cause the vanishing of this leading contribution to the scrambling time in specific instances. It is not a priori clear what the meaning of this observation would be for the dual theory, and we would like to understand this better in the future. 

For late (early) times in the evolution of the spacetime volume of the WDW patch, we found that there is a linear increase (decrease) in complexity at a rate governed by the cutoff, see eq.~\eqref{eq:late_time_rate_dS}. This behavior starts after (before) the top (bottom) joint of the WDW patch reaches future (past) infinity. Of course as in ref.~\cite{Jorstad:2022mls} which examined holographic complexity in empty dS space, the rate \eqref{eq:late_time_rate_dS} in this linear regime is governed by our choice of the regulator surfaces near the future and past timelike infinities -- see eqs.~\eqref{eq:definition_rmax_shocks} and \eqref{eq:rmax2}. The expression in  eq.~\eqref{eq:late_time_rate_dS} can be written as 
\beq
\frac{d \mathcal{C}_{V2.0}}{dt}  \approx 
\frac{T_{dS}}{T_{C1}} \, \frac{S_{C1}}{\delta^d}  \, ,
\label{laterate}
\eeq
where we are assuming $\delta \ll 1$. Here we are distinguishing $T_{dS}$, the temperature of empty dS space, and $T_{C1}$, the initial temperature of the cosmological horizon in the shockwave geometry. Of course, in the regime of small black holes (\ie $r_{h1}\ll r_{C1} $), these two temperatures are approximately the same and our result here matches that found in \cite{Jorstad:2022mls}. Of course, with a full nonperturbative description, this linear regime would arise because of the finiteness of the Hilbert space, and we expect that the prefactor $S_{C1}$ would be replaced by a factor exponential in the number of degrees of freedom.

Returning to the derivation of the scrambling time given in appendix \ref{app:formation}, the complexity of formation for the SdS$_3$ background initially decreases as a function of $t_w$, until there is a kink in correspondence of the transition from a conventional to a special configuration of the WDW patch, see fig.~\ref{fig:plot_CV20_form_SdS3}.
After the kink, the complexity of formation increases and can be approximated by a linear function with a scrambling time that shows the same features as the expression \eqref{eq:tpl_conclusions}. 
In particular, it is inversely proportional to the Hawking temperature of the cosmological horizon, and logarithmic in the parameters $\varepsilon$ and $(1-\rho)$. In fact, as shown in eq.~\eqref{eq:scrambling_formation}, the same behaviour is found in any number of dimensions $d\geq 3$ when we also consider the regime of small black holes.

Our entire discussion to this point describes the evolution in terms of the coordinate time $t$. It may be more natural to express the time evolution in terms of the proper time of a stationary observer sitting at the stretched horizon, \ie $t_{\rm prop}=\sqrt{f(r_{\rm st})}\, t$. As we are primarily concerned with the evolution after the shockwave passes the stretched horizon, this would simply introduce a constant overall prefactor in our results. For example, rather than being inversely proportional to $T_{C1}$, the scrambling time \eqref{smarter8a} measured in proper time units would be inversely proportional to $T_{C1,{\rm prop}}=T_{C1}/\sqrt{f(r_{\rm st})}$ which corresponds to the temperature experienced by the stationary observer.\footnote{Note that in general, a stationary observer actually sees a flux of radiation from both the cosmological and black hole horizons. However, if the observer is close to the cosmological horizon, her experience is dominated by the former flux with $T_{C1,{\rm prop}}$.}

\begin{figure}
    \centering
    \includegraphics[scale=0.4]{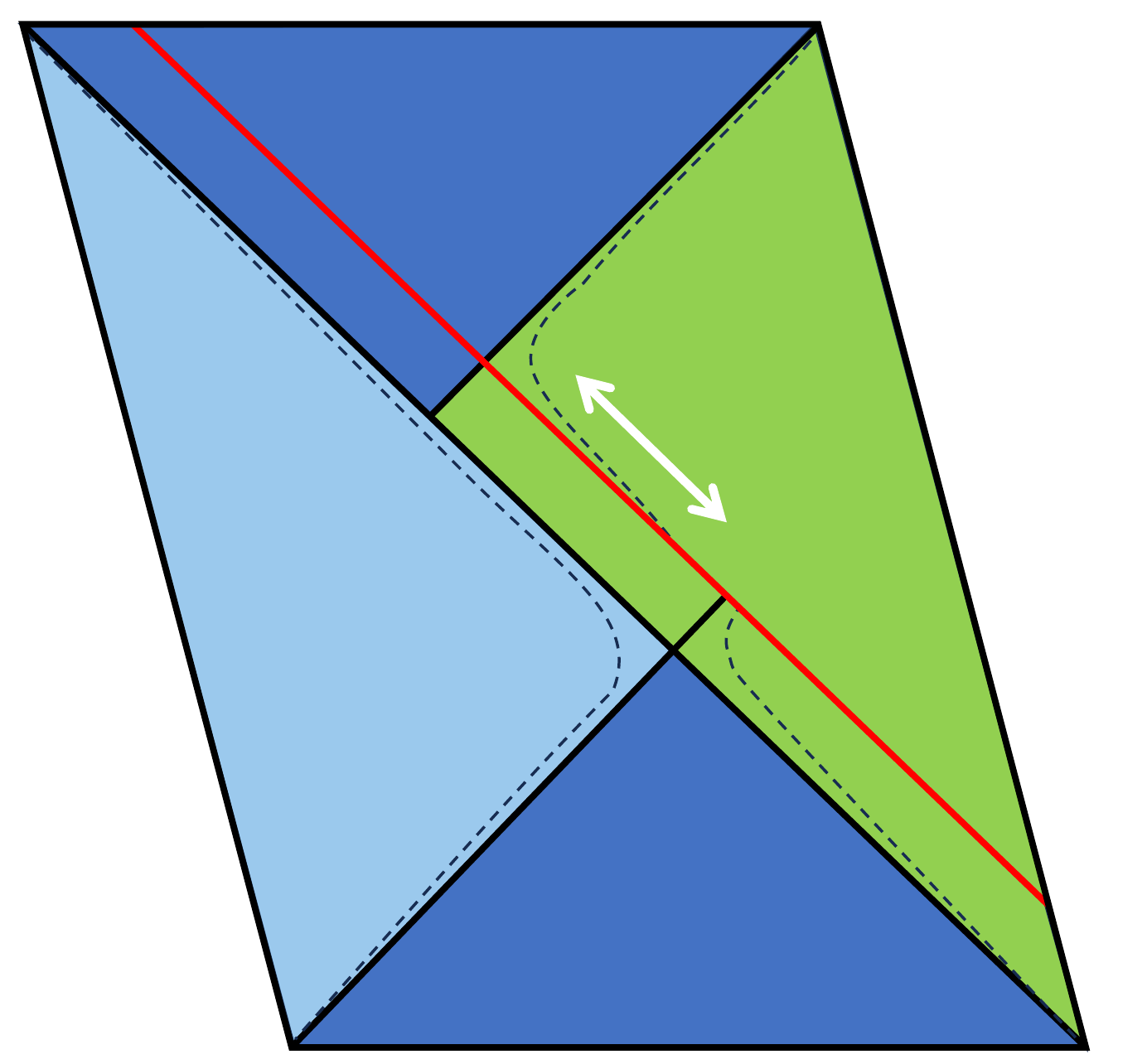}
    \caption{Pictorial representation of the time advance of a light rays crossing a shockwave (shown as the red line) in dS space, depicted for three dimensions for simplicity.
    The causal region for the observer at the left (right) pole is shaded in light blue (green). As illustrated, these two regions and hence the corresponding stretched horizons (shown as dashed lines) are in causal contact.
    This causal structure gives rise to the exceptional configurations discussed in section \ref{ssec:special_configurations_WDW}.}
    %\sch{The figure is a slightly more transparent way to draw the same Penrose diagram so that the radius doesn't jump. There are not really light rays in the figure, but you could think of one tracking the horizon from the left to illustrate the phenomenon of time advance/causal contact and so on. I think I agree with Stefano that a light ray crossing the shock will be a straight line in this diagram.} \rcm{update caption}}
    \label{fig:time_advance}
\end{figure}

In SdS with positive energy shockwaves, exceptional configurations of the WDW patch may arise, as we discussed in section \ref{ssec:special_configurations_WDW}. When the special configurations appear, there is a linear increase in the length of the plateau as a function of $t_w$. When the more conventional configuration of the WDW patch studied in section \ref{ssec:critical_times_dS} appears, the length of the plateau is approximately constant as a function of the insertion time $t_w$ and we found that it is the same constant for any energy of the shockwave $\varepsilon$. However, for large enough $t_w$, the special configurations always appear.

These exceptional configurations of the WDW patch arise because the Penrose diagram of dS space grows taller when a null pulse carrying positive energy is inserted in the bulk \cite{Gao:2000ga}. The corresponding time advance of a light ray crossing the shockwave brings the left and right causal patches of the geometry into causal contact.\footnote{We thank the referee emphasizing this point.} As shown in fig.~\ref{fig:time_advance}, a shockwave emerging from the right side allows signals from the left casual patch to enter that on the right. As commented in footnote \ref{footy178}, one then finds that irrespective of the detailed prescription defining the stretched horizons, these two surfaces on either side of the cosmological horizon are always in causal contact in the presence of shockwaves.
Furthermore, with such a structure, the two casual patches no longer provide a bipartition of a global time slice and hence the dual description will no longer  associate these regions with two halves of an entangled state. It
is, of course, an interesting research problem to better understand the relation of the dual holographic states on the stretched horizons -- see  \cite{Aalsma:2020aib,Aalsma:2021kle} for a related discussion.  Perhaps some inspiration for a dual holographic interpretation can be drawn from traversable wormholes in AdS \cite{Gao:2016bin,Maldacena:2017axo}, where a similar situation occurs.

For the purposes of the present work, we remark that as noted in the beginning of section \ref{sec:stretched_horizon_shocks}, the details of the swithcback effect and the corresponding scrambling time are not affected by the specific choice of the stretched horizon on either side of the cosmological horizon. The key feature (which all of the prescriptions noted there satisfy) is that the stretched horizon sits at a fixed radius after passing the shockwave.

\subsection{Future directions}

The investigations carried out in this work open the way for many other future developments:

%\paragraph{Future direction for this project: \sch{we may want to write this in a textual style rather than points eventually}}
%\SB{I elaborated all the points.}
\begin{enumerate}
    \item \textbf{Time-dependent regime of the stretched horizon.}
    We would like to study the spacetime volume of the WDW patch before the WDW patch crosses the shockwave. 
    In this regime the stretched horizon, when imposing constant redshift, is time-dependent. This feature complicates the derivation of the spacetime volume. 
    We would like to learn if this regime affects the plateau when the latter starts before the shock.
    \item \textbf{Other complexity conjectures.}
    In order to complete the holographic complexity picture, we would like to study the action of the WDW patch and the volume of maximal surfaces, which are the two conjectures originally proposed in \cite{Susskind:2014moa,Stanford:2014jda,Susskind:2014rva,Brown:2015bva,Brown:2015lvg}.
    Recently these proposals were extended to the \textit{complexity=anything} conjecture, which defines a class of geometrical quantities reproducing the linear growth of complexity and the switchback effect \cite{Belin:2021bga,Belin:2022xmt,Jorstad:2023kmq}.
    It would be interesting to study if these proposals all admit similar properties in dS space.
    \item \textbf{Circuit models.}
    The dual model suggested by static patch holography is a quantum-mechanical theory with a finite-dimensional Hilbert space in a maximally mixed state.
    However the details of the holographic dictionary are still unclear.
    We would like to study circuit models that might have similar properties to the holographic behaviour found in this work. In \cite{Susskind:2021esx,Lin:2022nss} a setting based on the epidemic spreading in a double-scaled SYK model was proposed as the dual theory to dS living on the stretched horizon. This model reproduces the hyper-fast growth of complexity for theories with a dS dual, see eq.~\eqref{eq:late_time_rate_dS}. We would like to study the effect on this model of a perturbation to see if circuit complexity shows similar traits as  the holographic complexity derived in our setup of shockwaves in SdS. 
    Furthermore, in recent years it was proposed that a concrete dual description of three-dimensional dS space can be described in terms of $T\Bar{T}$ deformations with an additional term which makes the cosmological constant positive \cite{Lewkowycz:2019xse,Coleman:2021nor,Silverstein:2022dfj}.
    It would be interesting to understand the relation between these two proposals.
    \item \textbf{Two-dimensional case.}
    In our work we only considered SdS$_{d+1}$ for $d\geq 2$. For $d=1$ SdS$_2$ is the same solution as empty dS$_2$. We would like to study the construction of dS$_2$ in the presence of shockwaves that are sent at an arbitrary time, by imposing the NEC condition.
     After the construction of the geometry, we would like to study holographic complexity following the steps that were done in this work.
    \item \textbf{General GR theorems.}
    We found a similar behavior of the spacetime volume of the WDW patch in shockwave geometries with either AdS or dS asymptotics.
    For example, in both cases we found a plateau of complexity and a similar scrambling time proportional to the logarithm of the relative size of the perturbation corresponding to a delay of the plateau. This is an interesting geometric property. We would like to study the effect of shockwaves on general geometries and learn whether there is a general theorem in GR that can explain the similarities, perhaps along the lines of the theorems presented in the volume case in reference \cite{Engelhardt:2021mju}.   
    \item \textbf{Centaur geometries and dual theory.}
    Centaur geometries provide a setting where dS space is connected to an asymptotically AdS region and the boundary is timelike, as in the standard application of AdS/CFT duality.
    Volume complexity has been studied in this setting in the two-dimensional case \cite{Chapman:2022mqd}, but the construction of the geometries in higher dimensions is non-trivial (see \cite{Anninos:2022ujl} for recent developments). 
    It would be interesting to study holographic complexity in these settings in the presence of a shockwave, to compare with the computation presented in this work.
    This will allow to achieve a better understanding of the dual theory, and to distinguish universal features of complexity.
    \item \textbf{Localized shockwaves.}
    Shockwave solutions along the cosmological horizon of the kind considered in \cite{Hotta_1993,PhysRevD.47.3323,Sfetsos:1994xa} can be localized, \ie the energy-momentum tensor has a non-trivial profile along the angular directions.
    We are interested in studying how these different shockwave perturbations influence holographic complexity and its spreading, and compare its qualitative features with the case of entanglement, considered in \eg \cite{Roberts:2014isa,Roberts:2016wdl,Mezei:2016zxg}. 
\end{enumerate}

\section*{Acknowledgements}

We are happy to thank Roberto Auzzi, Saskia Demulder and Dami\'an A. Galante for fruitful discussions and useful comments.
The work of SB, RB and SC is supported by the Israel Science Foundation (grant No.~1417/21), the German Research Foundation through a German-Israeli Project Cooperation (DIP) grant ``Holography and the Swampland'' and by  Carole and Marcus Weinstein through the BGU Presidential Faculty Recruitment Fund. 
SB is grateful to the Azrieli foundation for the award of an Azrieli fellowship.
Research at Perimeter Institute is supported in part by the Government of Canada through
the Department of Innovation, Science and Economic Development Canada and by
the Province of Ontario through the Ministry of Colleges and Universities. RCM is supported in part by a Discovery Grant from the Natural Sciences and Engineering Research Council of Canada,  by funding from the BMO Financial Group and by the Simons Foundation through the ``It from Qubit'' collaboration.

\appendix

\section{Comparison with reference \cite{Anegawa:2023dad}}
\label{app:comparison_jap}

As the present paper was being completed, we became aware of reference \cite{Anegawa:2023dad} which presents a related discussion on holographic complexity in de Sitter spaces with the insertion of null shockwaves. The latter 
mainly focuses on the special case of $d=2$, \ie of a three-dimensional bulk geometry. In this appendix, we compare our notation and results with those found in \cite{Anegawa:2023dad}. In table~\ref{tab:conventions}, the reader can find where in each of the papers, the details of the respective conventions were described.
In summary, most of the conventions coincide and the main difference between the two papers is that they focus on a different order of limits to study the critical times.
Below we specify the differences between the conventions in the two papers.

\begin{table}[ht]   
\begin{center}    
\begin{tabular}  {|c|c|c|} \hline  &  \textbf{Our conventions} & \textbf{Reference \cite{Anegawa:2023dad}}  \\ \hline
\rule{0pt}{4.9ex} \textbf{Geometry}  & section \ref{sec:geometric_preliminaries} & section 2.2   \\
\rule{0pt}{4.9ex} \textbf{Stretched horizon} & section \ref{sec:stretched_horizon_shocks}   &  section 2.2.1 and fig.~5   \\ 
 \rule{0pt}{4.9ex}  \textbf{Critical times} & sections \ref{ssec:critical_times_dS} and \ref{ssec:Plateau_of_complexity}  & section 3.3 \\[0.2cm]
\hline
\end{tabular}   
\caption{Summary of the sections where the conventions of our work and of reference \cite{Anegawa:2023dad} are stated.  } 
\label{tab:conventions}
\end{center}
\end{table}

\paragraph{Geometry.}
The conventions for the geometry and the null coordinates coincide with reference \cite{Anegawa:2023dad} after the mapping 
\beq
\mathcal{E} \leftrightarrow M \, , \qquad
\varepsilon \, \mathcal{E}_1 \leftrightarrow E \, ,
\eeq
between our (left-hand side) and their (right-hand side) notation.

\paragraph{Stretched horizon.}
The location  of the stretched horizon differs in the presence of a shockwave:
\begin{itemize}
    \item \textbf{Our conventions.} We define in section \ref{sec:stretched_horizon_shocks} various prescriptions, whose common feature is that the stretched horizon $r^{\rm st}_i \in [0, r_{C,i}]$ (here $i=1,2$) can approach the cosmological horizon of the corresponding part of the geometry, and is time-independent after crossing the shockwave.
    \item \textbf{Conventions of \cite{Anegawa:2023dad}.}
    The stretched horizon in reference \cite{Anegawa:2023dad} is not affected by the presence of the shockwave. It always varies along the interval $r^{\rm st} \in [0,\ell],$ with $\ell$ being the dS radius, independently of the cosmological horizon of the corresponding region.
\end{itemize}

\noindent
We define the left and right boundary times $(t_L, t_R)$ as the time coordinates running along the stretched horizons, while in reference \cite{Anegawa:2023dad} they are measured at the poles.
As a consequence, the constant value of the null coordinate $u_s$ for a shockwave inserted from the north pole is
\beq
u_s = \begin{cases}
    -t_w + r^*_2(r^{\rm st}_2) & \mathrm{our \,\, conventions} \\
    -t_w & \mathrm{reference \,\, [1]  }
\end{cases} 
\label{eq:conventions_tw}
\eeq
This is a minor change that only shifts the boundary times by a constant.

\paragraph{Critical times.}
The critical time $t_{\infty} = \ell \tau_{\infty}$ in eq.~(3.49) of reference \cite{Anegawa:2023dad} denotes the instant when the top joint $r_{m2}$ of the WDW patch (see section \ref{ssec:time_evo_WDW}) reaches future timelike infinity $\mathcal{I}^+$, in the absence of any cutoff regulator.
In our notation, this time corresponds to $t_{c2}$ defined in eq.~\eqref{eq:tc2} when $r_{\rm max,2} \rightarrow \infty$.
In fact, by summing  eqs.~\eqref{eq:identity_WDW_times2} and \eqref{eq:identity_WDW_times4} and using the identity $r^*(\infty) =0$, we obtain
\beq
t_{c2} = 2r^*_2(r_b) - 2r^*_1(r_b)  + r^*_1(r^{\rm st}_1) + r^*_2(r^{\rm st}_2) \, .
\eeq
This result is consistent with eq.~(3.49) of \cite{Anegawa:2023dad}, once we take into account the different prescriptions for the stretched horizon, the identity \eqref{eq:conventions_tw}, and map $r_b \leftrightarrow r_s$ between our and their conventions, respectively.

The critical times define the duration of the plateau regime $t_{\rm pl}$ in  eq.~\eqref{eq:def_plateau_time}.
In order to study the scrambling time and the linear approximation of the plateau for $t_w \gg L$, we take an ansatz in eq.~\eqref{eq:assumptions2_tc_dS}
which assumes
\beq
t_w \rightarrow \infty \, , \qquad
e^{- \frac{a_1}{L} t_w } \ll \varepsilon \qquad  (\mathrm{our \,\, conventions}) \, ,
\label{eq:double_scaling_us}
\eeq
and only afterwards we perform the limit $\varepsilon \rightarrow 0$.
The double scaling limit (2.58) in reference \cite{Anegawa:2023dad} focuses instead on the regime 
\beq
\varepsilon \rightarrow 0 \, , \qquad
t_w \rightarrow \infty \, , \qquad
\varepsilon  \, e^{\frac{a_1}{L} t_w } \quad \mathrm{fixed} \qquad
\mathrm{(reference \,\,[1] )} \, .
\label{eq:double_scaling_Norihiro}
\eeq
In order to investigate the limit \eqref{eq:double_scaling_Norihiro} in our analysis, we need to choose the following value for the energy of the shockwave:
\beq
\varepsilon = \frac{a_1^2}{2 G_N \mathcal{E}_1} \, e^{- \frac{a_1}{L} t_w} \, .
\label{eq:special_choice_epsilon}
\eeq
Plugging this value inside eq.~\eqref{eq:limit_scrambling_time_SdS3} gives
\beq
t_{\rm pl}|_{\mathrm{\eqref{eq:special_choice_epsilon}}} = 4(t_w - t_*) = \mathcal{O} (\varepsilon) \, ,
\eeq
which cancels the linear dependence in $t_w$ and the finite part and recovers the result~(3.63) of reference \cite{Anegawa:2023dad} (since $\beta \propto \varepsilon$). 
Therefore we observe that our and their analysis correspond to a different order of limits applied to a similar formula for the critical time.

\section{Ping-pong prescription for the constant redshift}
\label{app:ping_pong}

The procedure which defines the stretched horizon by imposing a constant cosmological redshift in section \ref{ssec:details_const_red} is subject to ambiguities, such as the location of the observer and the orientation of the light rays (ingoing/outgoing) used to fix the location of the stretched horizon. 
In the following, we will define the location of the stretched horizon by using the constant redshift prescription, but changing the boundary conditions compared to the analysis of section \ref{ssec:details_const_red}.
We will discuss the discrepancies and the similarities between the two cases.
Throughout this appendix, the terminology \emph{ingoing} and \emph{outgoing} will refer to light rays sent towards or outwards from the cosmological horizon, respectively.

\subsection{Details of the prescription}

\begin{figure}[ht]
    \centering
    \subfigure[]{\label{subfig:prescr1_SdS3altred} \includegraphics[scale=0.34]{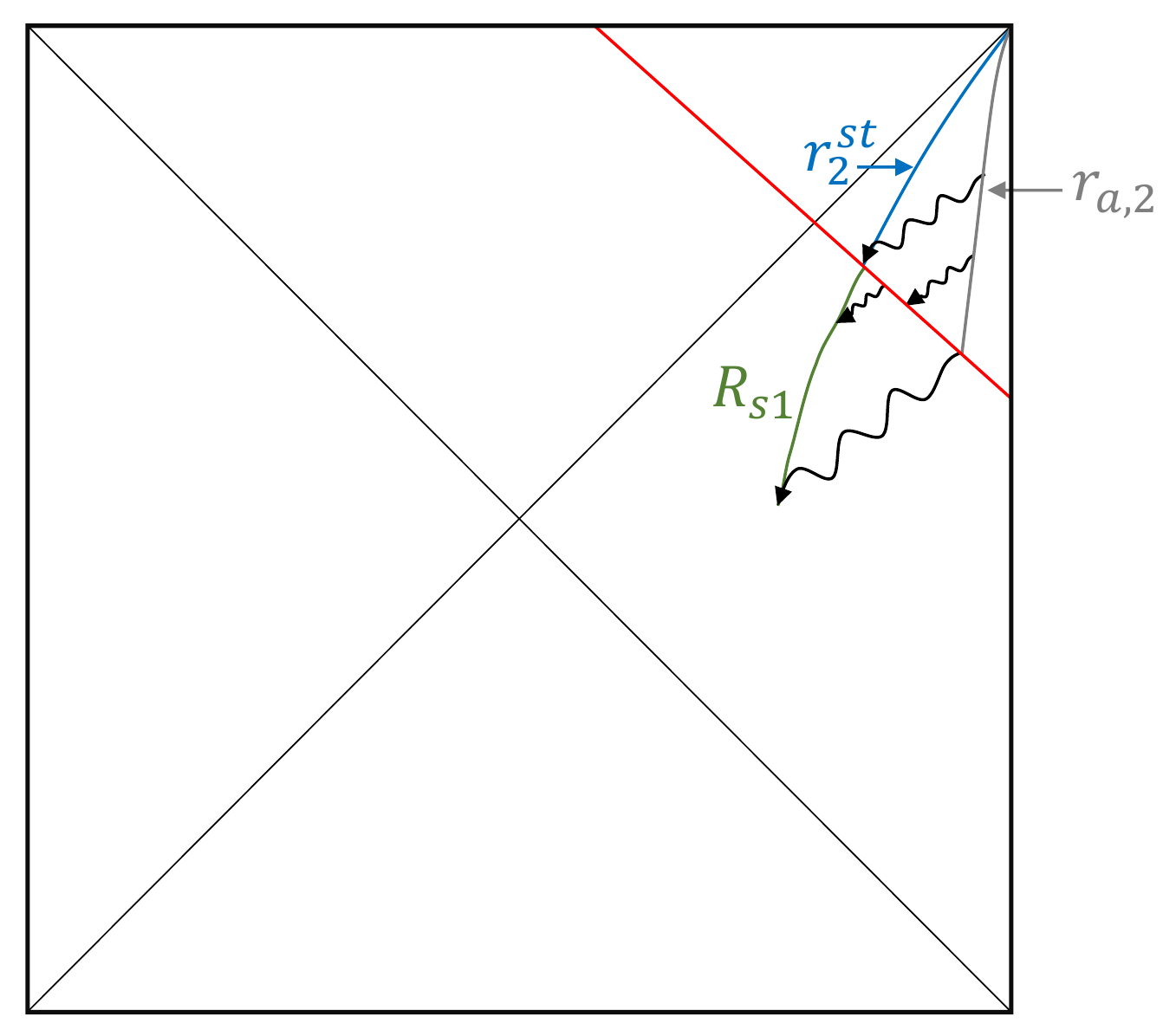} }
    \subfigure[]{\label{subfig:prescr2_SdS3altred} \includegraphics[scale=0.34]{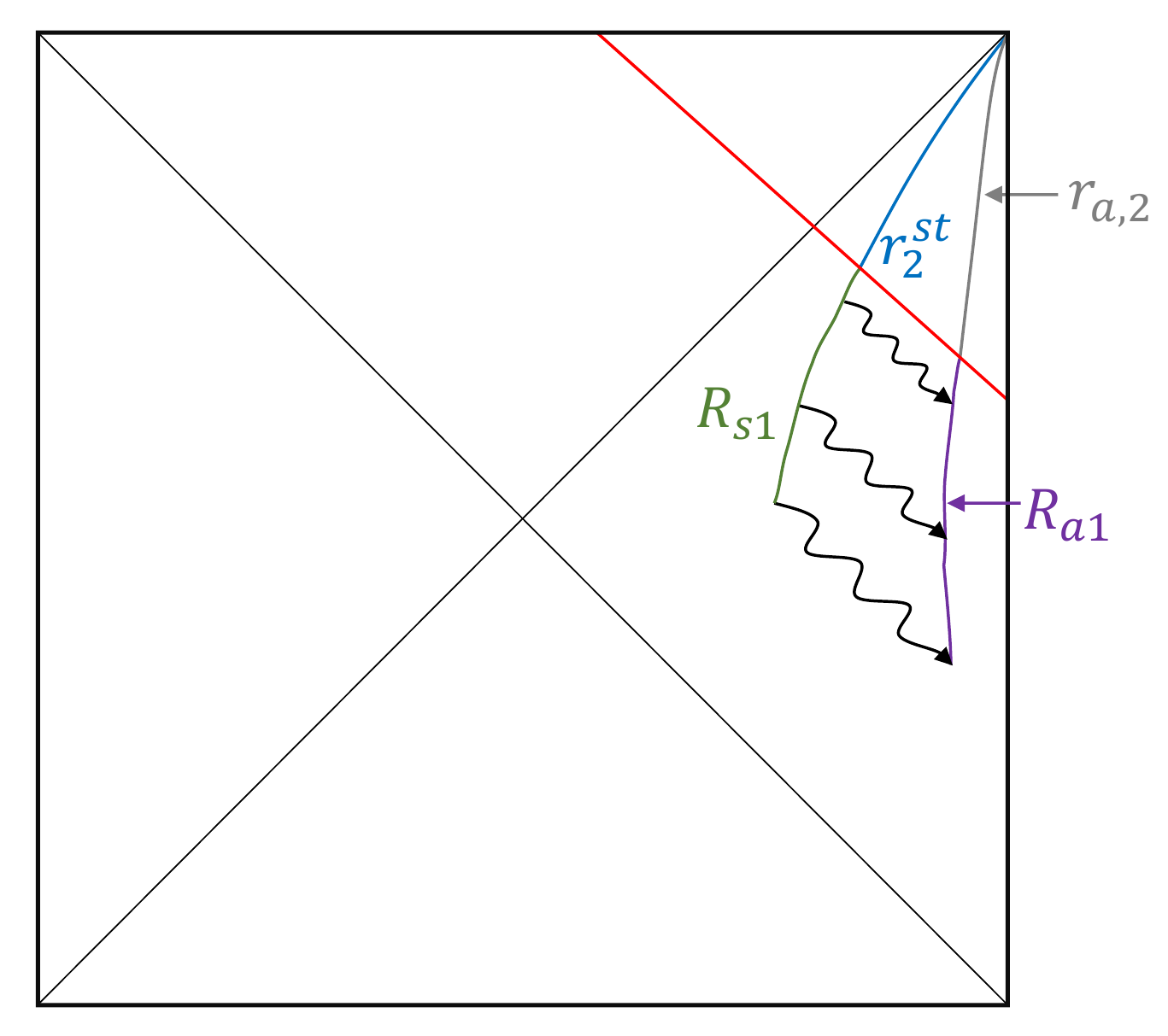} }
    \subfigure[]{\label{subfig:prescr3_SdS3altred} \includegraphics[scale=0.34]{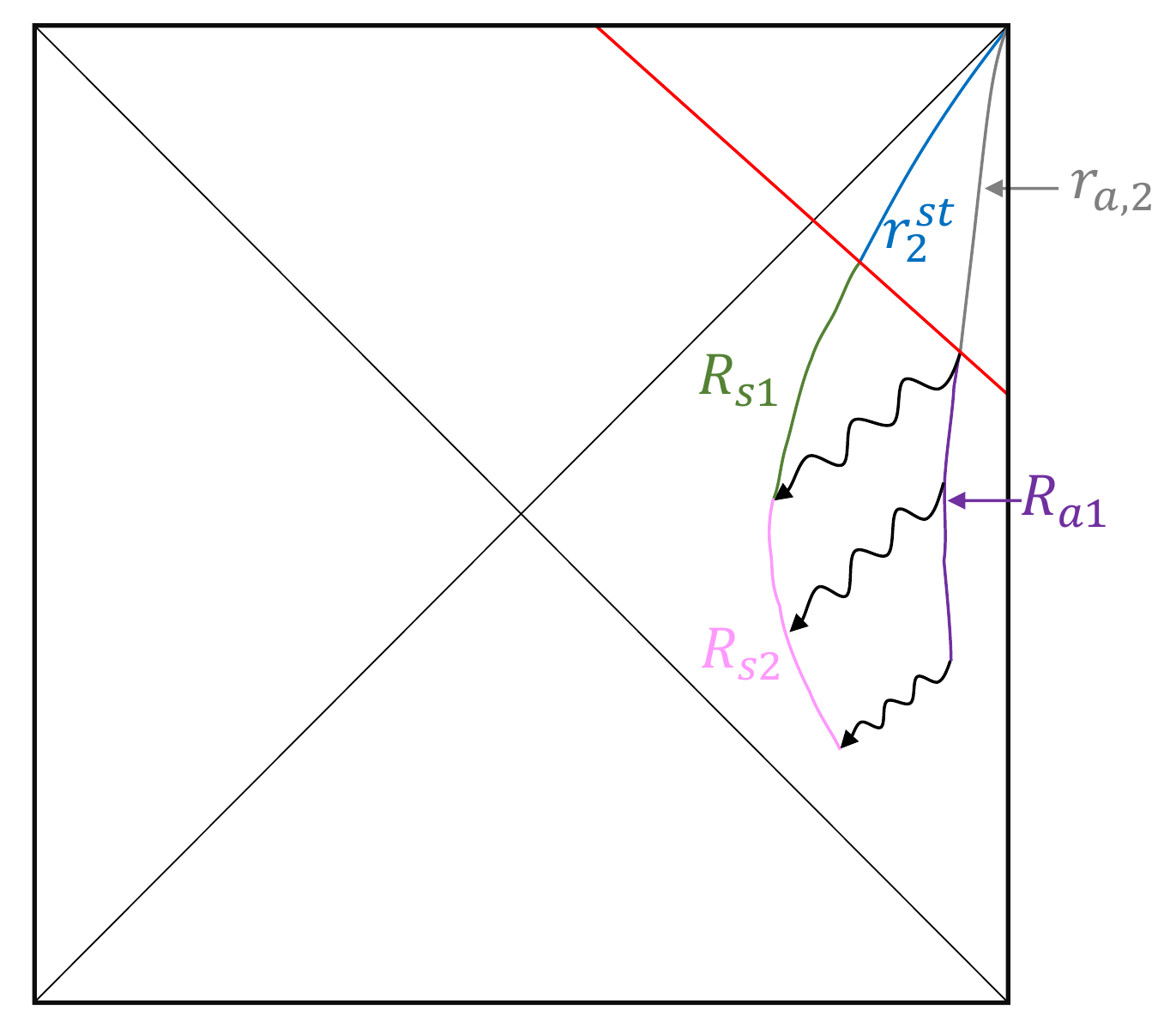} }
    \caption{Alternative prescription to define the stretched horizon in SdS$_3$ background. (a) First step: we start with time-independent curves located at $r^{\rm st}_2$ and $r_{a,2}.$ 
    Then we use ingoing light rays along the $v$ direction to define a time-dependent stretched horizon $R_{s,1}$ (green curve in the picture), requiring continuity when crossing the shockwave.
    (b) Second step: we send outgoing light rays along $u$ to reconstruct the time-dependent curve $R_{a,1}$, imposing continuity at the shockwave insertion. (c) Third step: using ingoing light rays at constant $v,$ we reconstruct another time-dependent part of the stretched horizon $R_{s,2}.$ The prescription is implemented recursively.
    The arrow along the light rays identifies if they are ingoing or outgoing. }
    \label{fig:newstr_SdS3}
\end{figure}

We consider for simplicity the setting with a transition between two SdS$_3$ black holes; the higher dimensional case is technically harder, but conceptually similar to the following analysis.
The procedure is schematically depicted in fig.~\ref{fig:newstr_SdS3}.
The idea is to define a (generically time-dependent) trajectory for two observers located along the curves $r=R_{s} (t_s)$ and $r=R_a(t_a)$ by using signals that they send and receive.
We refer to the former curve $R_s$ as the \emph{stretched horizon}, and to the latter $R_a$ as the \emph{auxiliary horizon}.
We will follow the time evolution backwards, starting from null future infinity and moving towards the past.
In the far future, we require that the trajectories are time-independent, \ie 
\beq
\lim_{t_s \rightarrow \infty} R_s (t_s) = r^{\rm st}_2 \, , \qquad
\lim_{t_a \rightarrow \infty} R_a (t_a) = r_{a,2}  \, ,
\label{eq:far_limits_stretched_ping_pong}
\eeq
and they satisfy $ r_{a,2} < r^{\rm st}_2 < r_{C2}$.
Contrarily to the method described in section \ref{ssec:details_const_red}, in this case the observers located along the two trajectories will repeatedly exchange the role of being the emitter and the receiver of light rays.
More precisely, we define the following steps:
\begin{enumerate}
    \item We send ingoing light rays at constant $v$ from the stretched horizon towards the auxiliary one.
    Once the boundary conditions $r^{\rm st}_2$ and $r_{a,2}$ are specified, we are able to reconstruct the time-dependent part of $R_s(t_s)$ represented in green in fig.~\ref{subfig:prescr1_SdS3altred}.   
    \item We proceed by sending outgoing light rays at constant $u$ from the green region of the stretched horizon $R_s(t_s)$ determined in point 1, in order to reconstruct the violet part of the auxiliary curve $R_a(t_a),$ see fig.~\ref{subfig:prescr2_SdS3altred}.
    \item We send ingoing light rays at constant $v$ from the violet part of $R_a(t_a)$ computed in step 2, and we rebuild the magenta part of the stretched horizon $R_s(t_s)$, as shown in fig.~\ref{subfig:prescr2_SdS3altred}.
    \item Finally, we apply steps 2 and 3 recursively.
\end{enumerate}

In this procedure, we impose that \emph{the stretched and auxiliary horizons are continuous when crossing the shockwave}.
This condition, combined with the requirement that the redshift is constant, is sufficient to uniquely determine the location of the two curves throughout the full backward time evolution.
Since we alternate the use of light rays at constant $u$ and $v$ and we repeatedly exchange the role of emitter and observer along the horizons, we will refer to this method as the \emph{ping-pong prescription}.

In the course of our analysis, we will find that the continuity requirement along the full time evolution is too strong, and leads to a solution for the stretched horizon inconsistent with our expectations.
Therefore, later on we will relax this this condition by allowing a jump of the stretched horizon when crossing the shockwave.
This will be studied in section \ref{app:ssec:results}.

\subsection{Implementation of the procedure}
\label{app:ssec:implementation_procedure}

In the following computations, the insertion time $t_w$ of the shockwave will be measured at the stretched horizon after the perturbation, according to eq.~\eqref{eq:def_us_appA}.
Furthermore, we specify that the time coordinate along the stretched and auxiliary horizons will always be measured before the shockwave insertion, so that they run continuously through all the backward time evolution.

\paragraph{Step 1.}
In the region after the shockwave, the curves are located at constant $R_s=r^{\rm st}_2$ and $R_a=r_{a,2}$, therefore the cosmological redshift \eqref{eq:redshift_timeindep} reads
\beq
1 + z_{\rm 0} = 
\sqrt{\frac{f_2(r_{a,2})}{f_2(r^{\rm st}_2)}} \, .
 \label{eq:step0_red_app}
\eeq
The non-trivial part of the procedure is to compute the stretched horizon in the green region of fig.~\ref{subfig:prescr1_SdS3altred}.
Using the general definition \eqref{eq:definition_redshift}, we find
 \beq
    1+z_{1}   = \sqrt{\frac{f_2(r_{o,2})}{f_1(R_{s,1} (t_{s,1}))}} \frac{1}{\sqrt{1-\frac{(R'_{s,1} (t_{s,1}))^2}{f_1(R_{s,1} (t_{s,1}))^2}}} \frac{dt_{a,0}}{dt_{s,1}} \, ,
    \label{eq:step1_red_app}
\eeq
where $t_{s,1}$ is the time measured along the trajectory $R_{s,1}$ and we denote the derivative as $ ' \equiv d/dt_{s,1}.$ 
We refer to $t_{a,0}$ as the time coordinate measured along the time-independent part of the auxiliary horizon $r_{a,2}$.
We relate the time coordinates along the stretched and auxiliary horizons by computing the constant value of the $v$ coordinate of an ingoing null ray crossing the shockwave in two different ways
\beq
\begin{aligned}
& t_{s,1} + r^*_1(R_{s,1} (t_{s,1})) = - t_w -r^*_2(r^{\rm st}_2) + 2r^*_1(r_{\rm sh}) \, , & \\
& -t_w - r^*_2(r^{\rm st}_2) + 2r^*_2(r_{\rm sh}) = t_{a,0} + r^*_2(r_{a,2}) \, , &
\label{eq:conditions1_red_app}
\end{aligned}
\eeq
where $r_\text{sh}$ is the intersection of the light ray with the shock. These relations allow to solve for $r_{\rm sh}$ and to express one time in terms of the other, \ie $t_{a,0}(t_{s,1})$. 

The configuration corresponding to the green curves depicted in fig.~\ref{subfig:prescr1_SdS3altred} applies during the time interval  $ \tilde{t}_{s1,c2} \leq t_{s,1} \leq \tilde{t}_{s1,c1} $ delimited by the critical times
\beq
\begin{aligned}
& \tilde{t}_{s1,c1} = -t_w - r^*_2(r^{\rm st}_2) + r^*_1(r^{\rm st}_2) \, , & \\
& \tilde{t}_{s1,c2} = -t_w -r^*_2(r^{\rm st}_2) - r^*_1(R_{s,1} (\tilde{t}_{s1,c2})) + 2 r^*_1(r_{a,2}) \, , &
\end{aligned}
\eeq
the latter identity being an implicit equation that we solve numerically.
The time $\tilde{t}_{s1,c1}$ is the instant when an ingoing ray at constant $v$ is received at the location of the shockwave, while $\tilde{t}_{s1,c2}$ corresponds to a null ray precisely emitted from the shock.
The light rays corresponding to these extremal configurations are the top and bottom ones depicted in fig.~\ref{subfig:prescr1_SdS3altred}, respectively.

By requiring $z_0=z_1$ in eqs.~\eqref{eq:step0_red_app} and \eqref{eq:step1_red_app} and extracting  $t_{a,0}(t_{s,1})$ from eq.~\eqref{eq:conditions1_red_app}, we then find a differential equation for $R_{s,1}(t_{s,1})$ with boundary condition $R_{s,1}(\tilde{t}_{s1,c1})=r^{\rm st}_2,$, which corresponds to the continuity of the stretched horizon when crossing the shockwave.

\vskip 5mm
\noindent
From now on, both the stretched and auxiliary horizons will always be time-dependent, and we will describe the procedure which recursively reconstructs their shape by evolving the system towards the past.

\paragraph{Recursive step 2: reconstruction of the auxiliary horizon.}
For $i\geq 1,$ we assume that the previous step in the procedure gave us the solutions $R_{s,i}(t_{s,1})$ and $t_{s,i}(t_{s,1})$, always expressed in terms of the time $t_{s,1}$ along the stretched horizon used in step 1 to parametrize the green curve in fig.~\ref{subfig:prescr1_SdS3altred}.
In this step we will reconstruct the region $R_{a,i}(t_{s,1})$ of the auxiliary horizon, and relate times along the auxiliary and stretched horizons as $t_{a,i}(t_{s,1})$.
When $i=1$, it means that we use the solution $R_{s,1}(t_{s,1})$ from step 1 to reconstruct the violet region of the auxiliary horizon in fig.~\ref{subfig:prescr2_SdS3altred}.

For $i \geq 1$, we use outgoing light rays at constant $u$ to find the identities
\begin{subequations}
\beq
t_{a,i} - r^*_1(R_{a,i}(t_{s,1})) = t_{s,i} (t_{s,1}) - r^*_1(R_{s,i}(t_{s,1}))  \, ,
\label{eq:condition2_app_red}
\eeq
\beq
    1+z_{2i}   = \sqrt{\frac{f_1(R_{a,i} (t_{s,1}))}{f_1(R_{s,i} (t_{s,1}))}} \frac{\sqrt{1-\frac{\le R'_{a,i} (t_{s,1}) \frac{dt_{s,1}}{dt_{a,i}} \ri^2}{f_1(R_{a,i}(t_{s,1}))^2}}}{\sqrt{1-\frac{\le R'_{s,i} (t_{s,1}) \frac{dt_{s,1}}{dt_{s,i}} \ri^2 }{f_1(R_{s,i} (t_{s,1}) )^2}}} \frac{dt_{a,i}}{dt_{s,1}} \frac{dt_{s,1}}{dt_{s,i}} \, ,
    \label{eq:step2_red_app}
\eeq
\end{subequations}
where the latter identity is the cosmological redshift \eqref{eq:definition_redshift}.
Then we perform the following steps:
\begin{itemize}
    \item Differentiate eq.~\eqref{eq:condition2_app_red} with respect to $t_{s,1}$ to get $dt_{a,i}/dt_{s,1}.$
    \item Plug $dt_{a,i}/dt_{s,1}$ inside eq.~\eqref{eq:step2_red_app} and solve the differential equation $z_0=z_{2i}$ for $R_{a,i}(t_{s,1}).$ 
    \item Plug the result back inside eq.~\eqref{eq:condition2_app_red} to get $t_{a,i}(t_{s,1}).$
\end{itemize}

Using this method, we expressed all the quantities as functions of the time $t_{s,1}$ defined along the green curve in fig.~\ref{fig:newstr_SdS3}.
Therefore the time interval characterizing this step is delimited by the same critical times $\tilde{t}_{s1,c2} \leq t_s \leq \tilde{t}_{s1,c1}$ used in step 1, and the boundary condition of the differential equation $z_0=z_{2i}$ is $R_{a,i}(\tilde{t}_{s1,c1})=R_{a,i-1}(\tilde{t}_{s1,c2}),$ corresponding to the continuity requirement for the auxiliary horizon along all the backward time evolution (including the transition at the shockwave).

\paragraph{Recursive step 3: reconstruction of the stretched horizon.}
For $i \geq 2,$ we assume that we computed in the previous step the functions $R_{a,i-1}(t_{s,1})$ and $t_{a,i-1}(t_{s,1}).$
We use this information to reconstruct another region $R_{s,i}$ and to relate times along the stretched horizon as $t_{s,i}(t_{s,1})$.
When $i=2$, it means that we reconstruct the magenta region in fig.~\ref{subfig:prescr3_SdS3altred}.

For any $i\geq 2,$ we send ingoing light rays at constant $v$ and we use the definition of redshift \eqref{eq:definition_redshift} to find
\begin{subequations}
\beq
t_{a,i-1}(t_{s,1}) + r^*_1(R_{a,i-1}(t_{s,1})) = t_{s,i} + r^*_1(R_{s,i}(t_{s,1})) \, ,
\label{eq:rays_constantv_step3_app}
\eeq
\beq
     1+z_{2i-1}   = \sqrt{\frac{f_1(R_{a,i-1}(t_{s,1}))}{f_1(R_{s,i}(t_{s,1}))}} \frac{\sqrt{1-\frac{\le R'_{a,i-1}(t_{s,1}) \frac{dt_{s,1}}{dt_{a,i-1}} \ri^2}{f_1(R_{a,i-1}(t_{s,1}))^2}}}{\sqrt{1-\frac{\le R'_{s,i}(t_{s,1}) \frac{dt_{s,1}}{dt_{s,i}} \ri^2}{f_1(R_{s,i} (t_{s,1}))^2}}} \frac{dt_{a,i-1}}{dt_{s,1}} \frac{dt_{s,1}}{dt_{s,i}}   \, .
    \label{eq:step3_red_app}
\eeq
\end{subequations}
We then proceed as follows:
\begin{itemize}
    \item Differentiate eq.~\eqref{eq:rays_constantv_step3_app} with respect to $t_{s,1}$ to get $\frac{dt_{s,i}}{dt_{s,1}}$.
    \item Plug $\frac{dt_{s,i}}{dt_{s,1}}$ inside eq.~\eqref{eq:step3_red_app} and then solve the differential equation $z_0=z_{2i-1}$ for $R_{s,i}(t_{s,1})$.
    \item Plug the result back in eq.~\eqref{eq:rays_constantv_step3_app} to find $t_{s,i}(t_{s,1}).$
\end{itemize}

In this way we expressed again all the quantities in terms of the time $t_{s,1}$ running along the green curve in fig.~\ref{fig:newstr_SdS3}.
Therefore, the critical times delimiting this configuration are still
$\tilde{t}_{s1,c1}, \tilde{t}_{s1,c2}$ and the boundary condition for the differential equation $z_0=z_{2i-1}$ reads $R_{s,i} (\tilde{t}_{s1,c1}) = R_{s,i-1} (\tilde{t}_{s1,c2})$.
This constraint imposes the continuity of the stretched horizon along all the past time evolution (including the intersection with the shockwave).

\subsection{Results}
\label{app:ssec:results}

\begin{figure}[ht]
    \centering
  \includegraphics[scale=0.82]{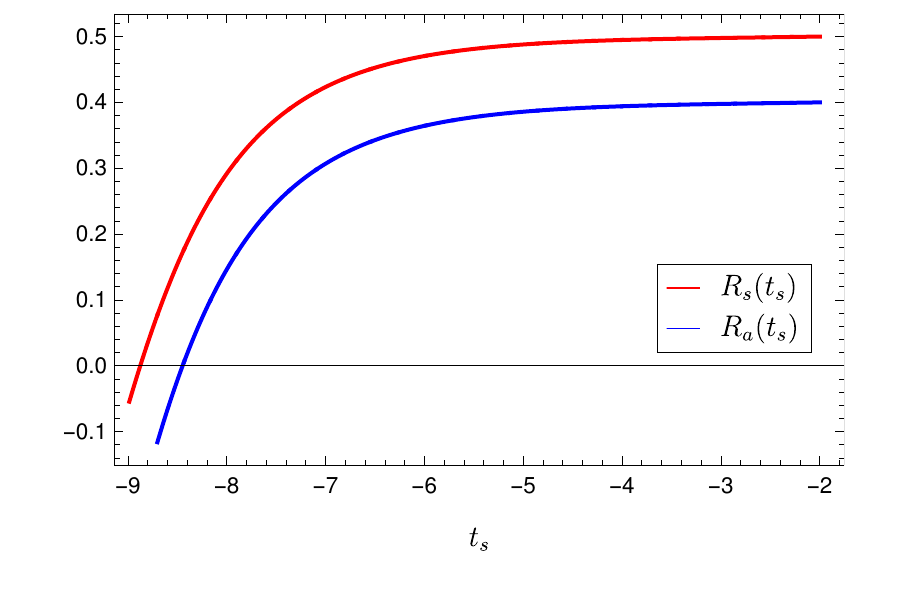} 
    \caption{Time-dependent part of the stretched $R_s$ and auxiliary $R_a$ horizons in the SdS$_3$ background, requiring continuity  when crossing the shockwave.
    We fix $r^{\rm st}_2 =0.5, r_{a,2}=0.4, 8 \pi G_N \mathcal{E}_1 =0.1, \varepsilon=0.05, L=1.$ }
    \label{fig:Rem_Rob_SdS3} 
\end{figure}

By applying the recursive procedure of section \ref{app:ssec:implementation_procedure} and imposing the continuity of the auxiliary and stretched horizons along all the time evolution, we obtain a numerical solution for their shape which is plotted in fig.~\ref{fig:Rem_Rob_SdS3}.
We find that the two horizons reach the north pole at finite and different times.

\begin{figure}[H]
    \centering
    \includegraphics[scale=0.82]{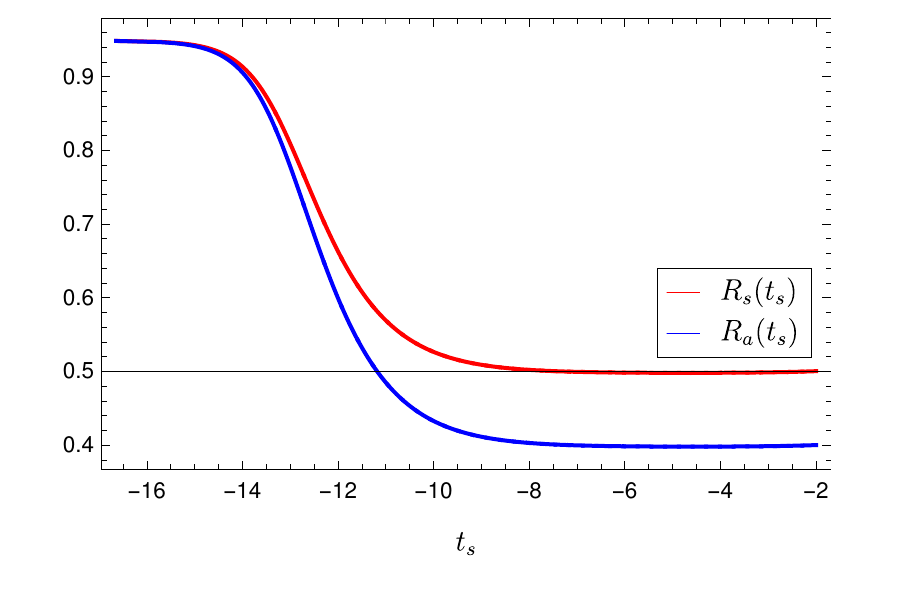} 
    \caption{Time-dependent part of the stretched $R_s$ and auxiliary  $R_a$ horizons in the SdS$_3$ background, with a jump $\delta \gtrsim \delta_{\rm cr}$ in the boundary condition according to eq.~\eqref{eq:jump_bdy_condition}. 
   }
    \label{fig:Rem_Rob_SdS3_jump}
\end{figure}

\begin{figure}[H]
    \centering
    \includegraphics[scale=0.78]{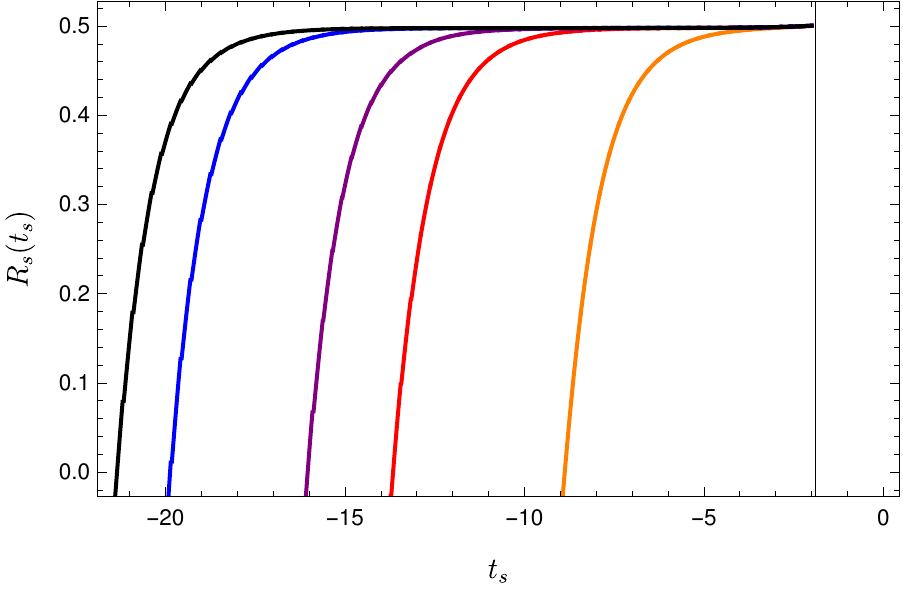}
    \includegraphics[scale=0.78]{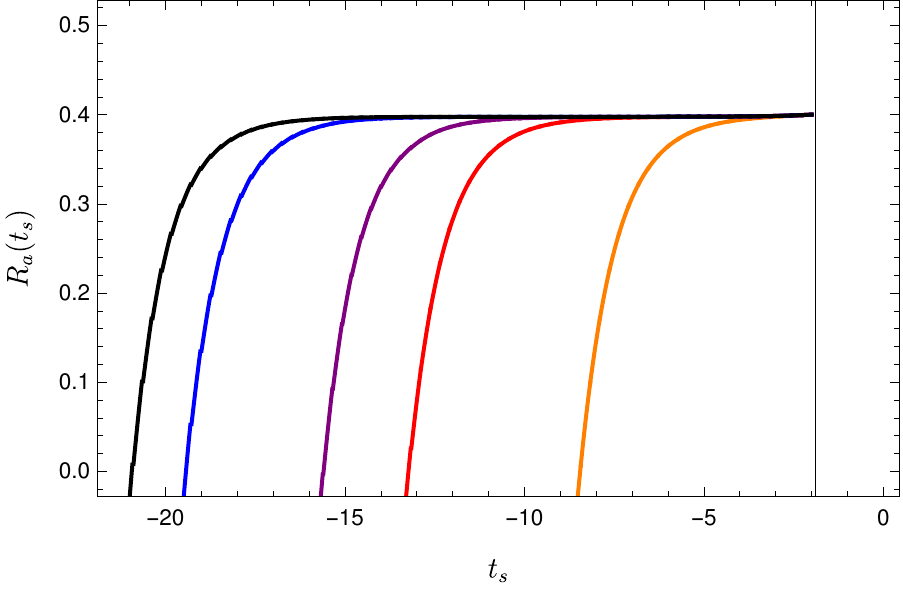}
    \caption{Time evolution of the stretched and auxiliary horizons for various values of the discontinuity $\delta$ in eq.~\eqref{eq:jump_bdy_condition}. The parameter $\delta$ increases for curves reaching the $r=0$ at earlier times (moving towards the left side of the plot). }
     \label{fig:Rem_Rob_longer_plateau}
\end{figure}

The previous result is somewhat unsatisfactory, because these curves do not reach past infinity.
Furthermore, one would expect from the constant redshift procedure of section \ref{ssec:details_const_red} that it should be possible (at least for certain choices of the parameters) to approach a trajectory at constant radial coordinate in the far past.
To this aim, we modify the boundary condition for the stretched horizon that we impose in step 1 of section \ref{app:ssec:implementation_procedure} when crossing the shockwave. In particular, we allow for a discontinuity $\delta >0$ of the form
\beq
R_s(t_{s1,c1}) = r^{\rm st}_2  ( 1+ \delta ) \, .
\label{eq:jump_bdy_condition}
\eeq
Studying various choices of $\delta$, we find three possibilities:
\begin{enumerate}
    \item When $\delta \lesssim \delta_{\rm cr} $, the solutions behave as in fig.~\ref{fig:Rem_Rob_SdS3}.
    For the choice of the parameters in the plot, we numerically find $\delta_{\rm cr} \simeq 0.000605...$
    \item When $\delta \gtrsim \delta_{\rm cr} $, both the stretched and auxiliary horizons approach the past cosmological horizon $r_{C1}$ as depicted in fig.~\ref{fig:Rem_Rob_SdS3_jump}.
    This result can be checked by using null coordinates (which are continuously defined in the relevant region of the Penrose diagram),
    showing that $u \rightarrow -\infty$, while $v$ reaches a finite value.
    \item The results in points 1 and 2 suggest the existence of a transition between the two regimes where the horizons either reach the north pole or the cosmological horizon in finite time.
    Indeed, it is possible to fine-tune the parameter $\delta \sim \delta_{\rm cr},$ such that there is a longer region where the two curves are approximately at constant radial coordinate, as shown in fig.~\ref{fig:Rem_Rob_longer_plateau}.
    Eventually, there is a particular value of $\delta$ when the two curves reach the pole at past null infinity.
\end{enumerate}

\noindent
The result described in the last point shows that it is possible to fine-tune the ping-pong prescription to find stretched and auxiliary horizons that are time-independent in the far past, as we required in eq.~\eqref{eq:far_limits_stretched_hor} for the constant redshift prescription.
This fact, combined with the time-independence of the curves after the shockwave, are then universal requirements that we can always impose when constructing the stretched horizon in shockwave geometries. 

\section{Complexity of formation for shockwaves in SdS}
\label{app:formation}

The complexity of formation is defined as the excess complexity in the presence of a shockwave, compared to that of empty dS space (without a shock), \ie
\beq
\Delta \mathcal{C} = \mathcal{C} \left({\rm SdS}_{\mathrm{shockwave}}\right)\big|_{t_L=t_R=0} - 
\mathcal{C} ({\rm dS})\big|_{t_L=t_R=0} \, .
\label{eq:general_formation}
\eeq
This corresponds to evaluating the complexity \eqref{eq:total_CV20_constred} at $t=0$ and then subtracting the CV2.0 result for dS without shocks.

The complexity of formation was extensively studied for AdS black hole backgrounds \cite{Chapman:2016hwi}  and for the AdS shockwave geometries (\ie AdS-Vaidya backgrounds) in \cite{Stanford:2014jda,Chapman:2018lsv}. In \cite{Chapman:2016hwi}, the authors considered the excess in complexity for a black hole background, compared to two copies of empty AdS. In this case, it was found that for boundary dimensions $d>2$, the difference in the complexities grows linearly with the thermal entropy at high temperatures. For the special case $d=2$, the complexity of formation was found to be a fixed constant. 
In \cite{Stanford:2014jda,Chapman:2018lsv}, the complexity of formation was studied for black hole backgrounds in the presence of shockwaves. Here the authors asked, how much more complex is the state at $t_L=t_R=0$ if a perturbation was inserted on the right side at the time $-t_w$ in the past, compared to such a state in the absence of the perturbation. The complexity interpretation, as well as the gravitational calculation, suggest that as $t_w$ increases, the complexity is roughly a constant (in fact it grows exponentially, but this exponential is suppressed by a tiny prefactor) up to the scrambling time, at which point it starts increasing linearly.

We would like to study the complexity of formation in SdS backgrounds in the presence of shockwaves and check if it has similar properties to those described above for AdS spacetimes. In particular, we would like to use it to read the scrambling time for the would-be boundary theory and compare it to the results obtained from the time dependence of complexity.

\subsection{General setting}

We compute the complexity of formation for the SdS$_{d+1}$ geometry perturbed by a shockwave with $d \geq 2$. We numerically observe that $ t_{c1} < 0 < t_{c2}, $ therefore the complexity of formation corresponds to the intermediate regime where the CV2.0 is finite and admits a plateau in the growth of the spacetime volume as a function of time.
The relevant quantities needed for the calculation are the following:
\begin{itemize}
    \item The special positions of the WDW patch, determined by eqs.~\eqref{eq:rs_analytic_SdS3}--\eqref{eq:rm2_analytic_SdS3} evaluated at $t_R=t_L=0.$
    \item The formal expression for the complexity given by eq.~\eqref{eq:total_CV20_constred} for the usual configurations of the WDW patch, and by eq.~\eqref{eq:total_CV20_alt5} when the joints move behind the stretched horizons.
\end{itemize}

\noindent
From these expressions, we notice that there always exists a sufficiently high value of $t_w$ such that the joints of the WDW patch move outside the cosmological horizon and reach the stretched horizon, thus producing the configuration in fig.~\ref{fig:alternative4_WDWpatch}.
Therefore, it is important to consider both regimes in the evaluation of the complexity of formation.
The standard computation corresponds to eq.~\eqref{eq:total_CV20_constred} evaluated at $t_R=t_L=0,$ that is

\beq
\begin{aligned}
 \mathcal{C}_{V2.0} (0) & =   \frac{\Omega_{d-1}}{G_N L^2}  \left[ \int_{r^{\rm st}_2}^{r_s} dr \, r^{d-1} \le 2 r^*_2 (r) - 2 r_2^* (r^{\rm st}_2)   \ri 
    +  \int_{r_s}^{r_b} dr \, r^{d-1}  t_w    \right. \\
     & \left. + \int_{r_{C1}}^{r_b} dr \, r^{d-1} \le - t_w + 2 r^*_1(r) - 
     r^*_1 (r^{\rm st}_1) - r^*_2 (r^{\rm st}_2)   \ri
     + \int_{r_s}^{r_{C1}} dr \, r^{d-1} \le  2 r^*_1(r) -2  r^*_1 (r_s)   \ri \right. \\
     & \left.  +  \int_{r^{\rm st}_1}^{r_{C1}} dr \, r^{d-1} \le 2 r^*_1 (r) - 2 r^*_1 (r^{\rm st}_1) \ri 
     + \int_{r_b}^{r_{ m2}} dr \, r^{d-1} \le  t_w +  2 r_2^* (r) - 2 r^*_2 (r_b) \ri \right. \\
     & \left.  
     + \int_{r_{C1}}^{r_{m1}} dr \, r^{d-1} \le t_w  + 2 r^*_1 (r) - 2 r^*_1 (r_{s}) - r^*_1(r^{\rm st}_1) + r^*_2 (r^{\rm st}_2) \ri 
    \right] \, .
   \label{eq:starting_formula_CVform_dS}
\end{aligned}
\eeq
When the configuration with the joints behind the stretched horizon is achieved, we use instead the expression \eqref{eq:total_CV20_alt5} at $t_R=t_L=0$, which reads
\beq
\begin{aligned}
\mathcal{C}_{V2.0} (0) & = \frac{\Omega_{d-1}}{G_N L^2} \left[   
\int_{r^{\rm st}_2}^{r_b} dr \, r^{d-1} \, \le t_w + 2 r^*_2(r^{\rm st}_2)  - 2 r^*_2 (r_b) \ri \right. \\
& \left. + \int_{r_b}^{r_s} dr \, r^{d-1} \, \le  t_w + 2 r^*_2(r^{\rm st}_2)  - 2 r^*_2 (r) + 2 r^*_1(r) - 2 r^*_1 (r_s) \ri  \right. \\
& \left. + \int_{r^{\rm st}_1}^{r_b} dr \, r^{d-1} \, \le  t_w+ r^*_1(r^{\rm st}_1)  + r^*_2(r^{\rm st}_2)  - 2 r^*_1 (r_s) \ri 
\right] \, .
\end{aligned}
\label{eq:total_CV20_alt5_form}
\eeq
We remark that the results \eqref{eq:starting_formula_CVform_dS} and \eqref{eq:total_CV20_alt5_form} are valid in any number of dimensions $d \geq 2$.

\subsection{Three-dimensional SdS space}

\begin{figure}[H]
    \centering
    \subfigure[]{\includegraphics[scale=0.82]{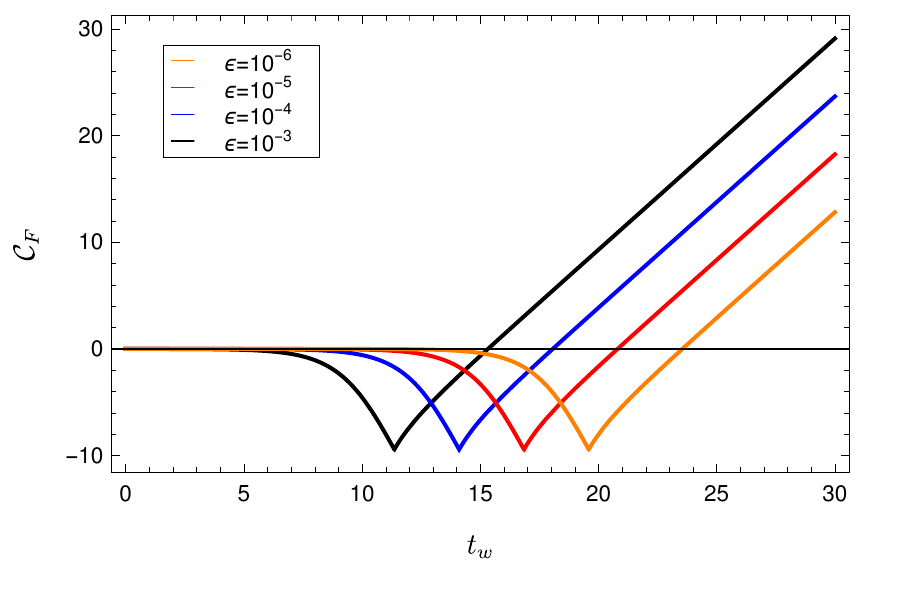}} 
    \caption{Complexity of formation as a function of $ t_w,$ for various choices of $\varepsilon$ at fixed $L=1,\rho=0.5, G_N\mathcal{E}_1=0.02.$ }
    \label{fig:plot_CV20_form_SdS3}
\end{figure}

Now we specialize to the three-dimensional case.
The relevant integrals to perform for SdS$_3$ are
\beq
\begin{aligned}
& \int dr \, r \, r^*_1(r) = \frac{L}{4 a_1} \left[   a_1^2 L^2 \log \left| \frac{a_1 L-r}{a_1 L+r} \right| + 2 a_1 L r + r^2  \log \left| \frac{a_1 L+r}{a_1 L-r} \right| \right] \, , & \\
& \int dr \, r \, r^*_2(r) = \frac{L}{4 a_2} \left[   a_2^2 L^2 \log \left| \frac{a_2 L-r}{a_2 L+r} \right| + 2 a_2 L r + r^2  \log \left| \frac{a_2 L+r}{a_2 L-r} \right| \right] \, . &
\label{eq:integrals_SdS3}
\end{aligned}
\eeq
Once this recipe is applied, the complexity of formation \eqref{eq:general_formation} according to the CV2.0 prescription is given by\footnote{One can check that once we set $\varepsilon=0,$ the complexity becomes independent of $t_w$. This is consistent with the fact that there is no shockwave in the geometry without deformation.}
\beq
\Delta \mathcal{C}_{V2.0} = \mathcal{C}_{V2.0;\,{\rm SdS}}\big|_{t=0 \atop \varepsilon\ne 0} - \mathcal{C}_{V2.0;\,{\rm dS}}\big|_{t=0 \atop \varepsilon= 0} \, ,
\label{eq:compl_CV20_form}
\eeq
where we subtract the dS$_3$ contribution.
We depict the dependence of the complexity of formation on $t_w$ for various choices of the parameters $\varepsilon$ in fig.~\ref{fig:plot_CV20_form_SdS3}. 

Due to the transition in the definitions of $r_{m1}$ and $r_{m2}$ in eqs.~\eqref{eq:rm1_analytic_SdS3} and \eqref{eq:rm2_analytic_SdS3} (taking $t_L=t_R=0$),  the functional dependence for small $t_w$ is smooth until it reaches a kink in correspondence with the transition $r_s=r_b$.
This is the case when the joints reach precisely the stretched horizon and the spacetime volume of the WDW patch vanishes.

\paragraph{Linear approximation.}

To get a better understanding of the complexity of formation as a function of the physical data, we perform the integrations explicitly (again focusing on $d=2$). 
The computation at a generic time is a straightforward but tedious exercise, which does not give particular physical insights.
It is more relevant to consider the limiting case $t_w \rightarrow \infty,$ which corresponds to a regime where the joints of the WDW patch move behind the stretched horizons. 

In this limit, the following simplifications occur
\beq
\lim_{t_w \rightarrow \infty} r_s = a_2 L \, , \quad
\lim_{t_w \rightarrow \infty} r_b = a_1 L \, , \quad
\lim_{t_w \rightarrow \infty} r_{m1} = r^{\rm st}_1 \, , \quad
\lim_{t_w \rightarrow \infty} r_{m2} = r^{\rm st}_2 \, .
\label{eq:simplifications_twinf_limit_CF_SdS3}
\eeq
As a consequence, the CV2.0 evaluated at $t=0$ becomes
\beq
\begin{aligned}
\lim_{t_w \rightarrow \infty} \mathcal{C}_{2.0V} (0) & = \frac{2 \pi}{G_N L^2} \left[   
\int_{r^{\rm st}_2}^{a_1 L} dr \, r \, \le t_w + 2 r^*_2(r^{\rm st}_2)  - 2 r^*_2 (a_1 L) \ri \right. \\
& \left. + \int_{a_1 L}^{a_2 L} dr \, r \, \le  t_w + 2 r^*_2(r^{\rm st}_2)  - 2 r^*_2 (r) + 2 r^*_1(r) - 2 r^*_1 (a_2 L) \ri  \right. \\
& \left. + \int_{r^{\rm st}_1}^{a_1 L} dr \, r \, \le  t_w+ r^*_1(r^{\rm st}_1)  + r^*_2(r^{\rm st}_2)  - 2 r^*_1 (a_2 L) \ri 
\right] \, .
\end{aligned}
\eeq
An analytic expression can be obtained after performing the explicit integrations according to eq.~\eqref{eq:integrals_SdS3}:
\beq
\mathcal{C}_{2.0V} (0) = \frac{\pi}{2 G_N}   \le  1 - \rho^2 \ri \le a_1^2 + a_2^2 \ri \le t_w - t_* \ri \, ,
 \label{eq:Cf_largetw_SdS3}
\eeq
with scrambling time 
\beq
\begin{aligned}
t_*   =  \frac{L}{a_1^2 + a_2^2 } \left[ 
 \frac{1}{2} \le a_1+ 2a_2 + \frac{a_1^2}{a_2} \ri  \log \le \frac{1-\rho}{1+\rho} \ri + \le a_1+a_2 \ri \log \le \frac{a_1+a_2}{a_2-a_1} \ri
\right] \, .
\end{aligned}
\label{eq:scrambling_time_Cf_SdS3}
\eeq
Note that these expressions are only valid when $t_w \gg L$. It is interesting that this expression is very similar to eq.~\eqref{eq:duration_plateau_SdS3} up to the precise form of the prefactors in front of the two logarithms.
In the limit $\varepsilon \rightarrow 0,$ we only keep the leading logarithmic divergence coming from the second term to find
\beq
t_* \underset{\varepsilon \ll 1}{\approx}   \frac{L}{a_1} \log \le \frac{a_1^2}{2 G_N \mathcal{E}_1} \frac{1}{\varepsilon} \ri   = 
 \frac{1}{2 \pi T_1} \log S_1  \, .
\label{eq:simplified_scrambling_time_Cf_SdS3}
\eeq
In particular, in the last equality, we recognize the Hawking temperature of the black hole before the shockwave insertion, and we required the energy of the shockwave perturbation to be proportional to a few thermal quanta.
This result matches the behavior found for the scrambling time in the same regime (\ie for tiny $\varepsilon$, the leading $\varepsilon$ divergence in the scrambling time matches) determined for the plateau regime in eq.~\eqref{eq:limit_scrambling_time_SdS3}. 

\subsection{Higher-dimensional SdS space}

It is difficult to perform an explicit computation of the complexity of formation in higher dimensions ($d \geq 3$), because the tortoise coordinate $r^*(r)$ becomes cumbersome and a closed form for the integrals in eqs.~\eqref{eq:starting_formula_CVform_dS} and \eqref{eq:total_CV20_alt5_form} is not always achievable.
However, we can attain an analytic expression for the scrambling time when the following assumptions hold:
\begin{itemize}
    \item The insertion time of the shock wave is large compared to the dS radius $t_w \gg L$.
    \item We perform the double-scaling limit \eqref{eq:double_scaling_limit}.
    \item We consider small black holes, such that $r_{h,i} / r_{C,i} \ll 1$ for $i=1,2$.
    % \item The stretched horizon after the shock wave is determined according to the continuous time prescription described in section \ref{ssec:details_cont_time}. In particular, this implies that $r_1^*(r^{\rm st}_1) = r_2^*(r^{\rm st}_2),$ see discussion around eq~\eqref{eq:relation_cont_time}.
\end{itemize}

\noindent
The previous assumptions lead to the following simplifications:
\beq
r_i^*(r) \approx \frac{r_{C,i}}{2} \log \left| \frac{r_{C,i} +r}{r_{C,i} - r}  \right|  \, , \qquad
r^{\rm st}_i \approx \rho r_{C,i} \,  \qquad (i=1,2) 
\label{eq:approx_tortoise_higherd}
\eeq
Equation~\eqref{eq:approx_tortoise_higherd} contains all the data entering the complexity of formation for large insertion times $t_w \gg L,$ as evaluated in eq.~\eqref{eq:total_CV20_alt5_form}.
At this point, the computation of the complexity of formation is straightforward and follows similar steps as those performed below eq.~\eqref{eq:integrals_SdS3}.
The result reads
\beq
\mathcal{C}_{2.0V} (0) \approx \frac{\Omega_{d-1} (r_{C1})^{d-2}}{2 G_N} \le 1-\rho^2 \ri \le t_w - t_* \ri \, ,
\eeq
with scrambling time
\beq
t_* \approx r_{C1} \log \le \frac{r_{C1}}{\beta r_{\rm cr}} \frac{1-\rho}{\varepsilon} \ri = 
\frac{1}{2 \pi T_{C1}}  \log \le \frac{r_{C1}}{\beta r_{\rm cr}} \frac{1-\rho}{\varepsilon} \ri  \, ,
\label{eq:scrambling_formation}
\eeq
where $\beta r_{\rm cr}$ is defined by the expansion \eqref{eq:expansion_radii_SdS4}.
Remarkably, this coincides with the result \eqref{rocker4} in the limit of small black hole horizons, \ie we replace
$r_{C1}-r_{h1}\approx r_{C1}$. 
Therefore in the regime described above, we find a precise matching in any bulk dimension between the scrambling time associated to the linear regimes ($t_w \gg L$) for either the duration of the plateau region or the complexity of formation.
A similar universal behaviour was found in the AdS case as well \cite{Chapman:2018lsv}.

\bibliographystyle{JHEP}

\bibliography{bibliography}

\end{document}